\documentclass[]{aa}
\usepackage{natbib,epsfig,txfonts,graphicx,afterpage}
\bibpunct{(}{)}{,}{a}{}{,} 

\def\HeI {He{\sc i}}
\def\HeII{He{\sc ii}}
\def\HeIII{He{\sc iii}}
\def\NIII{N{\sc iii}}
\def\NIV{N{\sc iv}}
\def\NV{N{\sc v}}

\def\PV{P{\sc v}}
\def\OVI{O{\sc vi}}
\def\Teff{$T_{\rm eff}$}
\def\logg{$\log g$}
\def\YHe   {$Y_{\rm He}$}

\def\Rsun {$R_{\odot}$}
\def\Rsune {R_{\odot}}
\def\Msun {${\rm M}_{\odot}$}

\def\LL {$\log (L/L_{\odot})$}

\def\Rstar{$R_{\ast}$}

\def\Rsuna{$\rm{R_{\odot}}$}

\def\Mdot{${\dot M}$}

\def\Dmom{D_{\rm mom}}
\def\Qr{Q_{\rm res}}

\def\MV {M$_{\rm v}$}

\def\vsini {$v \sin i$} 
\def\vmac {$v_{\rm macro}$} 

\def\vinf {$v_{\rm \infty}$}
\def\Vt {$v_{\rm{turb}}$}

\def\Ha {H$_{\rm \alpha}$}

\def\Bra{Br$_{\rm \alpha}$}
\def\Brg{Br$_{\rm \gamma}$}
\def\Pfg{Pf$_{\rm \gamma}$}

\def\kms {km~s$^{-1}$}
\def\mdu{$\cdot 10^{-6}\, {\rm M_{\odot}/yr}$} 
\def\Mdu1{$10^{-6}\, {\rm M_{\odot}/yr}$}

\def \um {$\mu$m }

\def\rarrow{$\rightarrow$}

\def\fcl{f_{\rm cl}}
\def\fv{f_{\rm v}}

\def\beq{\begin{equation}}
\def\eeq{\end{equation}}
\def\beqa{\begin{eqnarray}}
\def\eeqa{\end{eqnarray}}

\def \taur {\tau_{\rm Ross}}

\def \Rstare {R_\star}

\def \Teffe {T_{\rm eff}}

\def \vinfe {v_\infty}
\def \Mdote {\dot M}

\def\Lstar{$L_{\ast}$}

\def\Msunyr{\hbox{M$_\odot\,$yr$^{-1}$}}

\defcitealias{repo04}{REP04}
\defcitealias{mokiem05}{MOK05}
\defcitealias{repo05}{REP05}
\defcitealias{puls06a}{PUL06}

\begin{document}

\title{$L$-band spectroscopy of Galactic OB-stars\thanks{ Based on
observations collected at the European Organisation for Astronomical
Research in the Southern Hemisphere, Chile, under program ms ID 076.D-0149
(L-band ISAAC) and 266.D-5655(A) (UVES optical spectra)}}

\author{F. Najarro\inst{1}, M.M. Hanson\inst{2}\thanks{Visiting Astronomer
at the Infrared Telescope Facility, which is operated by the University of
Hawaii under Cooperative Agree\-ment no. NCC 5-538 with the National
Aeronautics and Space Ad\-minis\-tration, Science Mission Directorate, Planetary
Astronomy Program.} and J. Puls\inst{3}}

\offprints{F. Najarro}

\institute{
Centro de Astrobiolog\'{\i}a, (CSIC-INTA), Ctra. Torrej\'on  Ajalvir km\,4,
28850 Torrej\'on de Ardoz, Spain, \email{najarro@cab.inta-csic.es}
\and
Department of Physics, University of Cincinnati, PO Box 21001, Cincinnati,
Ohio, 45221-0011, USA, \email{hansonmm@ucmail.uc.edu} 
\and
Universit\"atssternwarte M\"unchen, Scheinerstr. 1, 
  D-81679 M\"unchen, Germany, \email{uh101aw@usm.uni-muenchen.de}
}

\date{Received; accepted }

\abstract {Mass-loss, occurring through radiation driven supersonic winds, is 
a key issue throughout the evolution of massive stars. 
Two outstanding problems are currently challenging the theory of 
radiation-driven winds: {{\it wind clumping}} and the {{\it weak-wind problem}}.}
{We seek to obtain accurate mass-loss rates of OB stars at different
evolutionary stages to constrain the impact of both problems in our current
understanding of massive star winds.}
{We perform a multi-wavelength quantitative analysis of a sample of ten
Galactic OB-stars by means of the atmospheric code {\sc cmfgen}, with special
emphasis on the $L$-band window. A detailed investigation is carried out on
the potential of \Bra\ and \Pfg\ as mass-loss and clumping diagnostics.}
{For objects with dense winds, \Bra\ samples the intermediate wind
while \Pfg\ maps the inner one. In combination with other indicators (UV,
\Ha, \Brg) these lines enable us to constrain the wind clumping structure and
to obtain ``true'' mass-loss rates. For objects with weak winds, \Bra\ emerges
as a reliable diagnostic tool to constrain \Mdot. The emission component at
the line Doppler-core superimposed on the rather shallow Stark absorption
wings reacts very sensitively to mass loss already at very low \Mdot\
values. On the other hand, the line wings display similar sensitivity to
mass loss as \Ha, the classical optical mass loss diagnostics.}
{Our investigation reveals the great diagnostic potential of $L$-band
spectroscopy to derive clumping properties and mass-loss rates of hot star 
winds. We are confident that \Bra\ will become the primary diagnostic tool
to measure very low mass-loss rates with unprecedented accuracy.}


\titlerunning{$L$-band spectroscopy of Galactic OB-stars}
\authorrunning{F. Najarro, M.M. Hanson \& J. Puls}

\maketitle

\section{Introduction}
\label{intro}

In the last decade, massive stars ($M_{\rm ZAMS} \ga 10 M_{\odot}$) have
(re-) gained considerable interest among the astrophysical community,
particularly because of their role in the development of the early Universe
(e.g., its chemical evolution and re-ionization) and as (likely) progenitors
of long gamma-ray bursters. Present effort concentrates on modeling various
dynamical processes in the stellar interior and atmosphere (mass loss,
rotation, magnetic fields, convection, and pulsation). Key in this regard is
the {\it mass loss} that occurs through supersonic winds, which modifies
evolutionary timescales, chemical profiles, surface abundances and
luminosities. A well-known corollary in massive star physics is that a
change of their mass-loss rates by only a factor of two has a dramatic
effect on their evolution \citep{meynet94}.

The winds from massive stars are described by the radiation-driven wind
theory \citep{cak, fa86, ppk}. Albeit its apparent success (e.g.,
\citealt{vink00, puls03}), this theory is presently challenged by two
outstanding problems (reviewed by \citealt{Puls08}), the {\it clumping} and
the {\it weak wind} problem.

\paragraph{The clumping problem.} During recent years, various
evidence\footnote{For details, we refer to the proceedings of the
international workshop on ``Clumping in Hot Star Winds'' \citep{Hamann08}.}
has been accumulated that hot star wind are not smooth, but clumpy, i.e.,
that they consist of density inhomogeneities which redistribute the matter
into clumps of enhanced density, embedded in an almost rarefied medium.

Theoretically, the presence of such {\it small-scale} 
structure\footnote{not to be confused with large scale structure which is
indicated by the ubiquitous presence of recurrent wind profile variability
in the form of discrete absorption components (DACs, e.g.,
\citealt{prinhow86, kaper96, LobelBlomme08}) and ``modulation features''
(e.g., \citealt{Fullerton97}).} has been expected since the first
hydrodynamical wind simulations
 \citep{ocr}, due to the presence of a
strong instability inherent to radiative line-driving. This can lead to the
development of strong reverse shocks, separating over-dense clumps from
fast, low-density wind material. Interestingly, however, the column-depth
averaged densities and velocities remain very close to the predictions of
stationary theory (see also \citealt{Feldmeier95}. For more recent results,
consult \citealt{run02, run05} (1-D) and \citealt{desowo03, desowo05}
(2-D)). At least for OB-stars, however, a {\it direct, observational}
evidence in terms of line profile variability has been found only for two
objects so far, the Of stars $\zeta$~Pup and HD\,93129A \citep{evers98,
Lepine08}. 

{\it Indirect} evidence for small-scale clumping, on the other hand, is
manifold, and is mostly based on the results from quantitative spectroscopy,
using NLTE model atmospheres. In order to treat wind-clumping in the present
generation of atmospheric models, the {\it standard} assumption of the
so-called ``microclumping model'' relates to the presence of optically thin
clumps and a void inter-clump medium.\footnote{The importance of a {\it
low-density} inter-clump medium for the production of \OVI\ has been
outlined already by \citet{Zsargo08}.} A consistent treatment of the
disturbed velocity field is still missing. The over-density (with respect to
the average density) inside the clumps is described alternatively by a
volume filling factor, $\fv$, or a clumping factor, $\fcl \ge 1$ , which in the
case of a void interclump-medium, are related via $\fv=\fcl^{-1}$. The most
important consequence of such a structure is that any mass-loss rate, \Mdot,
derived from density-squared dependent diagnostics (\Ha, \Bra\ or free-free
radio emission, involving recombination-based processes) using homogeneous
models needs to be scaled down by a factor of $\sqrt{\fcl}$.

Based on this approach, \citet{Crowther02, hil03} and \citet{bouret03,
bouret05} derived clumping factors of the order of 10{\ldots}50, with
clumping starting at or close to the wind base. From these values, a
reduction of (unclumped) mass-loss rates by factors 3{\ldots}7 seems to
be necessary (see also \citealt{repo04}).

Even worse, the analysis of the FUV \PV-lines by \citet{fulli06} seems to
imply factors of 10 or even more (but see also \citealt{Waldron10} who
argued that the ionization fractions of \PV\ could be seriously affected by
XUV radiation). However, as suggested by
\citet{Oskinova07}, the analyses of such optically thick lines might
require the consideration of wind ``porosity'', which reduces the
effective opacity at optically thick frequencies \citep{owo04}. Moreover,
the porosity in velocity space (= ``vorosity'') might play a role as well
\citep{owo08}. Consequently, the reduction of \Mdot\ as implied by the
work from Fullerton et al. might be overestimated. 

Indeed, \citet{Sundqvist10a}, relaxing \textit{all} the above {\it standard}
assumptions, showed that the microclumping approximation is not a suitable
assumption for UV resonance line formation under conditions prevailing in
typical OB-star winds. These results are supported for the case of B
supergiants by \citet{Prinja10}, who found that the observed
profile-strength ratios of the individual components of UV resonance line
doublets are inconsistent with lines formed in a ``microclumped'' wind
 (see also \citealt{Sundqvist10b}).
Resonance 
together with \Ha\ 
line profiles as calculated by \citet{Sundqvist10a, Sundqvist10b} from
2/3D, stochastic wind models allowing for optically thick clumps (=
``macroclumping'') and vorosity effects are compatible with mass-loss rates an
order of magnitude higher than those derived from the same lines but using
the microclumping technique. 

Low mass-loss rates as implied by the latter models would have dramatic
consequences for the evolution of and feed-back from massive stars (cf.
\citealt{so06}). \cite{hirschi08} concluded that evolutionary models could
``survive'' with \Mdot\ reductions of at most a factor of $\sim$2 in
comparison to the rates from \citet{vink00} (which translate to an
``allowed'' reduction of the {\it empirical} mass-loss rates of a factor of
about four) whilst factors around 10 are strongly disfavored. Furthermore,
such revisions would cast severe doubts on the theory of radiative driving,
since the present agreement between observations and theory would break down
completely.

Hence, a {\it reliable} knowledge of the amount of clumping (quantified
by the clumping-factor and its radial stratification)
is crucial to constrain the ``true'' mass-loss rate of the star. Since,
due to different oscillator strengths and cross-sections, the
corresponding formation depths vary from close to the base (\Ha) over
intermediate regions (\Bra, mid-IR continua) to the outermost wind
(radio), a {\it consistent} analysis of different diagnostic features
will provide severe constraints on the run of $f_{\rm cl}$ and \Mdot\
itself. To this end, we have started a project to exploit these
diagnostics, by collecting the required data and analyzing them in a
consistent way. First results with respect to constraints from the
IR/mm/radio-{\it continuum} combined with \Ha\ have been reported by
\citet{puls06a}, in particular regarding the radial stratification of the
clumping factor. They found that, at least in dense winds, clumping is
stronger in the lower wind than in the outer part, by factors of
4{\ldots}6, and that unclumped mass-loss rates need to be reduced at
least by factors 2{\ldots}3, in agreement with the results quoted above.

\paragraph{The weak wind problem.} From a detailed UV-analysis,
\citet{martins04} showed that the mass-loss rates of young O-dwarfs (late
spectral type) in N81 (SMC) are significantly smaller than predicted
theoretically (see also \citealt{bouret03} for similar findings), even
when relying on unclumped models (the presence of clumping would increase
the discrepancy). In the Galaxy, the same dilemma seems to apply,
particularly for objects with \LL\ $\la 5.2$ \citep{martins05b, Marcolino09},
including the O9V standard 10~Lac \citep{herrero02}, and maybe also 
$\tau$~Sco (B0.2V, see \citealt{repo05}), pointing towards very low
mass-loss rates, thus challenging our current understanding of
radiation-driven winds. Note that most mass-loss rates for (other) dwarfs
derived so far are only upper limits, due to the insensitivity of the usual
mass-loss estimator \Ha\ on (very) low mass-loss rates (see also
\citealt{mokiem06}). Present results based on UV studies may
suffer from effects such as X-rays, advection or adiabatic cooling, as 
discussed by \citet[ see also \citealt{hil08, Puls08}]{martins05b}.
Consequently, a detailed investigation by means of {\it sensitive
mass-loss diagnostics} in dwarfs {\it over a larger sample} is crucial to
confirm their very weak-winded nature.

\begin{table*}[t]
\caption{Sample stars and observing data in the {\it L}-band. ``S'' and
``I'' correspond to the IRTF/SpeX and VLT/ISAAC spectrograph, respectively,
where HD\,37128 and HD\,37468 have been observed with both instruments.}
\renewcommand{\arraystretch}{1.1}
\begin{center}
\begin{tabular}{l l c c c c | c c c c}
\hline
     & spectral  & sp. type  &       & obs.  & integr. & \multicolumn{4}{c}{previous investigations} \\ 
star & type      & reference &instr. & date  & time &\PV & opt & {\it H/K} & IR/radio \\
\hline
Cyg\,OB2 $\#$7             & O3 If$^*$ & MT91    & S  & 10, 11 Sep 05 & 360s, 400s  & - & 3,6 & 7 & 8\\
Cyg\,OB2 $\#$8A            & O5.5 I(f) & MT91    & S  & 11 Sep 05     & 400s     & - & 3,6 & 7 & 8\\
Cyg\,OB2 $\#$8C            & O5 If     & MT91    & S  & 10, 11 Sep 05 & 360s, 400s     & - & 3,6 & 7 & 8\\
HD\,30614 ($\alpha$ Cam)   & O9.5 Ia   & W72     & S  & 11 Sep 06     & 200s     & 1 & 4,5 & 7 & 8\\
HD\,36861 ($\lambda$ Ori A)& O8 III((f)) & W72   & I  & 08 Jan 06     &  816s(3.7\um), 1224s(3.9\um)     & 1 & 4   & - & 8\\
HD\,37128 ($\epsilon$ Ori) & B0 Ia     & WF90    & I  & 08 Jan 06     &  102s(3.7\um), 204s(3.9\um)     & - & 2   & 7 & -\\
                           &           &         & S  & 11 Sep 05     & 80s     & \\
HD\,37468 ($\sigma$ Ori)   & O9.5 V    & CA71    & I  & 08 Jan 06     &  306s(3.7\um), 612s(3.9\um);     & - & -   & 7 & -\\
                           &           &         & S  & 11 Sep 05     & 320s     & \\
HD\,66811($\zeta$ Pup)     & O4 I(n)f  & W72     & I  & 08 Jan 06     & 204s(3.7\um), 408s(3.9\um)     & 1 & 4,5 & 7 & 8\\
HD\,76341                  & O9 Ib     & M98     & I  & 08 Jan 06     & 826s(3.7\um), 1656s(3.9\um)     & - & -   & - & -\\
HD\,217086                 & O7 Vn     & W73     & S  & 10, 11 Sep 05 & 420s, 400s  & 1 & 5,6 & 7 & -\\
\hline
\end{tabular}
\end{center}
\smallskip 
\footnotesize{ 
Spectral references: CA71, \citet{ca71}; M98, \citet{mason98}; MT91,
\citet{massey91}; W72, \citet{wal72}; W73, \citet{wal73}; WF90,
\citet{walfitz90}.\\
Previous investigations refer to 
(1) the analysis of the \PV $\lambda\lambda$\,1118/28 doublet by \citet{fulli06}; 
(2) \citet[ unblanketed analysis]{kudetal99}; 
(3) \citet{herrero02}; 
(4) the \Ha\ mass-loss analysis by \citet{markova04} based on stellar 
parameters calibrated to the results from optical NLTE analyses by \citet{repo04}; 
(5) \citet{repo04};
(6) \citet{mokiem05}; 
(7) the {\it H/K}-band analysis by \citet{repo05}; and
(8) the combined \Ha, IR, mm and radio continuum analysis by \citet{puls06a}.\\
}
\label{runs}
\end{table*}

\medskip 
\noindent 
In this paper, we intend to show that IR spectroscopy, in particular in the
$L$-band, is perfectly suited to investigate {\it both} problems, due to the
extreme sensitivity of \Bra\ on mass-loss effects.
\begin{enumerate}

\item[i)] For objects with large \Mdot, this line samples the intermediate 
wind (because of the larger oscillator strength of \Bra\ compared to \Ha\
and \Brg), thus enabling us to derive constraints on the (local) clumping
factor, and, in combination with other indicators (UV, \Ha, \Brg,
Pf$_{\gamma}$), to derive ``true'' mass-loss rates. We are aware that our
UV-analysis might be hampered by macroclumping/vorosity effects (see above),
but lacking suitable methods to include these effects into our
NLTE treatment, we consider corresponding constraints as suggestive
only.  

\item[ii)] For objects with very weak winds, \Bra\ provides not just 
upper limits but {\it reliable} constraints on \Mdot. The relevance of \Bra\
has been pointed out already by \citet{ah69}, who predicted that even for
hydrostatic atmospheres the (narrow) Doppler-cores should be in emission,
superimposed on rather shallow Stark-wings. As we will show in the
following, this emission component (and the line wings!) react sensitively on
\Mdot, particularly for very weak winds (see also \citealt{najarro98}).
\end{enumerate}

To accomplish our objective(s), we have performed a pilot study with the
high resolution IR spectrograph ISAAC attached to the 8.1m Unit 1 telescope of the
European Southern Observatory, Very Large Telescope (VLT), and the
intermediate resolution spectrograph SpeX at the NASA Infrared Telescope 
Facility (IRTF), and secured high S/N $L^{(')}$-band spectra of 
selected Galactic OB stars. These spectra will be
analyzed in the course of the present paper that is organized as follows:
In Sect.~\ref{obs}, we describe our stellar sample, the observations and the
data reduction. Sect.~\ref{codes} summarizes the relevant features of the
used NLTE model atmosphere, and describes our implementation of clumping. In
Sect.~\ref{dense} we concentrate on those objects with dense winds within
our sample, and investigate their clumping properties by combining the
$L$-band spectra with other diagnostics. The complementary objects with thin
winds are considered in Sect.~\ref{thin}, after the principal line formation
mechanism of \Bra in thin winds has been discussed, and additional problems
have been illuminated. In Sect.~\ref{discussion}, finally, we discuss our
results (particularly with respect to wind-momentum rates), and summarize our
results and present our future perspectives in Sect.~\ref{summary}.

\section{The stellar sample, observations and data reduction}
\label{obs}

The specific targets comprise a subsample of the northern (SpeX) and
southern (ISAAC) objects of the large sample of Galactic OB-stars which has
been observed and analyzed in the optical (see Table~\ref{runs}) and in the
$H/K$-band (\citealt{repo05}, based on the material presented by
\citealt{hanson05}). Our subsample covers mostly supergiants from O3 to B0,
and has been augmented by one weak wind candidate (HD\,37468, O9.5V) and
three comparison objects (HD\,217086, O7Vn; HD\,36861, O8III((f)),
HD\,76341, O9Ib) which should behave as theoretically predicted. Note that
for all objects UV archival data are available as well, that four of them
have been analyzed with respect to \PV\, (cf. Sect.~\ref{intro}), and that
all O-supergiants plus the O8 giant have been investigated in \Ha\ and the
IR, mm- and radio-continuum by \citet{puls06a} regarding their clumping
properties. Two of our objects have been observed by SpeX {\it and} ISAAC,
to enable a comparison of the data obtained by both instruments.
Table~\ref{runs} summarizes our target list, important observing data and
the previous investigations of our objects in the individual wavelength
bands.

\subsection{The IRTF/SPEX Sample of Stars}
Our first spectra were obtained in September, 2005 at the 3.0~m IRTF 
on Mauna Kea, Hawaii, using SpeX
\citep{spex}. Spex is a medium-resolution, 0.8 to 5.4 micron spectrograph
built at the Institute for Astronomy (IfA) and is available for general use
to the public through qualified time on the IRTF. Two nights were granted to
our program. While the second night was good, the first night experienced
intermittent cirrus, sometimes seriously reducing the signal in the
instrument. Our spectra were taken in the cross-dispersed mode of SpeX,
providing full spectral coverage from 0.8 to 5.5$\mu$m with a resolution of
$\lambda / \delta \lambda \sim$ 2000 with the narrowest slit (0.3'').
Luckily, the seeing was never greater than about 0.8'' and was typically
closer to 0.6''. About every hour during the night, wavelength arcs and flat
fields were taken.  Spectra were also obtained of telluric standards.  These
telluric standards were selected to share the same airmass and approximate
sky location as the target stars, are observed to considerably higher
counts, and were chosen to be early A-dwarfs with low $v$sin$i$.

A clear advantage to using the IRTF/SpeX system in cross-dispersed mode is
the ease with which the spectra can be reduced. A sophisticated IDL-based
reduction package called SpeXtool has been developed by \citet{spextool}
which incorporates calibration frames and readily produces final wavelength
and flux calibrated spectra.  While the data is cross-dispersed, there is
sufficient room in the 15" slit to allow two uniquely-observed positions.
This provides for a traditional A position - B position for background
subtraction.  SpeXtool also takes advantage of the telluric correction
methods recently developed by \citet{vacca03} which includes a
high-resolution model of Vega.  Because Vega has an extremely low \vsini,
the A-dwarfs selected for tellurics in our study (HD\,219190 and HD\,33654)
were also chosen to have very low \vsini\ to ensure the best match in
profile shape. The Appendix of \citet{hanson05} provides graphic evidence of
why the \vsini\ match between model and star is so critical for very high
signal-to-noise spectroscopic work like this. 

We were particularly lucky for this program as Vega was observable at the
start of the night during our run.  We took full advantage of this for the
purpose of carefully confirming the integrity of the high signal-to-noise
telluric spectra derived from all the A-dwarf standard stars observed
throughout the night compared against the theoretically-perfect telluric
spectrum determined from our direct Vega observations and derived by the
Vega models provided in the SpeXtool package.

\begin{figure*}
\resizebox{\hsize}{!} {\includegraphics{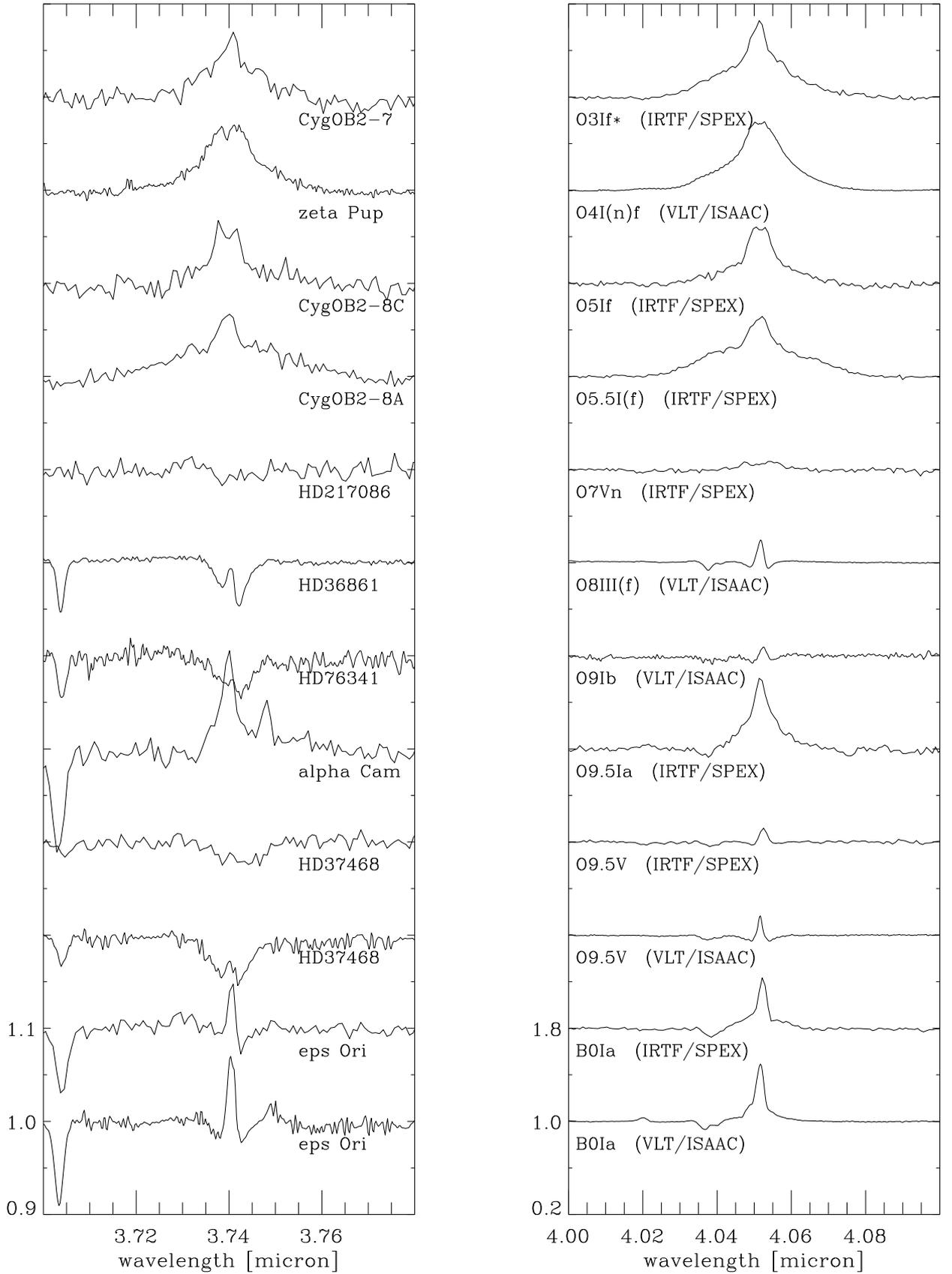}} 
\caption{VLT/ISAAC and IRTF/SpeX $L$-band spectra for our sample stars in the
two important regions centered around \Pfg(left) and \Bra(right). Note the
\HeI3.703$\mu$m line which is present at later spectral types. Two objects
(HD\,37468 and $\epsilon$~Ori) have been observed by both instruments.} 
\label{lband}
\end{figure*}

While spectra were obtained throughout the full spectral range of 0.8 to 5.4
microns with SpeX, we are presenting just the two (most interesting) narrow
spectral regions centered at \Pfg\ and \Bra\ in Figure~\ref{lband}.

\subsection{The VLT/ISAAC Sample of Stars}
We were granted one night, in visitor mode, on 8 January 2006 on VLT1 (Antu)
with ISAAC \citep{Moorwood98}.  The weather was at times marginal, with
highly variable seeing and sometimes cloud cover too dense to observe.  To
achieve the highest resolution, again we had the slit set to 0.3''. This
proved extremely challenging when the seeing dropped below 2.0''.  However,
during that single night, we also experienced a few extended moments of
reasonable weather and seeing. By sticking to the brightest sources in our
sample we were able to observe several stars with sufficiently high count
rates, even at 4.05$\mu$m, to achieve the high signal-to-noise needed for
our analysis.

For reduction, we closely adhered to the advise found in the ISAAC Data
Reduction Guide 1.5 \citep{Amico02} and updates found on the ESO ISAAC
website.  An all encompassing reduction package is not available for the
ISAAC instrument such as is available with SpeX.  We used fits manipulation
routines available from the IRAF\footnote{IRAF is distributed by the
National Optical Astronomy Obser\-va\-tory, which is operated by the
Association of Universities for Research in Astronomy (AURA) under
cooperative agreement with the National Science Foundation.} software
package. The reduction starts with a simple ESO provided configuration file
to remove electrical ghosts (provided in the 'eclipse' package).  From there
reduction steps involved dark subtraction, linearity corrections and flat
fielding, all accomplished using scripts written in IRAF.  ISAAC has ample
slit length for multiple positions in the slit.  However, we observed all of our 
targets and our telluric standards in the near exact same two positions 
in the slit, noted as position A and position B. 

The ISAAC instrument shows a pronounced curvature and distortion with
wavelength on the array.  Perhaps more importantly, the shape of this
curvature is also a function of position along the slit.  Before extracting
a 1-D spectrum from the 2-D image, this needs to be corrected.  ESO has
provided scripts in the eclipse package that use the arc images taken
throughout the night to create a distortion correction that can be applied
to the 2-D images. For typical applications, this allows stars observed
anywhere in the slit, to line up properly in wavelength space.  However,
these corrections will not be good enough for our observations.  

This is the reason only two positions were used in the slit. Two positions
are needed to remove the background in the 2-D images.  The distortion in
the 2-D images was corrected using the ESO eclipse program for this purpose.
But we were careful to use a single solution applied to all 2-D images taken
during the night, so there was no introduction of very small, but differing 
wavelength solutions.   Once the 2-D image was fully processed and had 
been converted to a 1-D spectrum, the two slit positions, A and B, were 
never used together in processing again. Target star A slit positions were 
only processed with telluric star A slit position and the two slit positions, 
A and B, were kept separate in processing in much the same way two 
differing grating tilts would proceed separately.  At our resolution and 
signal-to-noise, subtle misalignment of grating solutions in this wavelength
regime, because of the numerous, very deep and narrow telluric features 
would create strong beat phenomena when 1-D spectra are divided from each 
other in later processing.

To remove telluric absorption, our strategy is one of bootstrapping all telluric
observations off each other (see \citealt{hanson05}) to come up with a
consistent set of telluric-free spectra.  We started with a synthetic spectrum 
of the Earth's atmosphere for the airmass
covering our observations using ATRAN \citep{Lord92}. This model telluric
spectrum was used to divide out (remove) the telluric component in all of our telluric
A-stars to first order.  We then fit the remaining hydrogen lines in the 
A-stars. We also went through this exercise using several
of the OB-stars. While we did not derive the hydrogen profiles of these 
OB stars in this manner, the relatively narrow width of the OB hydrogen 
lines allowed us to use their spectra to constrain the very broad wing 
component of the A-star and to ensure a proper continuum for 
the hydrogen line fit in the A-star.  Then we returned to
the original full A-star spectrum, removed the fitted hydrogen lines for
that star and created what was the best estimate for the telluric features
through that spectral range towards that star. In this way, the telluric
spectrum was individually solved for numerous A-star sight-lines with a
similar airmass range. These can be checked against each other and then
averaged to reduce any possible errors introduced in the hydrogen line fits
from a single A-star. From this, a final few telluric spectra as a function
of airmass during the night was derived, and the appropriate telluric
spectrum could be removed from the OB target star.  A rough, but independent
check was made by dividing our raw OB star spectra by an ATRAN derived
telluric spectrum to ensure there were no obvious mistakes introduced in our
method above.

\begin{figure}
\resizebox{\hsize}{!} {\includegraphics{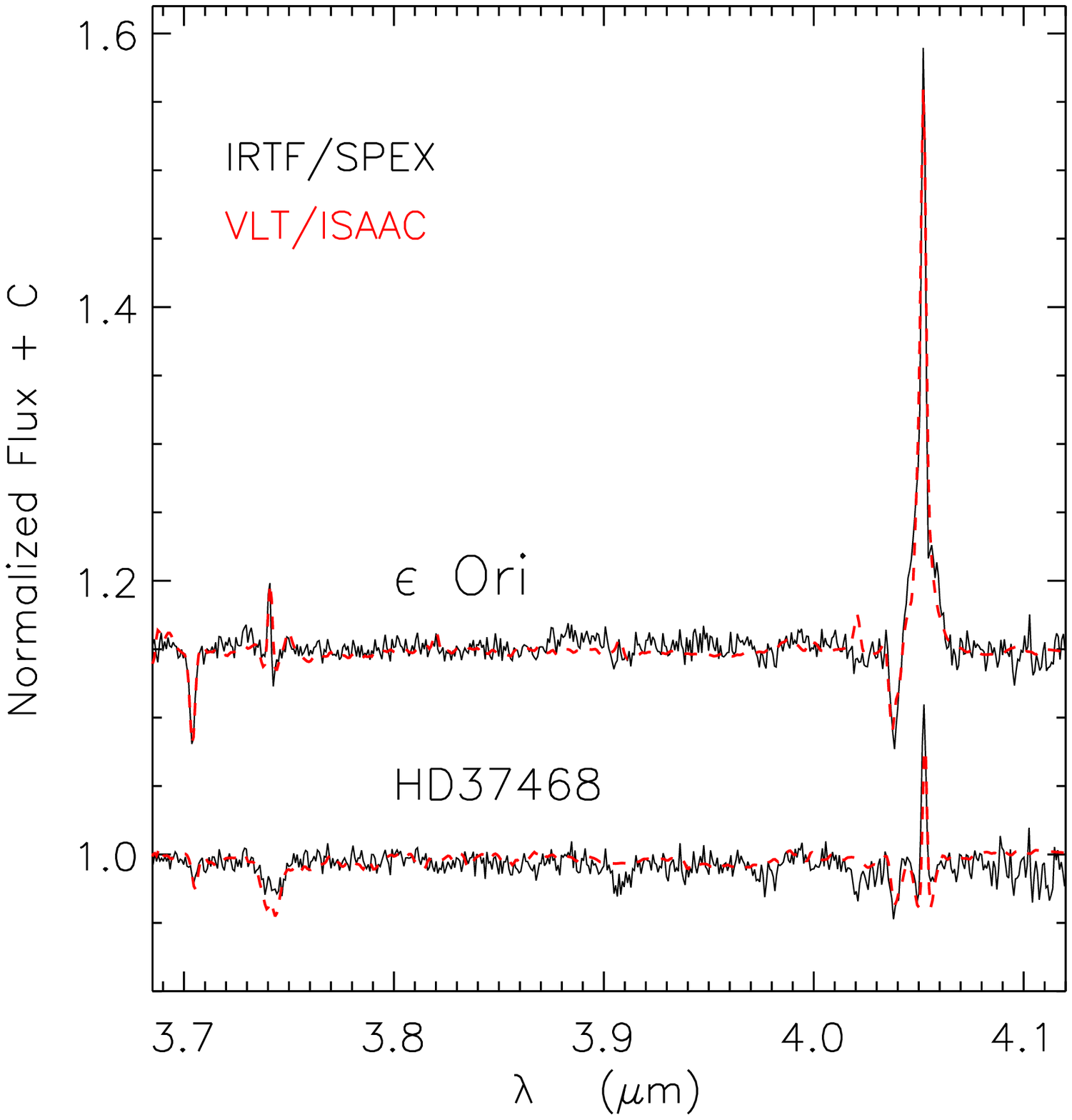}} 
\caption{Comparison between IRTF/SpeX (solid) and VLT/ISAAC (dashed) 
$L$-band spectra for the two objects (HD\,37468 and $\epsilon$~Ori) observed
in both runs. The VLT/ISAAC spectra have been reduced to the resolution of the
IRTF/SpeX instrument. Note the excellent agreement between both datasets.} 
\label{compa-spex-isaac}
\end{figure}

The validity of the reduction procedure for each dataset is confirmed in
Fig.~\ref{compa-spex-isaac}, where a comparison between IRTF/SpeX and
VLT/ISAAC $L$-band spectra for the two objects (HD\,37468 and
$\epsilon$~Ori) observed in both runs is presented. In
Fig.~\ref{compa-spex-isaac} the VLT/ISAAC spectra have been reduced to the
resolution of the IRTF/SpeX instrument (note the larger S/N ratio of the
former). The excellent agreement between both datasets clearly supports the
different data reduction procedures employed for each run. Given the higher
resolution of the VLT/ISAAC data, we made use of these observations for our
quantitative analysis.

\section{IR diagnostics} 
\label{codes} 
To model the infrared spectra of our sample of objects we have utilized {\sc
cmfgen}, which is an iterative, non-LTE, line-blanketed model
atmosphere/spectrum synthesis code developed by \citet{hil98}. It solves the
radiative transfer equation in the co-moving frame and in spherical geometry
for the expanding atmospheres of early-type stars. The model is prescribed
by the stellar radius, \Rstar, the stellar luminosity, \Lstar, the mass-loss
rate, \Mdot, the velocity field, $v(r)$ (defined by \vinf\ and $\beta$), the
volume filling factor characterizing the clumping of the stellar wind, and
elemental abundances. Following \citet[ see also \citealt{paul94}]{hil98} we
include X-rays characterizing the X-ray emissivity in the wind by two
different shock temperatures, velocities and filling factors.  We refer to
\citet{hil98,hil99} for a detailed discussion of the code. 

%
\begin{table*}
\caption{Stellar and wind parameters as adopted/derived in the present
analysis. \Teff\ in kK, \Rstar\ in \Rsuna, all velocities in \kms, \Mdot\ in
\Mdu1 and modified wind-momentum rate, $\Dmom = \Mdote \vinfe
(\Rstare/\Rsune)^{0.5}$, in cgs. The gravitational acceleration, \logg, is
the effective one, i.e., {\it not} corrected for centrifugal forces. The
volume filling factor, $\fv$, corresponds to the parameter CL$_1$ in
Eq.~\ref{eq:clump} and describes the maximum degree of clumping reached in
the stellar wind. However, since our analysis bases on stratified clumping
factors, the usual scaling of $\Mdote \propto \sqrt{\fv}$ does not or only
approximately apply in most cases. CL$_2$ (\kms) indicates the onset of
clumping in the wind. The horizontal line separates objects
displaying significant wind emission in \Ha, i.e., dense winds (see
Sect.~\ref{dense}), from objects with a pure absorption \Ha\ profile, i.e.,
thin winds (Sect.~\ref{thin}).}
%
\begin{center}
\tabcolsep1.3mm
\begin{tabular}{l l c c c r r c r r c r r r r r c}
\hline
star            & sp.type     & \MV       & \Teff & \logg & \Rstar & \YHe
& \LL & \vsini & \vmac & \vinf & \Mdot & $\beta$ & $\fv$ & CL$_2$ & \Vt &$\log \Dmom$ \\
\hline
Cyg\,OB2 $\#$7  & O3 If$^*$   & -5.91$^1$ & 45.1  & 3.75  & 14.7   & 0.13 & 5.91& 95    & 65   & 3100 & 1.2   & 1.05    &0.03& 100 & 10 & 28.95\\
HD\,66811       & O4 I(n)f    & -6.32$^2$ & 40.0  & 3.63  & 18.9   & 0.14 & 5.92& 215   & 95   & 2250 & 2.1   & 0.90    &0.03& 180 & 10 &  29.11\\
Cyg\,OB2 $\#$8C & O5 If       & -5.61$^1$ & 37.4  & 3.61  & 14.3   & 0.10 & 5.56& 175   & 90   & 2800 & 2.0   & 1.30    &0.10& 550&  20 &29.13\\
Cyg\,OB2 $\#$8A & O5.5 I(f)   & -6.91$^1$ & 37.6  & 3.52  & 26.9   & 0.10 & 6.12& 110   & 80   & 2700 & 3.4   & 1.10    &0.01& 500&  10 & 29.48\\
HD\,30614       & O9.5 Ia     & -7.00$^2$ & 28.9  & 3.01  & 32.0   & 0.13 & 5.81& 100   & 75   & 1550 & 0.50  & 1.60    &0.01&  25& 17.5 &  28.44\\
HD\,37128       & B0 Ia       & -6.99$^3$ & 26.3  & 2.90  & 34.1   & 0.13 & 5.70& 55    & 60   & 1820 & 0.46  & 1.60    &0.03& 30 &  15   &28.49\\
\hline
HD\,217086      & O7 Vn       & -4.50$^2$ & 36.8  & 3.83  & 8.56   & 0.1  & 5.08& 350   & 80    & 2510 & 0.028 &  1.2   &0.10& 30 &  10 & 27.11\\
HD\,36861       & O8 III((f)) & -5.39$^4$ & 34.5  & 3.70  & 13.5   & 0.11 & 5.37& 45     & 80    & 2175 & 0.28  &  1.3  &1.0 & -  & 7.5 & 28.15\\
HD\,76341       & O9 Ib       & -6.29$^4$ & 32.2  & 3.66  & 21.2   & 0.1  & 5.64& 63     & 80    & 1520 & 0.065 &  1.2  &1.0 & -  & 7.5 & 27.46\\
HD\,37468       & O9.5 V      & -3.90$^4$ & 32.6  & 4.19  &  7.1   & 0.1  & 4.71& 35     & 100   & 1500 & 0.0002&  0.8  &1.0 & -  & 5   &24.70\\
\hline
\end{tabular}
\end{center}
\smallskip
$^1$ \citet{mokiem05},
$^2$ \citet{repo04},
$^3$ \citet{kudetal99},
$^4$ from the calibration provided by \citet{martins05a}.
\label{para}
\end{table*}

Given the large range of stellar parameters and the variety of luminosity
classes covered by our O-star sample, the location of the $\taur=2/3$ radius
for these objects will vary from being placed in the deep hydrostatic layers
(dwarfs) up to the upper layers where the wind takes off (supergiants).
Noting that not only the IR lines but also the IR continuum, through
bound-free and free-free processes, will have different formation depths as
a function of wavelength, the role of the hydrostatic structure and the
transition region between photosphere and wind becomes crucial to interpret
the stellar spectra. 

{With this in mind, we computed 
CMFGEN models with a photospheric structure modified  
following the approach from \citet{sph97}
\footnote{In this approach we compute the density from the hydrostatic
equation and
the velocity from the continuity equation.},
smoothly connected to a beta velocity law. 
In our approach the
Rosseland mean from the original formulation was replaced by the more
appropriate flux-weighted mean. Several comparisons using ``exact''
photospheric structures from  {\sc tlusty} \citep{hub95} showed
excellent agreement with our method. Likewise, model atoms were expanded
and optimized to make use of the IR metal lines arising from high
lying levels.}

To investigate the role of clumping, we follow the conventional approach of 
microclumping and assuming a void inter-clump matter. In this case, and as
already outlined in Sect.~\ref{intro}, the volume filling factor is just the
inverse of the clumping factor, $\fv=\fcl^{-1}$, and the clumping factor
itself quantifies the overdensity of the clumps with respect to the averaged
density, $\langle \rho \rangle = \Mdote/(4 \pi r^2 v(r))$. Moreover, the
radial stratification of the volume filling factor is described by the
clumping law introduced by \cite{naj09}:
\begin{equation}
\label{eq:clump}
\fv(r) = CL_1 + ( 1 -CL_1 ) \, {\rm e}^{-\frac{v(r)}{CL_2}} + ( CL_4 - CL_1) \, 
{\rm e}^{-\frac{\vinfe-v(r)}{CL_3}}
\end{equation}
where CL$_1$ and CL$_4$ are volume filling factors and CL$_2$ and CL$_3$ are
velocity terms defining locations in the stellar wind where the clumping
structure changes. CL$_1$ sets the maximum degree of clumping reached in the
stellar wind (provided CL$_4$ $>$ CL$_1$) while CL$_2$ determines the
velocity of the onset of clumping. CL$_3$ and CL$_4$ control the clumping
structure in the outer wind. From Eq.\ref{eq:clump} we note that as the wind
velocity approaches \vinf, so that $(\vinfe-v(r)) \leq$ CL$_3$, clumping
starts to migrate from CL$_1$ towards CL$_4$. If CL$_4$ is set to unity, the
wind will be unclumped in the outermost region.  Such behavior was already
suggested by \cite{nug98} and was utilized by \cite{fig02} and \cite{naj04}
for the analysis of the WNL stars in the Arches Cluster. Recently,
\cite{puls06a} have found a similar behavior from \Ha\ and IR/mm/radio
studies for OB stars with dense winds. Furthermore, our clumping
parametrization seems to follow well the results from hydrodynamical
calculations by \cite{run02}. From Eq.~\ref{eq:clump} we note that if the
term including CL$_3$ and CL$_4$ is neglected or if CL$_3 \rightarrow$ 0, we
recover the law proposed by \cite{hil99}. For the present study (except for
$\zeta$~Pup, see below), we have set CL$_4$=1, i.e., the outer wind
regions are assumed to be unclumped.

Observational constraints are set by the $L$-band spectra presented above
and UV, high-resolution optical and $H$ and $K$-band spectra collected by
our group as well as by optical, IR and radio continuum measurements from
literature/archival data. The individual sources are quoted in the
corresponding figure captions.
In this paper we concentrate on the strong diagnostic potential provided by
the infrared $K$- and $L$-bands to determine mass-loss rates and trace wind
clumping as an alternative to \Ha, the classical mass loss indicator. Thus,
we defer a detailed full wavelength analysis and discussion of these
objects to a forthcoming paper.

Table~\ref{para} displays the stellar parameters obtained for our sample,
whereas a detailed comparison with results from previous investigations is
provided in Appendix~\ref{detcomp}. We obtain uncertainties of $\sim$1000~K
for the effective temperature of the objects (see Appendix~\ref{detcomp} for
a thorough discussion) while typical errors of 0.1~dex are estimated for
\logg. For objects with dense wind, we estimate our mass-loss accuracy to
be better than 25\%, with the corresponding \Mdot/$\fv^{0.5}$ = const
scaling for the error on the clumping factor. For objects with thin winds we
consider 0.5~dex as a conservative error on the mass-loss rate estimate (see
Sect.~\ref{mdot_reliability}).

Projected rotational speeds, \vsini, have been derived via the
Fourier-transform technique as developed by \citet{simon06} (based on the
original method proposed by \citealt{Gray73, Gray75}), applied to weak metal
lines (and partly \HeI-lines) available in the spectra. The remaining
discrepancies between synthetic and observed lines\footnote {Such
discrepancies were already detected and described by \citet{Rosenh70, CE77,
LDF93, Howarth97}.} were ``cured'' by convolving the spectra with an
additional, radial-tangential macro-turbulent velocity distribution (entry
\vmac\ in Table~\ref{para}). The derived values are of similar order as
found in alternative investigations \citep{Ryans02, simon06, simon07,
Lefever07, markova08}, indicating highly supersonic speeds in photospheric
regions that would be difficult to explain. Recently, however, \citet[ see
also \citealt{Lucy76}]{Aerts09} interpreted such extra-broadening in terms
of {\it collective} effects from hundreds of non-radial gravity-mode
oscillations, where the individual amplitudes remain sub-sonic.
\footnote{First observational evidence in support of this scenario has been
provided by \citet{simon10}.} They pointed out that the rotational
velocity could be seriously underestimated whenever the line profiles are
fitted assuming a macroturbulent velocity rather than an appropriate
expression for the pulsational velocities, or if a Fourier technique is
applied to infer the rotational velocity. If this were true, our values for
\vsini\ (and also those from the quoted investigations) would provide only
lower limits. For the present investigation, however, this is of minor
concern, since our main interest is to obtain a correct {\it shape} of
the profiles (irrespective of the responsible process), to enable meaningful
fits.


\section{Objects with dense winds, constraints on the clumping factor}
\label{dense}
In this section we discuss our results for the objects of our sample 
displaying dense winds, and compare them with previous studies carried out at
optical (\citealt{repo04}, hereafter REP04; \citealt{mokiem05}, MOK05) and
near-infrared wavelengths (\citealt{repo05}, REP05) as well as the combined
\Ha/IR/mm/radio analysis by \citet[][ hereafter PUL06]{puls06a}. For further
details, see Appendix A.

\begin{figure}
\resizebox{\hsize}{!}{
\includegraphics{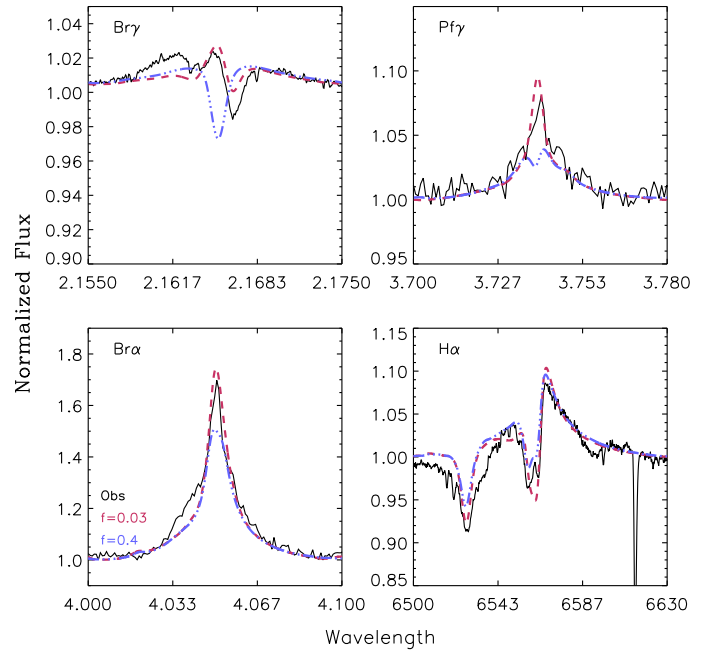}}
\caption{Model fits to IR and \Ha\ observations of Cyg\,OB2~\#7 (solid, black).
Two different models corresponding to clumping values of $\fv$=0.03 (dashed, red)
and $\fv$=0.4 (dashed-dotted, blue) are displayed (see text).
The \Ha\ spectrum was obtained with ISIS at the William Herschel Telescope on La Palma
(kindly provided by A. Herrero).}
\label{fitiroptcyg7}
\end{figure}

\begin{figure*}
\begin{minipage}{9.0cm}
\hspace{-.1cm}\resizebox{\hsize}{!}
{\includegraphics
{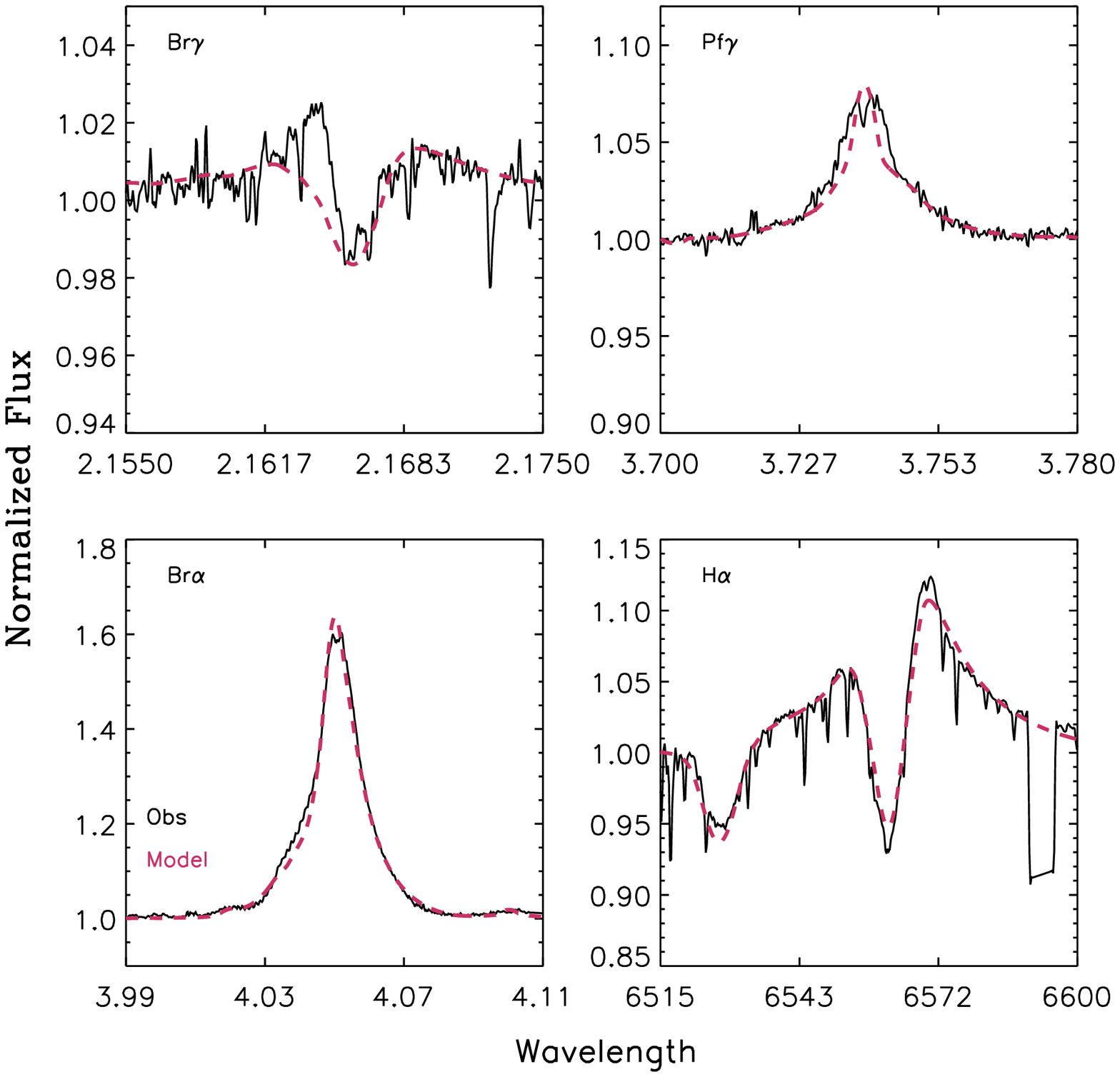}}
\end{minipage}
\hfil
\hspace{-.2cm}\begin{minipage}{9.3cm}
\resizebox{\hsize}{!} 
{\includegraphics
{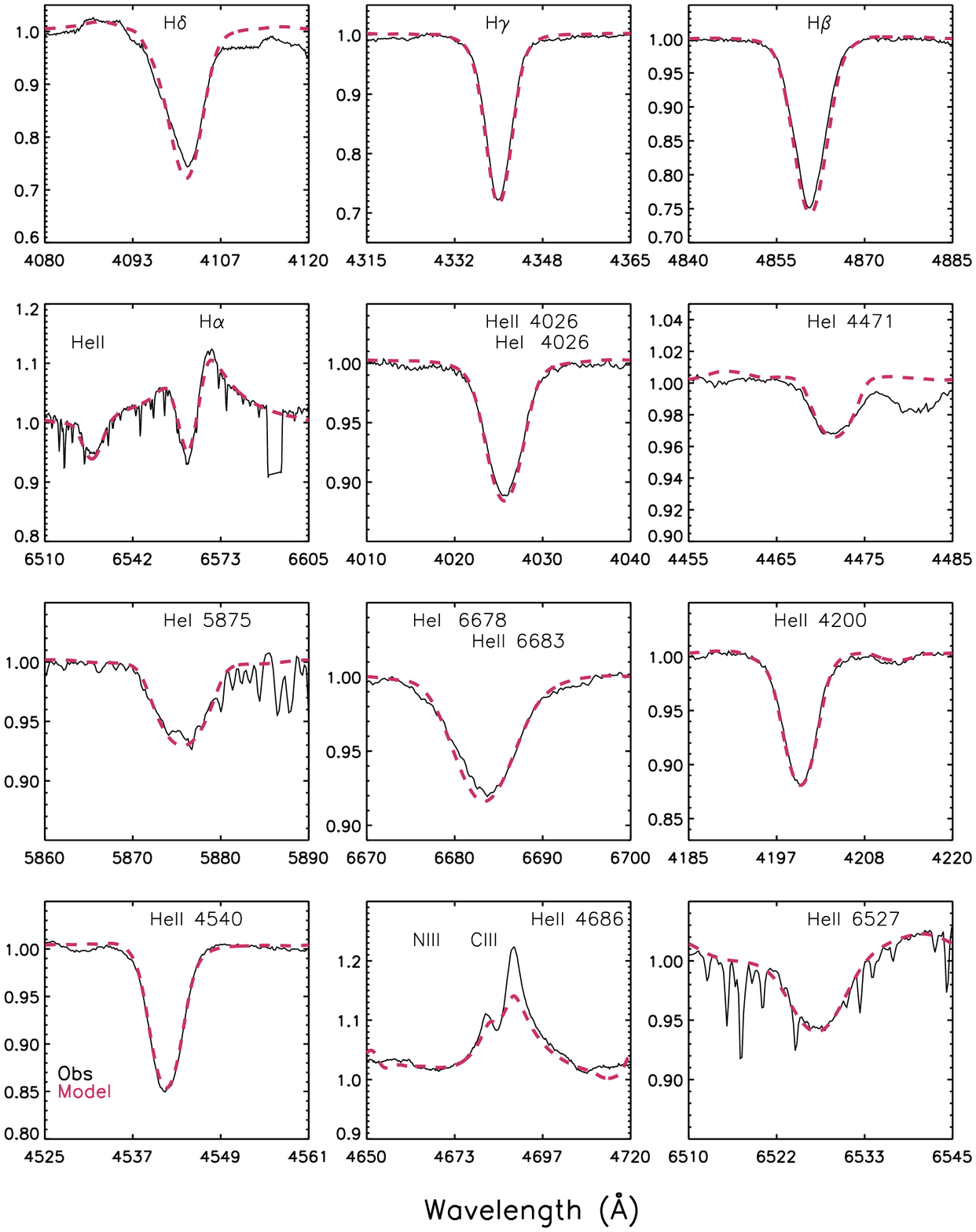}}
\end{minipage}
\caption{Model fits (dashed) to IR and optical observations of $\zeta$ Pup (solid). Optical
data were retrieved from the ESO archive, programme 266.D-5655(A)}
\label{fitiroptzetpup}
\end{figure*}

\paragraph{Cyg\,OB2~\#7.}
\label{par-cyg7}
Our derived main stellar parameters for Cyg\,OB2~\#7 (see Table~\ref{para})
agree very well with those obtained from optical (\citetalias{mokiem05}) and
$H$ and $K$-band (\citetalias{repo05}) analyses with respect to the
effective temperatures (within less than 1000~K). This is a very encouraging
result since our temperature determination relies not only on the
\HeII/\HeI\ equilibrium (e.g., \citetalias{mokiem05,repo05}) but also on the
\NV/\NIV/\NIII\ equilibria. Our results confirm the consistency between both
criteria and the validity of the weak \HeI\ lines as diagnostics in this
high temperature regime. Likewise, our \logg\ value lies in between the ones
derived by \citetalias{mokiem05} and \citetalias{repo05}, while our
unclumped mass-loss rates ($\sim$ 7.8 \mdu) are roughly 30\% lower. We
attribute this discrepancy to our lower He abundance (0.13 vs 0.21 in
\citetalias{mokiem05}) and higher $\beta$ (1.0 vs 0.8).  Interestingly, our
models favor a large clumping starting relatively close to the photosphere
to provide {\it{consistent}} simultaneous fits to the UV\footnote {remember
our caveat regarding the impact of macroclumping on UV resonance lines, as
stated in Sect.~\ref{intro}.}, optical and IR observations of this object
(see Fig.~\ref{fitiroptcyg7}). However, such a strong clumping -- via the
corresponding lower mass-loss rate -- tends to produce much too deep cores
in the optical H and \HeII\ lines. For comparison, a less clumped model
($\fv$=0.4), better matching the optical lines, is also displayed in 
Fig.~\ref{fitiroptcyg7}. Note, however, that such a model leads to severe
mismatches with the $L$-band and UV (not displayed here) spectra.  From
Fig.~\ref{fitiroptcyg7} we see that in the case of strong winds \Brg\ and
\Pfg\ provide stronger response to clumping than \Bra.  \citetalias{puls06a}
also found a strong clumping in the inner wind of this object. Their average
clumping factors lie between the ones we obtain in our IR and optical
analysis.  We would like to stress that while the optical and IR spectra of
Cyg\,OB2~\#7 provide strong constraints on CL$_1$ and CL$_2$, the UV spectra
and submillimeter and radio observations constitute crucial diagnostics to
determine CL$_4$ and CL$_3$. Indeed, our UV and submillimeter data
\citep{naj08} support the presence of constant clumping, at least up to
mid-outer wind regions where the millimeter continua of Cyg\,OB2~\#7 are
formed. However, radio observations by \citetalias{puls06a} show that
clumping may start to vanish at the outermost regions of the stellar wind
(note that radio continua form at much larger radii).
The expected emission of our models with constant clumping severely
overestimate the upper limits provided by the observations by
\citetalias{puls06a} of Cyg\,OB2~\#7. This demonstrates the need of
multiwavelength observations to constrain the run of the clumping structure.

\paragraph{$\zeta$ Puppis.}
\label{par-zetpup}
Our derived main stellar parameters agree fairly well with those presented
by \citetalias{repo04} and \citetalias{repo05}. We find, however, a slightly
lower He abundance and a much higher (by 50\%) ``unclumped'' mass-loss rate. 
We attribute this discrepancy in \Mdot\ to our clumping parametrization for 
this object. Given the large number of spectroscopic and continuum
constraints at nearly all wavelengths available for $\zeta$~Pup, we
performed a detailed clumping study aiming to constrain as accurately as
possible the run of the clumping factor throughout the wind and compare
it with recent results from \citetalias{puls06a}. Thus, we made use of all
the clumping parameters presented in Eq.\ref{eq:clump} and obtained
CL$_1$=0.03, CL$_2$=180., CL$_3$=600 and CL$_4$=0.17 as a 
best fit (see Fig.~\ref{firuncl}).  
With this parametrization, maximum clumping ($\fv$=0.03) is reached only in a
very narrow, $\Delta R \approx 0.1$~\Rstar, region around r=1.5~\Rstar.
Thus, the classical \Mdot/$\fv^{0.5}$ = const scaling for constant clumping
does not hold and causes the above discrepancy regarding unclumped \Mdot\
values. Our best model (see Fig.\ref{fitiroptzetpup}) is able to reproduce
satisfactorily not only the $L$- and $K$-band spectra but also the optical
lines and the full UV to radio energy distribution of the object. We stress
the almost perfect fit reached for \Ha.  That quality of fit can only be achieved
for models with stratified clumping as otherwise the observed absorption and
emission components cannot be fitted simultaneously (see also
\citetalias{puls06a}).
 We note, however, that since we aimed at 
a compromise solution, our model is unable to fit the blue wing of \Brg. Thus, while
a different clumping would yield a better fit to \Brg, it would also worsen 
significantly the model fits to other lines.

\begin{figure}
\resizebox{\hsize}{!}
{\includegraphics
{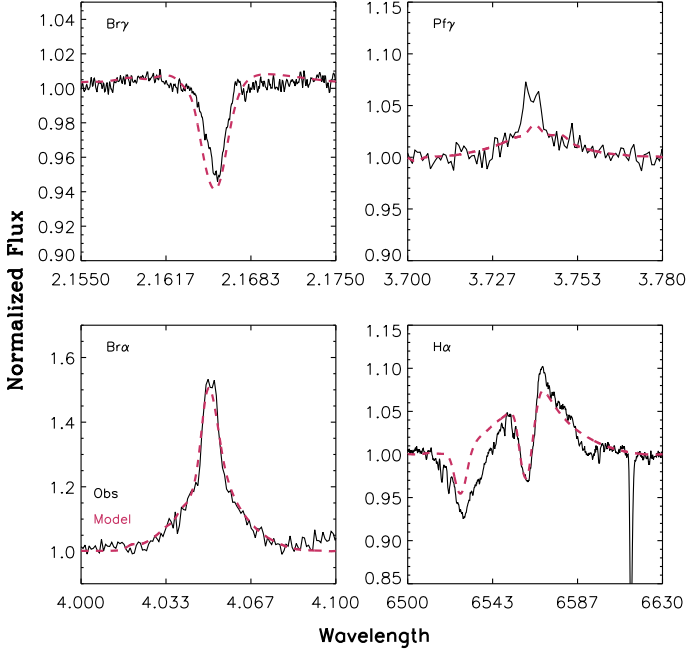}}
\caption{Model fits (dashed) to IR and \Ha\ observations of Cyg\,OB2~\#8C (solid).
\Ha~observations as in Fig.~\ref{fitiroptcyg7}.}
\label{fitiroptcyg8c}
\end{figure}

\begin{figure}
\resizebox{\hsize}{!}
{\includegraphics
{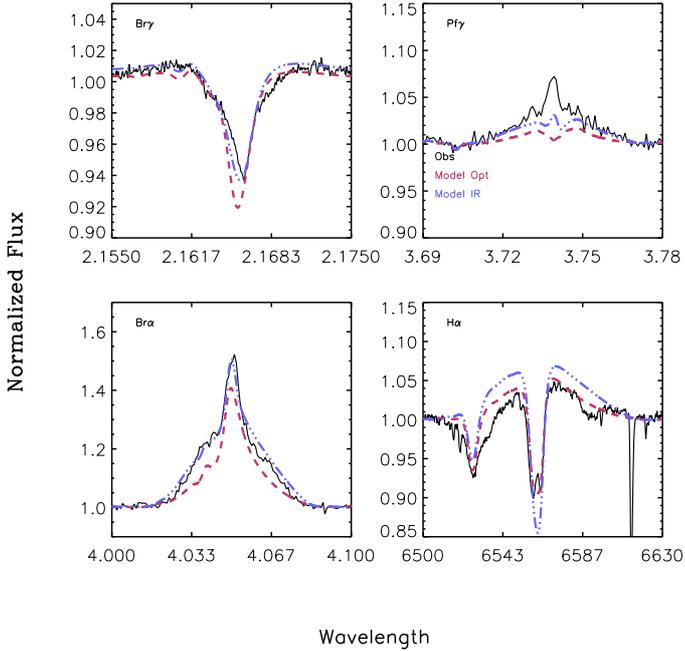}}
\caption{Model fits to IR and \Ha\ observations of Cyg\,OB2~\#8A. 
 (solid, black).
 Best model fitting only the optical (dashed, red) is displayed together with 
 that providing the best fit to the IR data (dashed-dotted, blue)  (see text).
\Ha~observations as in Fig.~\ref{fitiroptcyg7}.}
\label{fitiroptcyg8a}
\end{figure}

In their analysis of the clumping structure of O-star winds,
\citetalias{puls06a} pointed out that the derived clumping values for
$\zeta$~Pup are strongly dependent on the assumptions regarding the run of
the He ionization.  Interestingly, their maximum clumping factors ($\sim$5
for \HeIII\ recombining at v=0.86\vinf and $\sim$11 for a wind with
completely ionized He, when assuming an unclumped outer wind) are reached in
the same wind region (i.e. around r=1.5~\Rstar) as in our models. When
scaled to a similar outer clumping factor as derived in this work, the
agreement is even more striking (see Fig.~\ref{firuncl}).  Thus, both
studies reach the same conclusions concerning the run of the clumping
factor. As for Cyg\,OB2~\#7, our investigation shows the high sensitivity of
\Brg\ and \Pfg\ to clumping (see Fig.~\ref{fitiroptzetpup}) and the enormous
potential of IR spectroscopy to constrain the structure of stellar winds.

\paragraph{Cyg\,OB2~\#8C.}
\label{par-cyg8c}
The derived stellar parameters differ considerably from those obtained by
\citetalias{mokiem05}.  Our temperature is more than 4000~K lower while our
gravity lies only 0.12~dex below. We are confident in our temperature
determination since, as discussed above, we make use of both helium and
nitrogen ionization equilibria (see Fig.\ref{fitiroptcyg8c}). In fact, our
model reproduces satisfactorily the optical and IR spectra of this object.
We suggest that the discrepancy is closely related to the lower \vsini \
derived by \citetalias{mokiem05} (\vsini=145\,\kms vs. our 175\,\kms + 90\,\kms
macroturbulence) and the strong reaction of \HeI~4471 in this parameter
domain. The difference in derived \vsini\ values can be attributed to the
fact that our spectra are of higher quality than those used by
\citetalias{mokiem05}. We stress that our best compromise solution underestimates the emission core of
 \Pfg.

%
Compared to the other strong wind objects discussed in this section, this
object requires a lesser degree of clumping. Further, our models imply an
onset of clumping which is located at larger velocities than for the rest of
our supergiants. \citetalias{puls06a} also derived a low degree of clumping.
However, a detailed comparison with their results is not possible as their
analysis of this object remained rather unconstrained due to the lack of
sufficient flux measurements. 

\paragraph{Cyg\,OB2~\#8A.}
\label{par-cyg8a}
Unlike for the case of Cyg\,OB2~\#8C, our analysis of Cyg\,OB2~\#8A yields
excellent agreement with the stellar parameters obtained by
\citetalias{mokiem05}. \citet{deBecker04} report this object to be a
O6I~+~O5.5III binary system. This can be clearly inferred from the absolute
magnitude of the system displayed in Table\ref{para}. In fact, no single
best fit could be obtained to fit simultaneously the optical and IR
observations of Cyg\,OB2~\#8A (see Fig.\ref{fitiroptcyg8a}). Our preferred
model, which reproduces better the IR spectra, is characterized by strong
clumping ($\fv$=0.01) that leads, as in the aforementioned case of
Cyg\,OB2~\#7, to somewhat too strong absorption cores in the optical
hydrogen lines.  We note that the inclusion of clumping nicely removes the
discrepancy in \Brg\ found by \citetalias{repo05}. On the other hand, a
model with $\fv$=0.1 tuned to optimize the optical (see
Fig.\ref{fitiroptcyg8a}) severely underestimates the emission components in
the IR lines. The analysis of \citetalias{puls06a} was hindered by the
non-thermal nature of Cyg\,OB2~\#8A's radio emission. Nevertheless, their
combined \Ha\ + IR photometry analysis yields a wind structure which is
significantly less clumped than inferred in this paper. 
Both results could be only reconciled if the outer wind would be strongly
clumped.

\begin{figure*}
\begin{minipage}{9.0cm}
\hspace{-.1cm}\resizebox{\hsize}{!}
{\includegraphics
{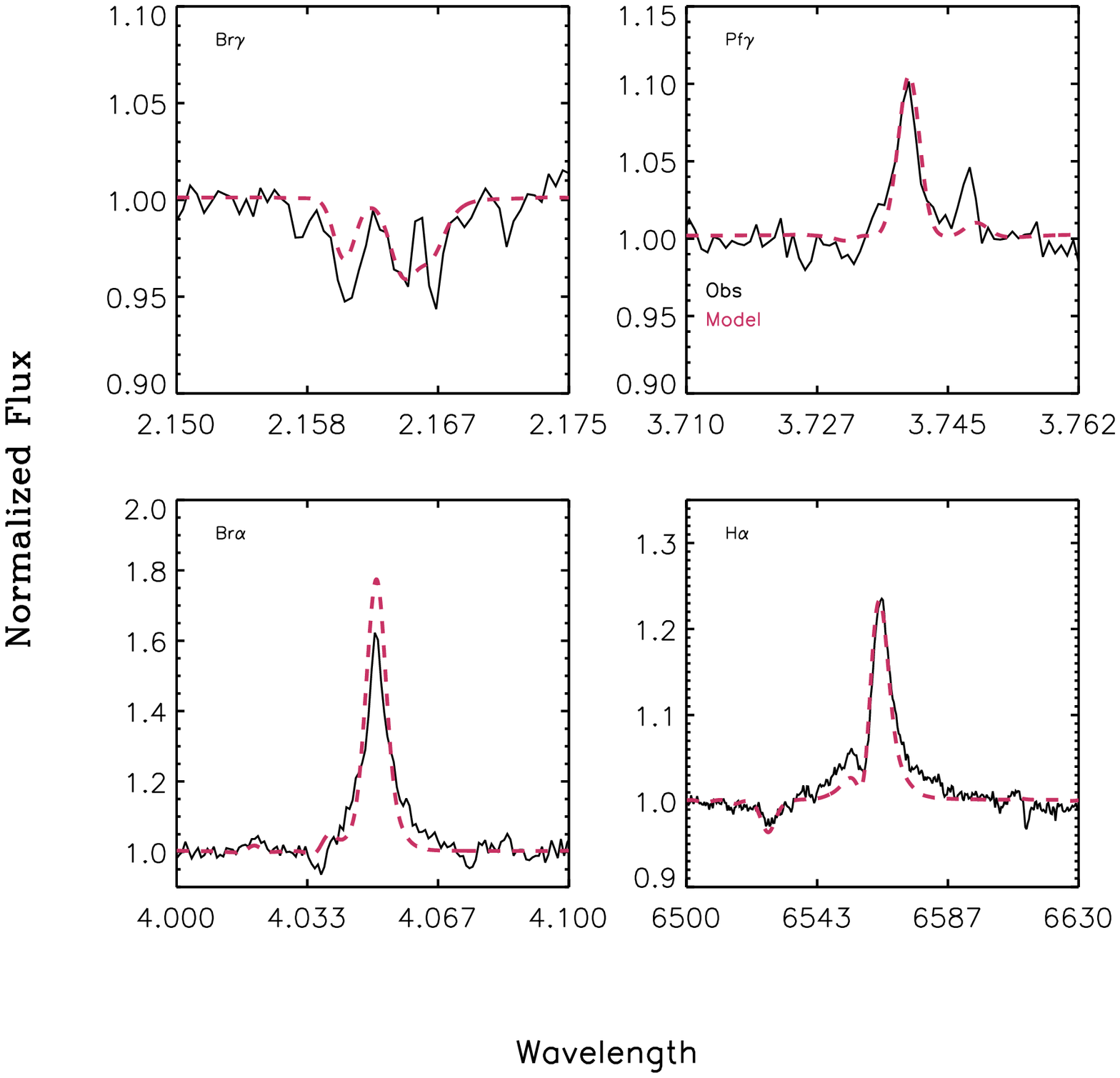}}
\end{minipage}
\hfil
\hspace{-.2cm}\begin{minipage}{9.3cm}
\resizebox{\hsize}{!} 
{\includegraphics
{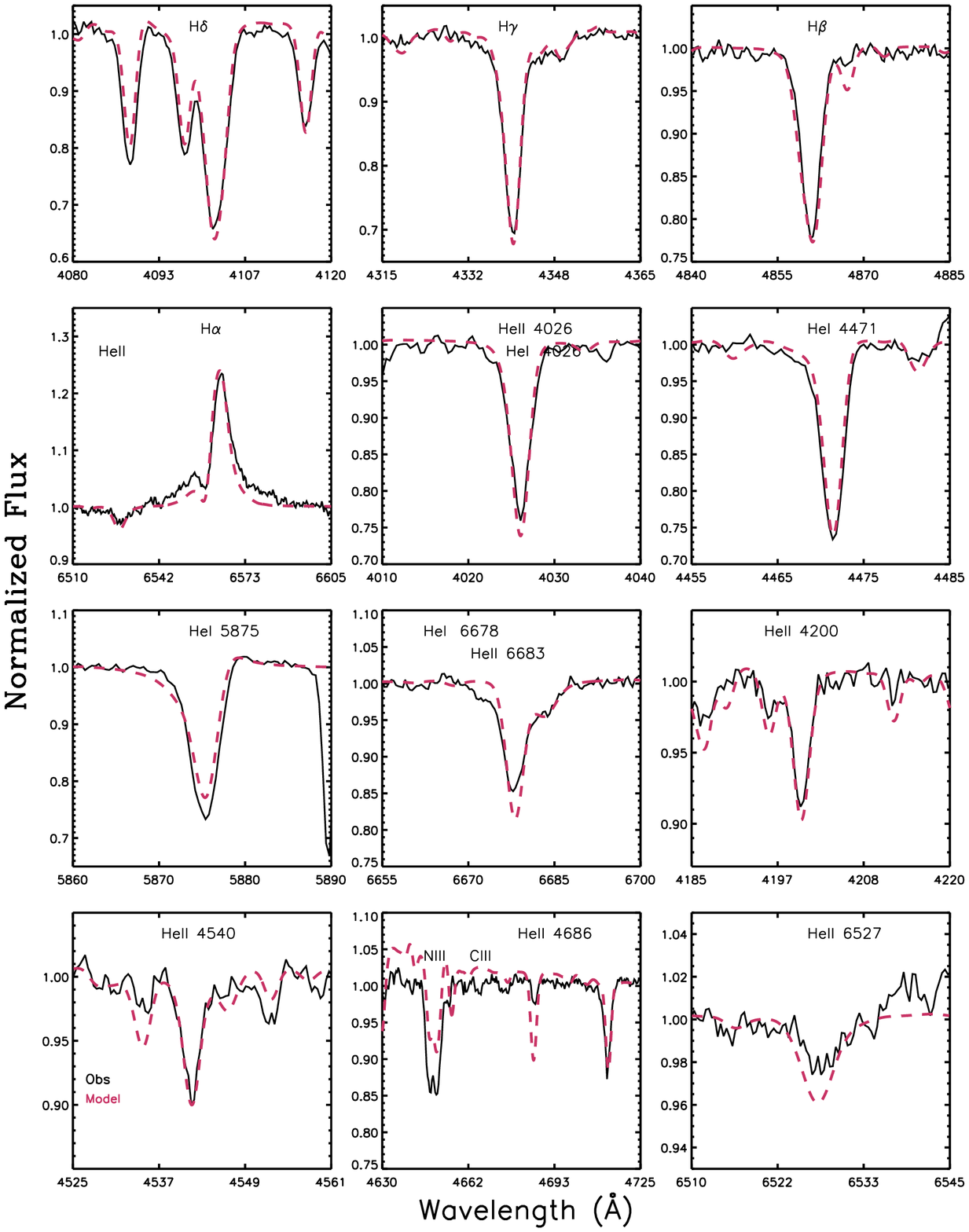}}
\end{minipage}
\caption{Model fits (dashed) to IR and optical observations of $\alpha$ Cam (solid). 
Optical data are from the Indo-U.S. Library of Coud\'e Feed Stellar Spectra
\citep{valdes04}.}
\label{fitiroptacam}
\end{figure*}

\paragraph{$\alpha$ Cam.}
\label{par-acam}
We find excellent agreement with the stellar parameters derived by
\citetalias{repo05}. A very high degree of clumping, starting very close to
the photosphere, is required to match \Pfg. In fact, this line turns into
the most sensitive clumping diagnostic at the base of the wind in O
supergiants. Our result agrees qualitatively with \citetalias{puls06a}'s
conclusions on the run of the clumping factor. They found a moderate degree
of clumping in the inner and mid wind regions. Figure~\ref{fitiroptacam}
shows that our model can reproduce satisfactorily the IR and optical spectra
of HD\,30614. 

\begin{figure}
\resizebox{\hsize}{!}
{\includegraphics
{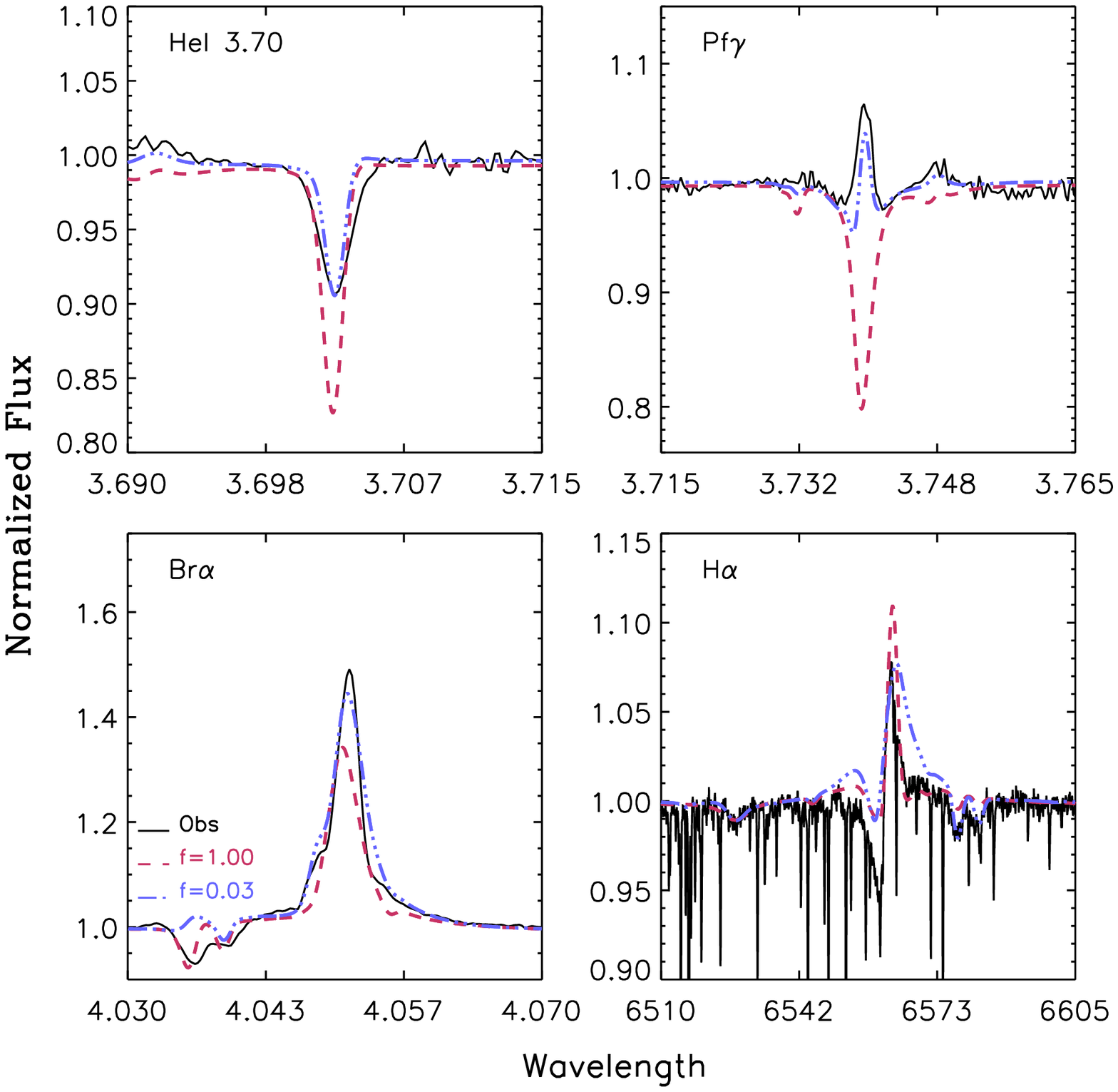}}
\caption{Model fits to IR and \Ha\ observations of $\epsilon$ Ori
 (solid, black).
 Two different models corresponding to no clumping (dashed, red)
 and a clumping value $\fv$=0.03 (dashed-dotted, blue) are displayed (see text).
The \Ha\ 
profile was kindly provided by N. Przybilla.}
\label{fitiropepsori}
\end{figure}

\begin{figure}
\resizebox{\hsize}{!}
{\includegraphics{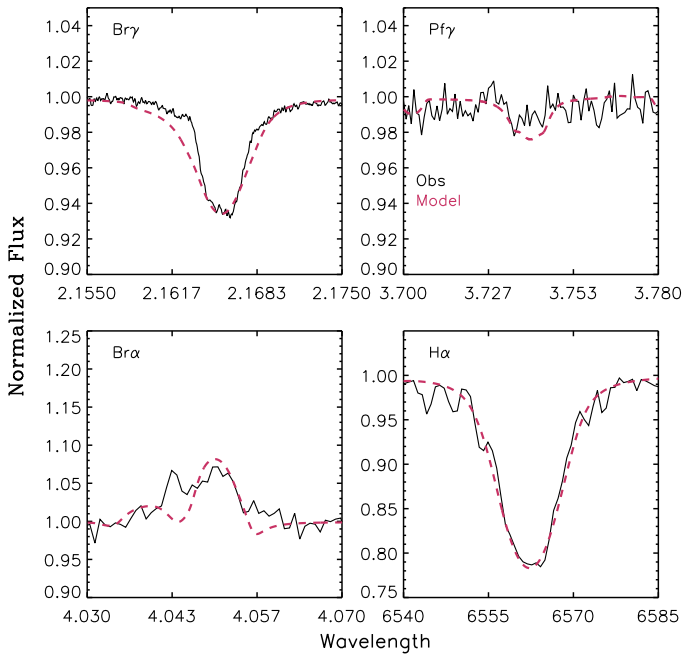}}
\caption{Model fits (dashed) to IR and \Ha\ observations of HD\,217086 (solid). 
\Ha\ observations as in Fig.\ref{fitiroptcyg7}.}
\label{fithd217}
\end{figure}

\begin{figure*}
\begin{minipage}{8.8cm}
\hspace{-.1cm}\resizebox{\hsize}{!}
{\includegraphics{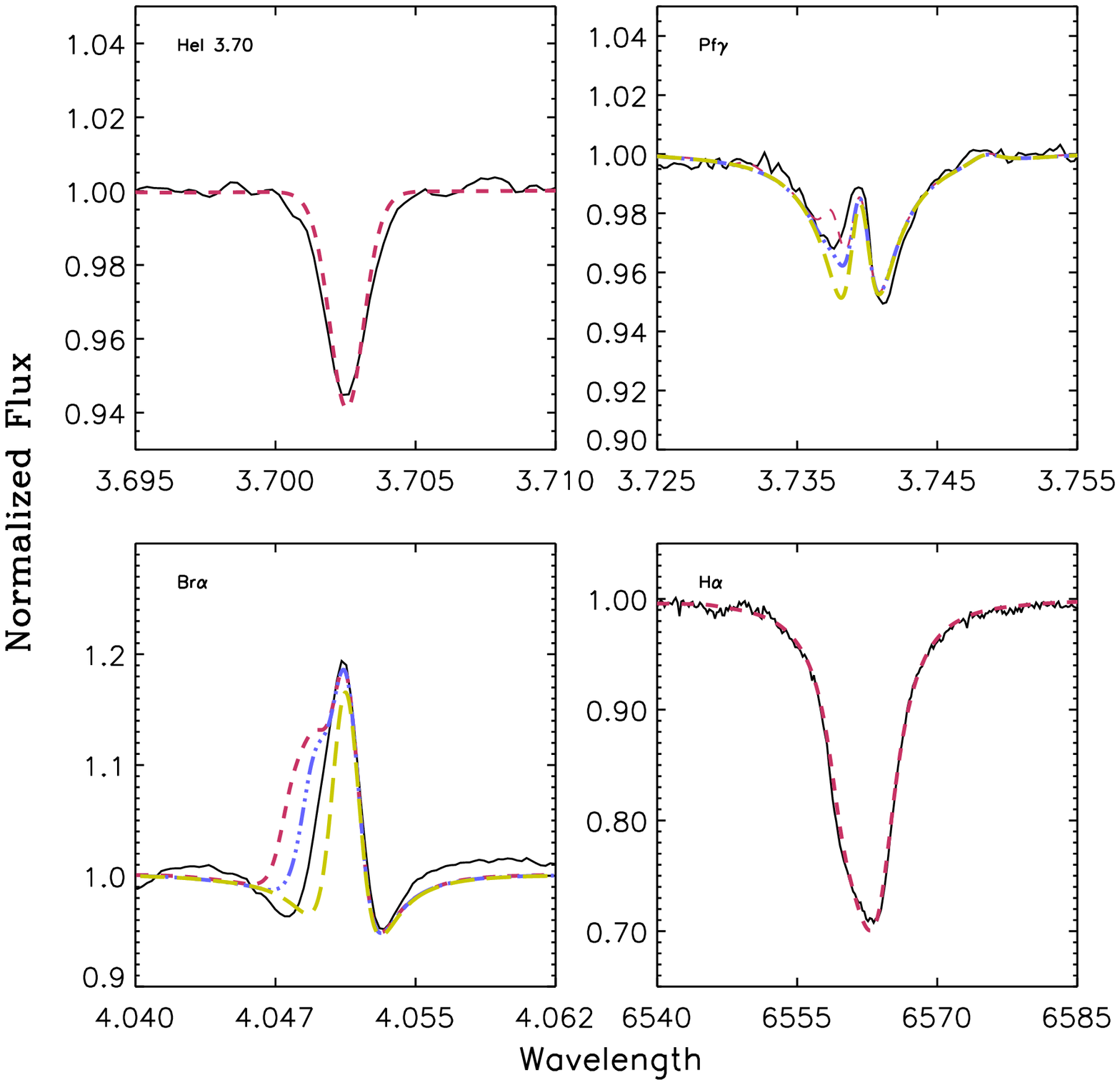}}
\end{minipage}
\hfil
\begin{minipage}{9.0cm}
\hspace{-.4cm}
\resizebox{\hsize}{!} 
{\includegraphics{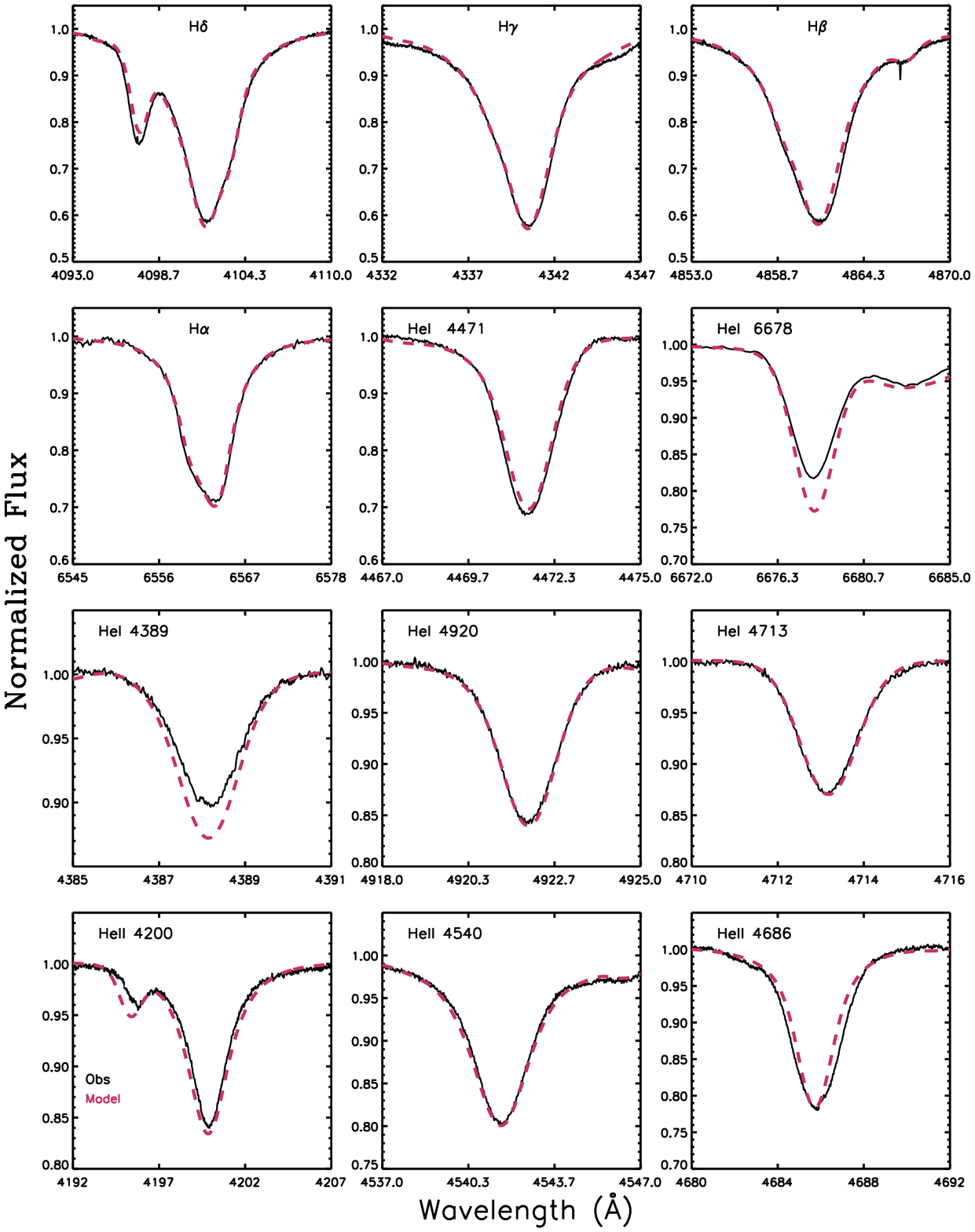}}
\end{minipage}
\begin{minipage}{8.8cm}
\hspace{-.1cm}\resizebox{\hsize}{!}
   {\includegraphics{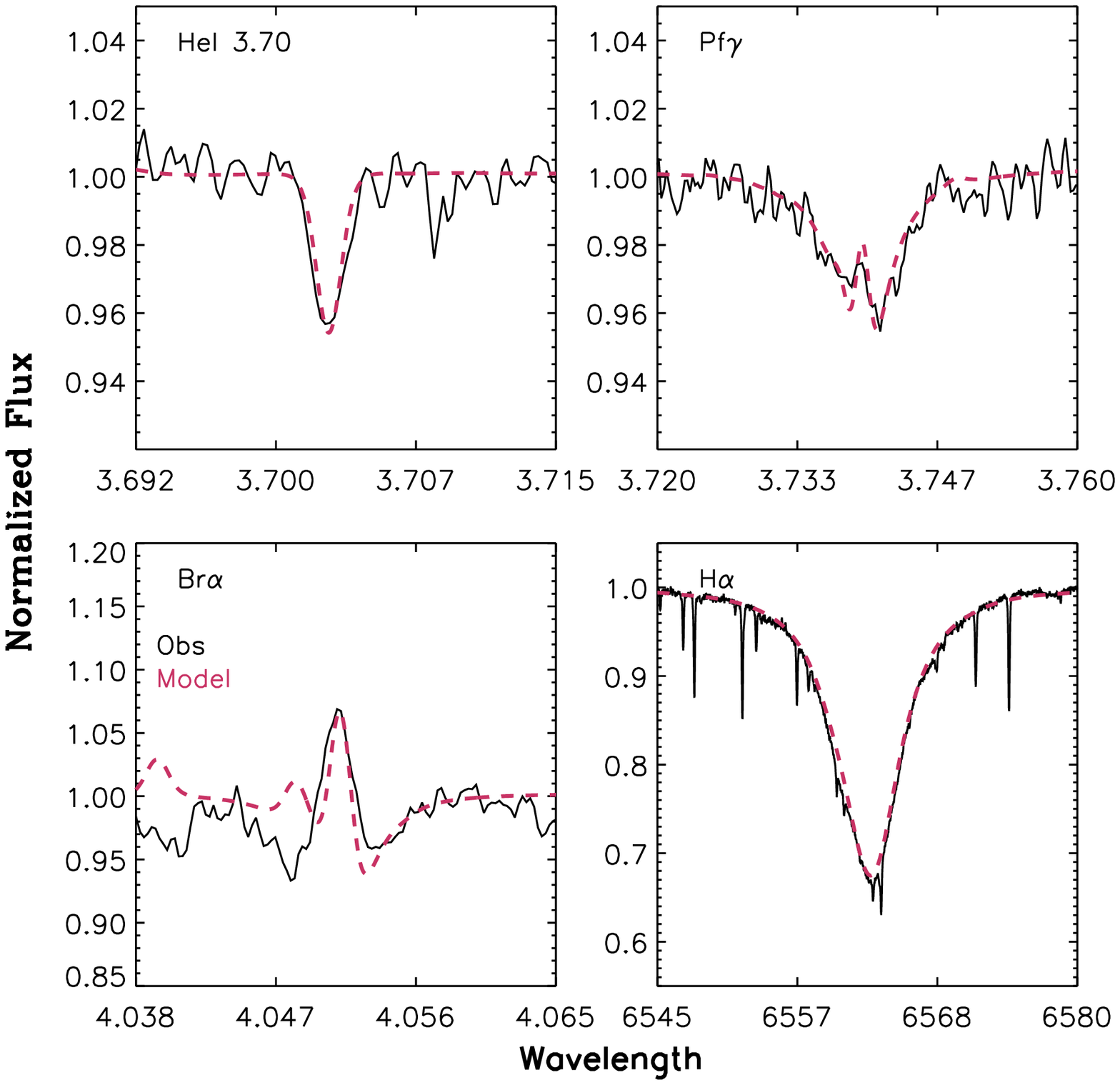}}
\end{minipage}
\hfil
\hspace{-.2cm}\begin{minipage}{9.0cm}
\resizebox{\hsize}{!} 
   {\includegraphics{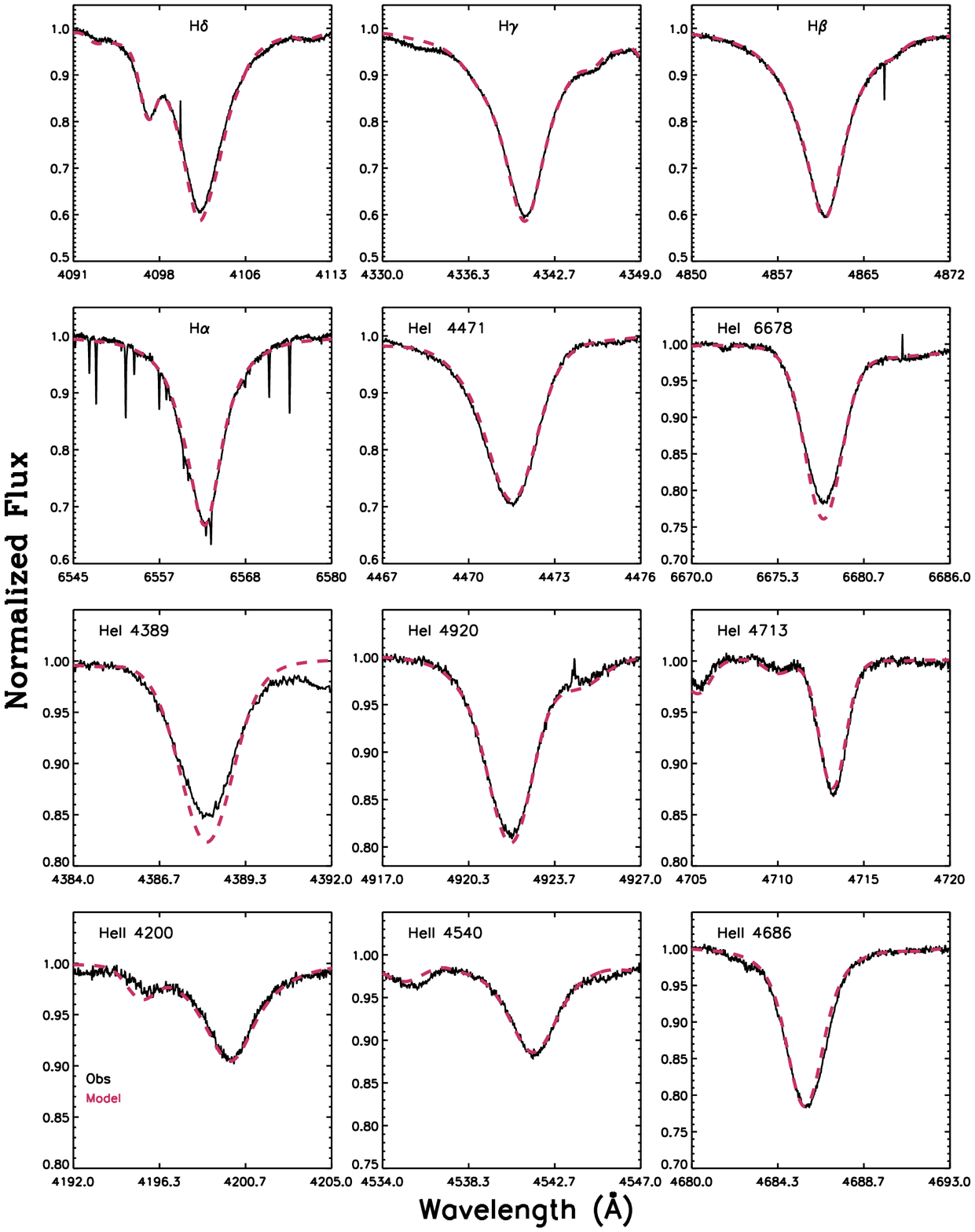}}
\end{minipage}
\caption{Model fits (dashed) to IR and optical observations (solid) 
of HD\,36861 (top) and
HD\,76341 (bottom). 
For HD36861 we show as well model fits to \Pfg\ and \Bra\ computed 
without \HeI\ components (dashed-dotted), and with
no He at all (long dashed, see text).
Optical data were retrieved from the UVES POPS Catalog
archive, programme 266.D-5655(A).}
\label{fitiropthd368hd763}
\end{figure*}

\paragraph{$\epsilon$ Ori.}
\label{par-eori}
%
%
Several spectroscopic studies of $\epsilon$~Ori using {\sc cmfgen}
\citep[e.g.][]{searle08} and {\sc fastwind} \citepalias{repo05}, yielding
similar parameters, have recently appeared in the literature. While the {\sc
cmfgen} study made use of UV and optical data, the {\sc fastwind} one used 
infrared spectroscopic observations alone. Interestingly, 
\citetalias{repo05} could derive only an upper limit on the effective
temperature due to the absence of \HeII\ IR diagnostic lines. Our results,
arising from a full UV to IR investigation and displayed in Table\ref{para},
revise down the effective temperature by roughly 1000~K. A striking result
revealed by Fig.\ref{fitiropepsori} is the requirement of strong clumping to
match the IR spectra. Figure~\ref{fitiropepsori} demonstrates the failure of
an unclumped ($\fv=1$) wind to reproduce the $L$-band lines. As for most of 
the previous stars, such strong clumping leads to overestimated HI
and \HeI\ line cores. In the case of $\epsilon$~Ori, however, this mismatch
could be also due to the intrinsic line profile variability, as
the IR and optical observations where taken at different epochs.


\section{Objects with thin winds}
\label{thin}

\paragraph{HD\,217086.}
\label{par-hd217}
%
%
Our derived stellar parameters for this fast rotating dwarf are in excellent
agreement with those obtained by \citetalias{repo04} and \citetalias{repo05}
by means of optical and IR spectroscopy. While their IR analysis could
provide an upper limit to the mass-loss rate being a factor of two lower
than the optical one, no firm determination of this parameter could be
assessed. Our $L$-band data (see Fig.~\ref{fithd217}) clearly show how \Bra\
constitutes a much more powerful \Mdot\ diagnostic for low density winds
than the previously used \Ha\ and \Brg\ lines. 
Our UV through IR study yields a less clumped wind than for the case of the
   supergiants. Without the UV we could not have broken the \Mdot - clumping
   degeneracy though.

\paragraph{HD\,36861.}
\label{par-hd368}
%
This object displays the strongest wind within our sample of stars with low
density winds, as to be expected from its O8 III((f)) spectral type
classification. Our derived effective temperature is hotter than the one
adopted by \citetalias{puls06a} (based on the calibrations used by
\citealt{markova04}). Interestingly, no clumping is required by our models
to match the IR and optical lines. This result is consistent with one of the
two possible solutions found by \citetalias{puls06a}.
Figure.~\ref{fitiropthd368hd763} (upper two panels) shows the excellent
agreement of our models with the observations. Exceptions are two of the
optical \HeI\ singlets (related to the so-called singlet problem,
\citealt{naj06}) and the He components of the IR H lines. Our models (see
Fig.~\ref{fitiropthd368hd763}, upper left) tend to show the \HeI\ components
in emission.\footnote{As a guideline, we plot as well the line profiles for
\Bra\ and \Pfg\ computed without \HeI\ components (dashed-dotted) and with
no He at all (long dashed).} Since this discrepancy appears only in objects
with low density winds where the \HeI\ hydrogenic components form at higher
densities, we suggest that realistic broadening functions should be
developed and used to replace the assumed pure Doppler profiles. 

\paragraph{HD\,76341.}
\label{par-hd763}
%
As for HD\,36861, no clumping is required to reproduce optical and IR (also
UV, not shown here) spectra of HD\,76341 (Figure~\ref{fitiropthd368hd763},
lower two panels). Again we note the problem with the \HeI\ components in
\Bra, but also the enormous potential of this line to determine mass-loss rate
in thin winds when \Ha\ struggles to react to changes in \Mdot. 

\begin{figure}
\resizebox{\hsize}{!}
{\includegraphics{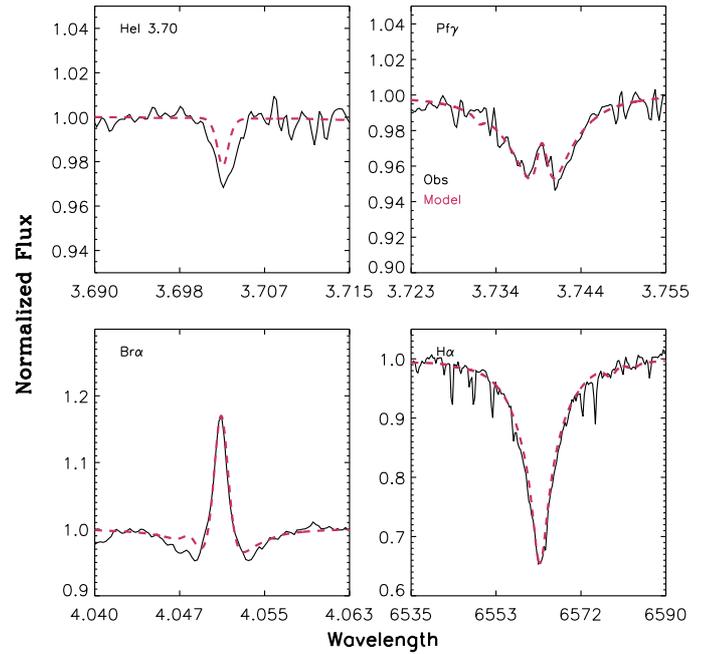}}
\caption{Model fits (dashed) to IR and \Ha\ observations of HD\,37468 (solid). 
The \Ha~profile was obtained with IDS at the INT Telescope on 
La Palma (kindly provided by S. Sim{\'o}n-D{\'{\i}}az).}
\label{fitiropthd374}
\end{figure}

\paragraph{HD\,37468.}
\label{par-hd374}
This is the object with the thinnest wind in our sample. Our analysis has
made use of available UV and optical spectra as well. Once more, no clumping
is required and we obtain \Mdot = 2 $\cdot 10^{-10}\, {\rm M_{\odot}/yr}$ as our current best estimate (see
Fig.~\ref{fitiropthd374}). The need of correct broadening functions for the
\HeI\ lines is again evident in the \Bra\ complex. Even though our \Mdot\
determination appears perfect, we stress that in
this regime of extremely low mass-loss rates the resulting synthetic \Bra\
profile can be very sensitive to the data set used for the hydrogen
collisional bound-bound processes (see below). 

In the following, we will discuss the formation of the specific shape of the
\Bra\ profile in these very thin winds in considerable detail.

\subsection{Theoretical considerations}
\label{thin_theory} 
As has been extensively discussed by \citet{mih78}, \citet{kud79},
\citet{najarro98}, \citet{PB04} and \citet{len04}, the low value of
$h\nu/kT$ in the IR leads to the fact that even small departures from LTE
become substantially amplified (in contrast to the situation in the UV and
optical).  This can be immediately seen from the line source-function in the
Rayleigh-Jeans limit,
\begin{equation}
S_L/B_{\nu} \approx (1 + \delta/(h \nu/kT))^{-1}, \qquad \delta = b_l/b_u -1
\label{sline}
\end{equation}
where $b_l$ and $b_u$ are the NLTE departure coefficients for the lower and
upper level, respectively. For temperatures at 30~kK, the value of $h\nu/kT$
is 0.24 at \Brg\ and 0.11 at \Bra. Thus, under typical thin-wind conditions
(where in the line forming region the lower level becomes underpopulated
compared to the upper one, see below) the line source-function can
easily exceed the continuum, or can become even ``negative'', i.e.,
dominated by induced emission. E.g., for the case of $b_l/b_u = 0.95$,
$S_L/B_{\nu} \approx 1.83$ at \Bra, whereas for a ratio of 0.9 already a
value of $S_L/B_{\nu} \approx 11$ is present, and lasering sets in at a
ratio of 0.89.

From these examples, it is immediately clear that the synthesized profile, 
particularly \Bra, reacts very sensitively to this ratio, where the major
effect regards the height of the narrow emission peak that we will
use to constrain the mass-loss rate. Thus, we must also check the
influence of uncertainties in atomic data and
atmospheric parameters that can influence this ratio and might weaken our
conclusions.

To investigate the general formation mechanism and the above problems, we
have calculated a large number of models exploring the sensitivity of \Bra\
on various effects, and will comment on those in the following, by means of
our model of HD\,37468 with atmospheric parameters as outlined in
Table~\ref{para}. 

\begin{figure*}
\resizebox{\hsize}{!} {\includegraphics{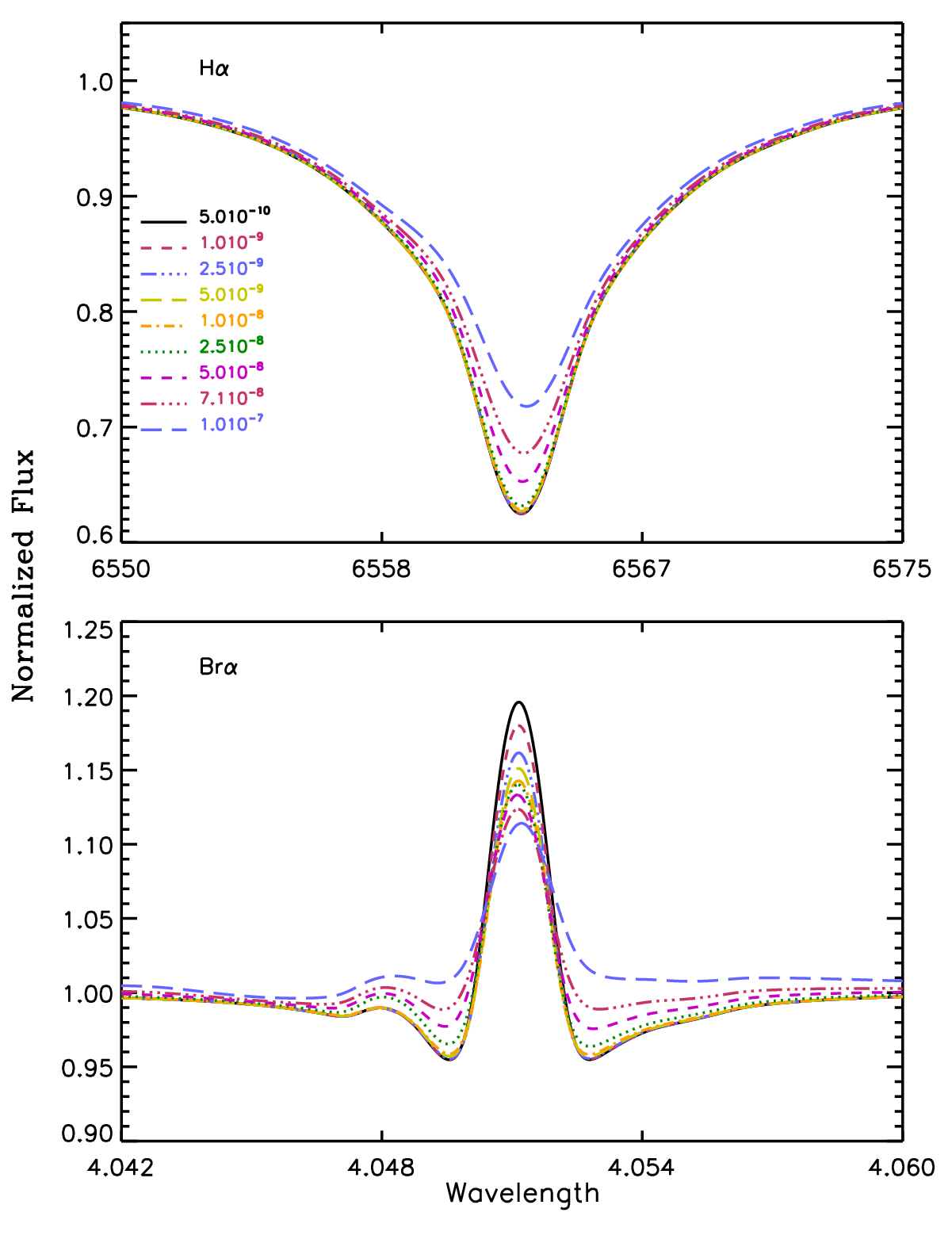}\includegraphics{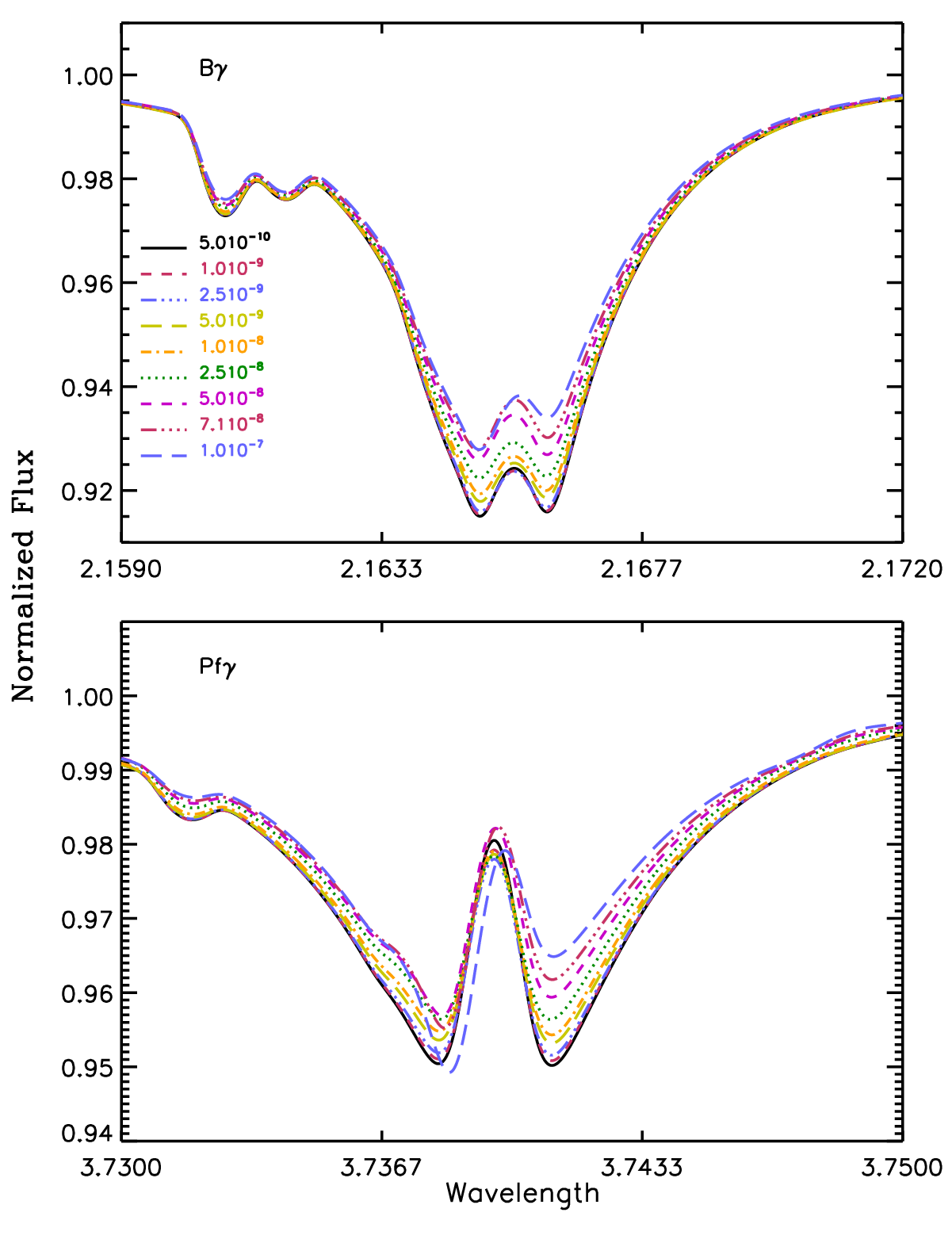}}
\caption{\Ha, \Bra, \Brg, and \Pfg\  profiles for our model of HD\,37468 and
different mass-loss rates (see text).}
\label{comphabra}
\end{figure*}

\paragraph{General behavior.} In Fig.~\ref{comphabra}, we compare the
reaction of
 \Ha, \Bra, \Brg, and \Pfg\ 
on different mass-loss rates, varied
within \Mdot\ = 5$\cdot 10^{-10}$ (solid black) and 1$\cdot 10^{-7}\, {\rm
M_{\odot}/yr}$ (long-dashed, blue). Whereas a clear reaction of \Ha\ is
found only for \Mdot\ $\ge$ 5$\cdot 10^{-8}\, {\rm M_{\odot}/yr}$, \Bra\
remains sensitive at even the lowermost values.  For increasing mass-loss,
the height of the emission peak in \Bra\ {\it decreases}, whereas the wings (in
absorption for lowest \Mdot) become more and more refilled, going into
emission around $10^{-7}\, {\rm M_{\odot}/yr}$.
 From Fig.~\ref{comphabra}
we note that also the \Brg\ and \Pfg\ line profiles are more sensitive than
\Ha\ in the thin wind regime as their
wings typically require a factor of 5 lower \Mdot\ to 
start displaying reactions to mass-loss.

\begin{figure}
\resizebox{\hsize}{!} 
   {\includegraphics[angle=90]{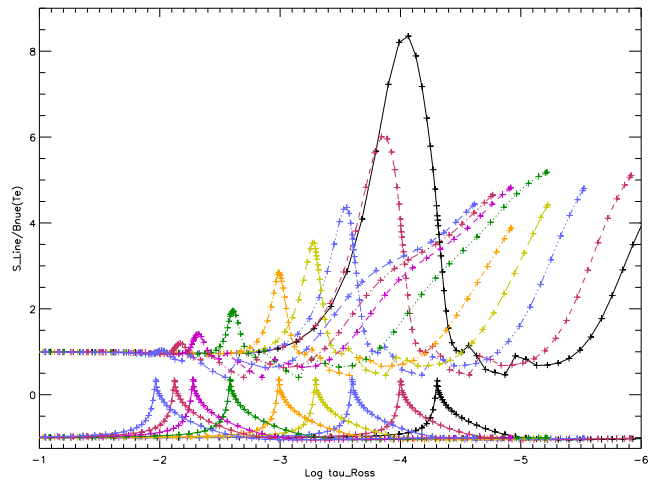}}
\caption{Line source function of \Bra\ (in units of the local Planck function)
and velocity gradient (vertically scaled) as a function of $\taur$, for the
same \Mdot-sequence as in Fig.~\ref{comphabra}. Note that the transition
region between photosphere and wind moves towards lower $\taur$ for
decreasing \Mdot, i.e., the model with the lowest mass-loss is located at
the right of the figure.} 
\label{compsltaur}
\end{figure}

The refilling of the wings with increasing mass-loss can be explained by the
increasing influence of the bound-free and free-free continuum (i.e.,
the typical continuum excess in stellar winds becomes visible) as well as a
certain ``conventional'' wind emission. However, at line
center the line processes always dominate and the peak height depends on
the location (in $\tau$) where the wind sets in.
This is shown in Fig.~\ref{compsltaur} which
 relates the strength of the \Bra\ source function
 at each point of the atmosphere with the gradient of 
velocity field and reveals whether the photosphere,
transition region or wind control the resulting \Bra\ line profile.

\paragraph{Depopulation of the $n=4$ level in the outer photosphere.} As
outlined in the introduction, already \citet{ah69} found in one of their
first NLTE-models a stronger depopulation of $n=4$, compared to $n=5$. They
argued as follows: Whenever the density becomes so low that collisional
coupling plays no role, the decisive processes are recombination and
cascading, as in nebulae. Because the decay channel 4 \rarrow\ 3 is very
efficient in this process, level 4 becomes stronger depopulated than level
5, and the core of \Bra\ goes into emission. 

\begin{figure*}
\begin{minipage}{8.8cm}
\resizebox{\hsize}{!}
   {\includegraphics{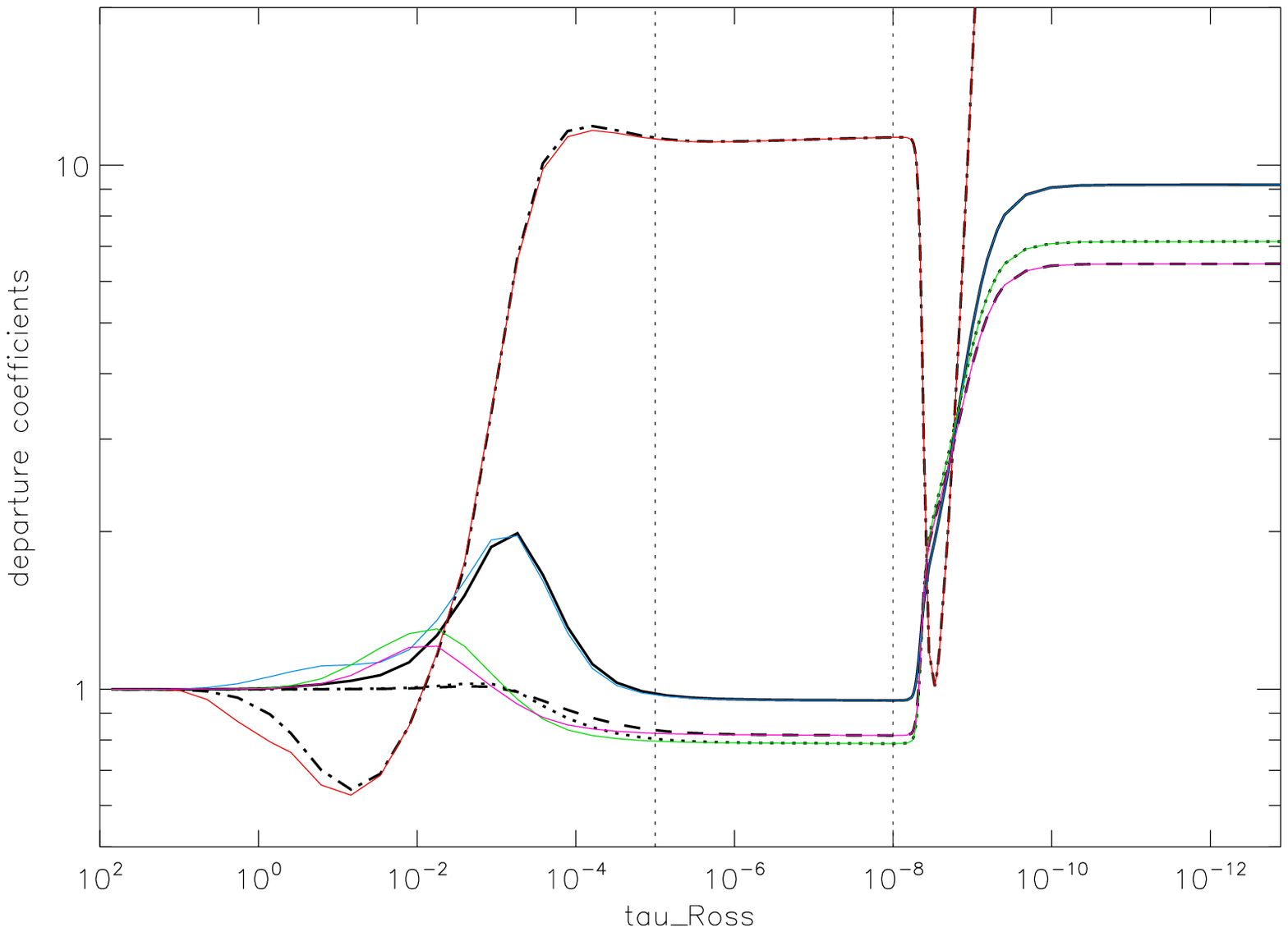}}
\end{minipage}
\hfill
\begin{minipage}{8.8cm}
   \resizebox{\hsize}{!}
      {\includegraphics{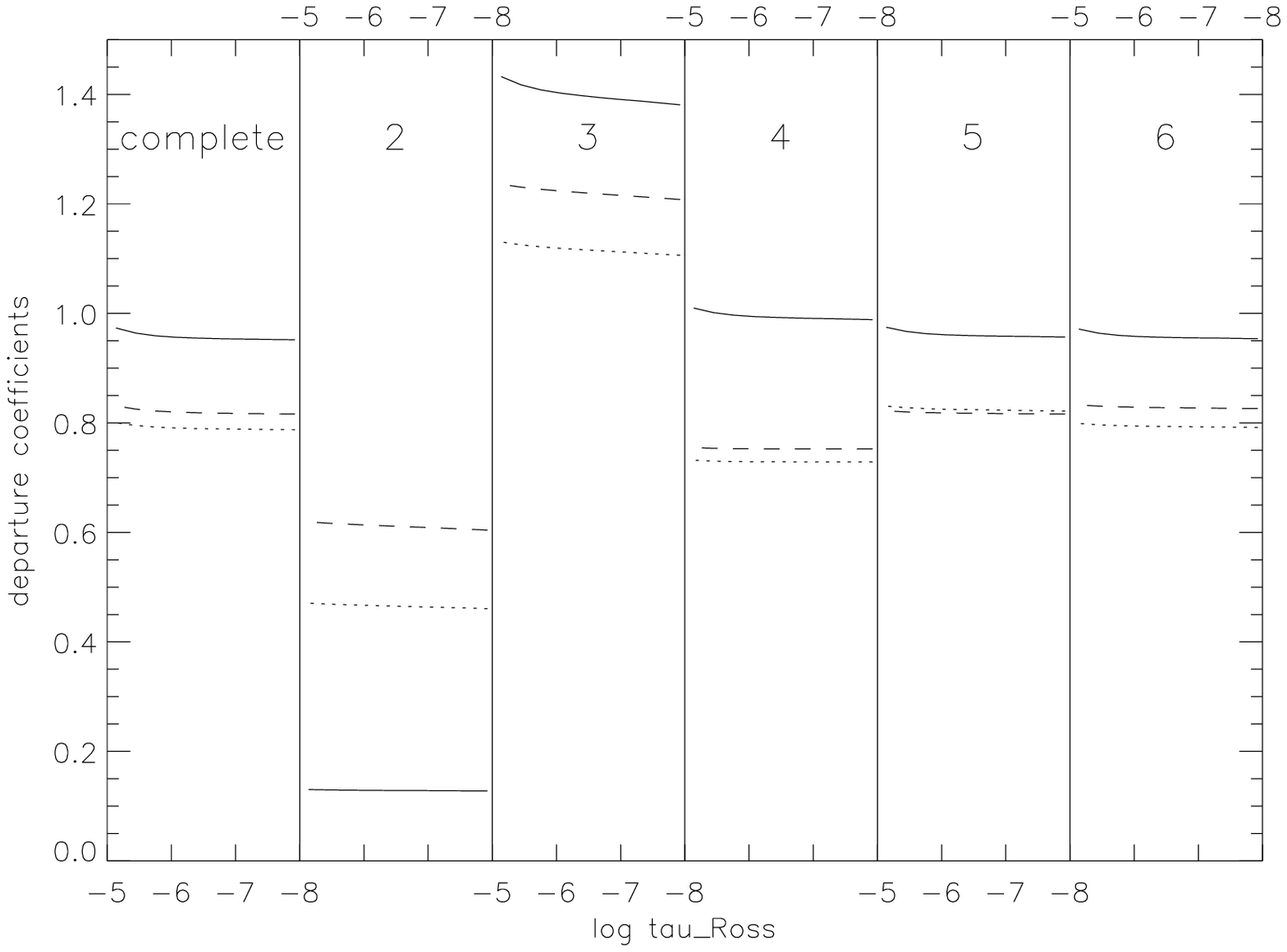}}
\end{minipage}
\caption{Underpopulation of the hydrogen $n=4$ level (compared to $n=5$) in
the outer photospheres of late-type O-dwarfs with very thin winds (pure
H/He model atmosphere).
\newline
{\bf Left}:
Departure coefficients for $n=1,2,4,5$ (dashed-dotted, solid, dotted,
dashed), accounting for {\it all} processes (radiative and collisional).
Overplotted are the corresponding departure coefficients (red, blue, green,
magenta) as resulting from a NLTE solution discarding the collisional
processes. Note the underpopulation of $n=4$
compared to $n=5$ in the outer photosphere (responsible for the line
core emission in \Bra), which is no longer present in the wind.
\newline
{\bf Right}: Departure coefficients of $n=2,4,5$ (solid, dotted, dashed) in
the outermost photosphere (corresponding to the region embraced by dotted
lines on the left), as resulting from the {\it complete} NLTE solution and various
approximations, the latter all without collisions: {\it 2}: nebula approximation (Case A); {\it 3}: as {\it 2},
but including excitation/induced deexcitation from resonance lines (roughly
Case B); {\it 4}: as {\it 3}, but including ionization from excited states;
{\it 5}: as {\it 4}, but including excitation/deexcitation from all lines with
lower level $n \le 3$; {\it 6}: as {\it 4}, but including excitation/deexcitation from
all lines with lower level $n \le 5$ (see text).}
\label{nebula}
\end{figure*}

Since their findings refer to different conditions (\Teff\ = 15,000 K), and
since it is important to understand the dependence of the depopulation on
the precision of the involved processes, we investigated the process in more
detail. As it turned out that line-blocking/blanketing effects have a minor
influence on the principal results for hydrogen (only the absolute peak
height of \Bra\ is affected, but not its systematic behavior), we used a
pure hydrogen/helium model (with parameters as derived for HD\,37468, but
with a very weak wind, \Mdot = $5\cdot 10^{-12}$ \Msun/yr) for this purpose,
in order to allow for a multitude of calculations. The results of our
investigation are displayed in Fig.~\ref{nebula}. In the optically thin part
of the photosphere ($10^{-4} < \taur < 10^{-8}$), all departures remain
roughly constant, where the ground-state (dashed-dotted) is overpopulated by
a factor of 10, the 2nd level (solid) is roughly at LTE and, indeed, $b_4$ 
(dashed) is smaller than $b_5$ (dotted). In contrast, the
groundstate is strongly overpopulated in the wind ($\propto r^2$ for constant
temperature, see below), whereas the excited levels are overpopulated by
factors between 10 and 5, with a different order than in the photosphere,
i.e., $b_5 < b_4 < b_2$.

In the following, we will investigate the influence of various effects that
determine this pattern, by solving the NLTE rate equations and omitting
certain rates, but using a fixed radiation field (which is legitimate at
least in the optically thin part of the atmosphere). 

At first, we checked the influence of the collisional rates, by leaving them
out everywhere. The result of this simulation is displayed on the left of
Fig.~\ref{nebula}. Obviously, below $\log \taur = -5$ collisions do not play
any role, since  the departure coefficients with (in black) and without
(in color) collisions are identical. In the lower photosphere, of course
all departures are affected by collisions, but particularly for level 4 and
5 a difference is visible until $\log \taur = -5$, where these levels remain
thermalized until $\log \taur = -2$. Since the major part of the line core
forming region of \Bra\ is just in the range between $10^{-2} < \taur <
10^{-4}$, we have to conclude that collisions {\it do} play a certain role
in the formation of \Bra, and will return to this point later on. 

Before doing so, however, we try to understand the depopulating processes in
the outer photosphere (opposed to the conditions in the wind) by
concentrating on that part which is not affected by collisions, in order to
gain insight into the dominating mechanism. To this end, we display the
results of various approximations on the right of Fig.~\ref{nebula}, by
considering the outermost photosphere (indicated by dotted lines on the left
panel). 

The first approximation ({\it ``2''}) follows the suggestion by
\citet{ah69}, i.e., we solved the rate-equations for a Case A nebula
approximation, i.e., allowed for ground-state ionization, spontaneous decays
and radiative recombination into the excited levels. In this case, we derive
departures with $1 > b_5 > b_4 > b_2$ (i.e., level 4 is more strongly depopulated
than level 5), but far away from the {\it complete} solution. Moreover, the
ground-state (not displayed) is roughly consistent with the exact solution,
whereas the departures of the excited levels in the wind do {\it not} differ
from the conditions in the outer photosphere. 

In simulation {\it 3}, we switched on the resonance lines, by including the
corresponding excitation and induced deexcitation rates (roughly
corresponding to Case B nebula conditions).  Immediately, all departures
{\it in the wind} obtain values very close to the complete model (i.e., $b_2
> b_4 > b_5 > 1$), whereas in the outer photosphere the ``correct'' order is
achieved ($b_2 > b_5 > b_4$), though at a much too high level.

This is cured by simulation {\it 4}, where ionization from the excited
states is switched on. By this process, all levels are depopulated again,
and the corresponding solution looks very close to the exact one. (The
departures in the wind remain unaffected, since these rates are {\it almost}
unimportant because of the strongly diluted radiation field).

One might conclude now that the remaining missing rates
(excitation/deexcitation between excited levels) are negligible.
Unfortunately, this is not the case, at least if one is interested in the
precise ratio of $b_5/b_4$, which is of major importance in our
investigation. Though the line transitions between the excited states are
optically thin, the mean line intensity is still large enough to be of
influence. This becomes clear by comparing simulation {\it 5} with {\it 6}.
In the former, we have included the excitation/deexcitation from all lines
with lower level $n \le 3$ (i.e., level 4 and 5 are only affected by line
transitions in terms of resonance lines and spontaneous decay), with the
effect that now $b_4 \ga b_5$, i.e., the emission core of \Bra\ would
disappear. Only if all bound-bound processes for lines with at least a lower
level of $n \le 5$ are included (simulation {\it 6}), the ``exact result''
is recovered (i.e., higher levels contribute indeed almost only by
spontaneous decay). 

\begin{figure*}
\begin{minipage}{5.0cm}
\hspace{-.3cm}\resizebox{\hsize}{!}
      {\includegraphics[angle=0]{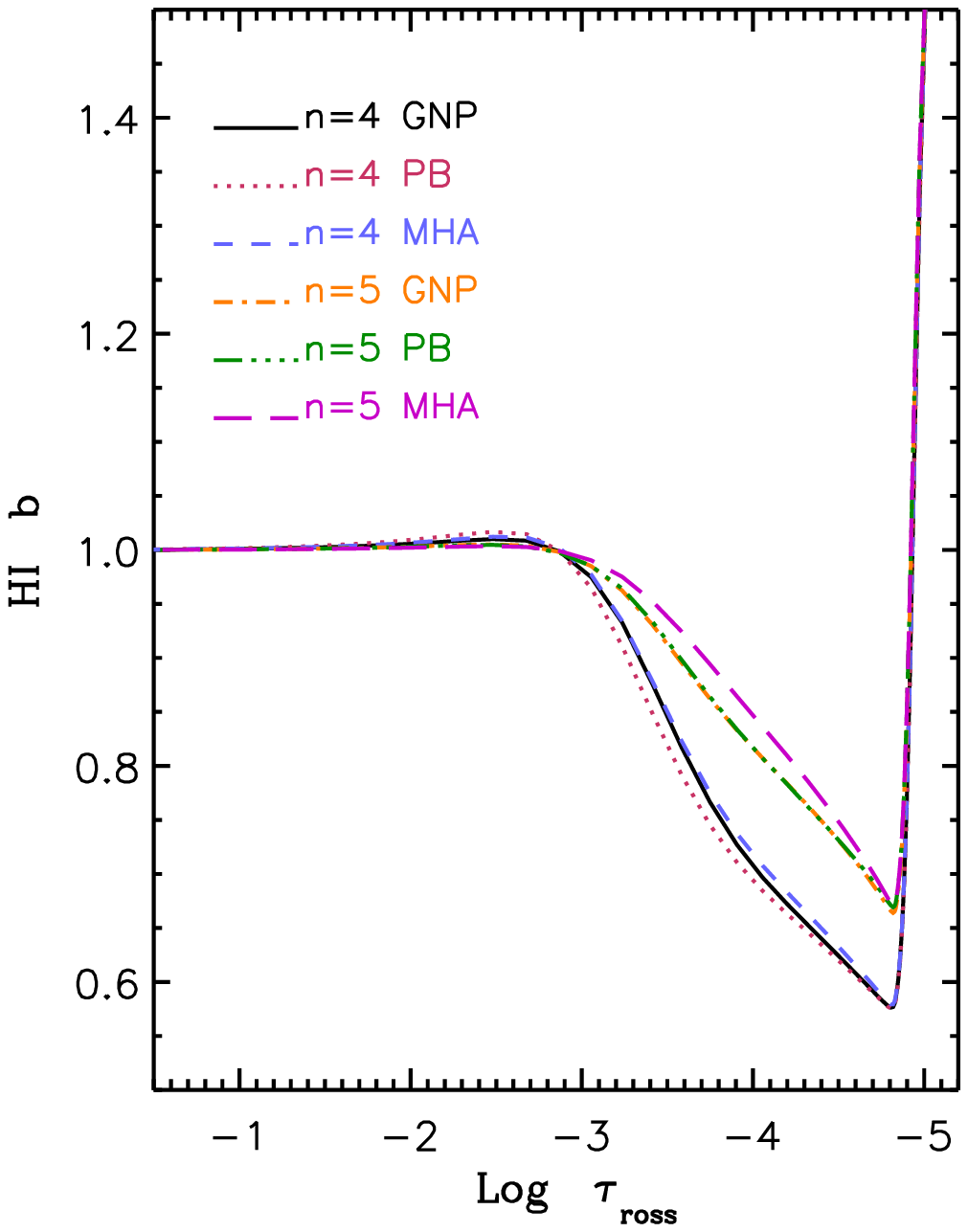}}
\end{minipage}
\hspace{-1.cm}
\begin{minipage}{5.0cm}
\resizebox{\hsize}{!}
      {\includegraphics[angle=0]{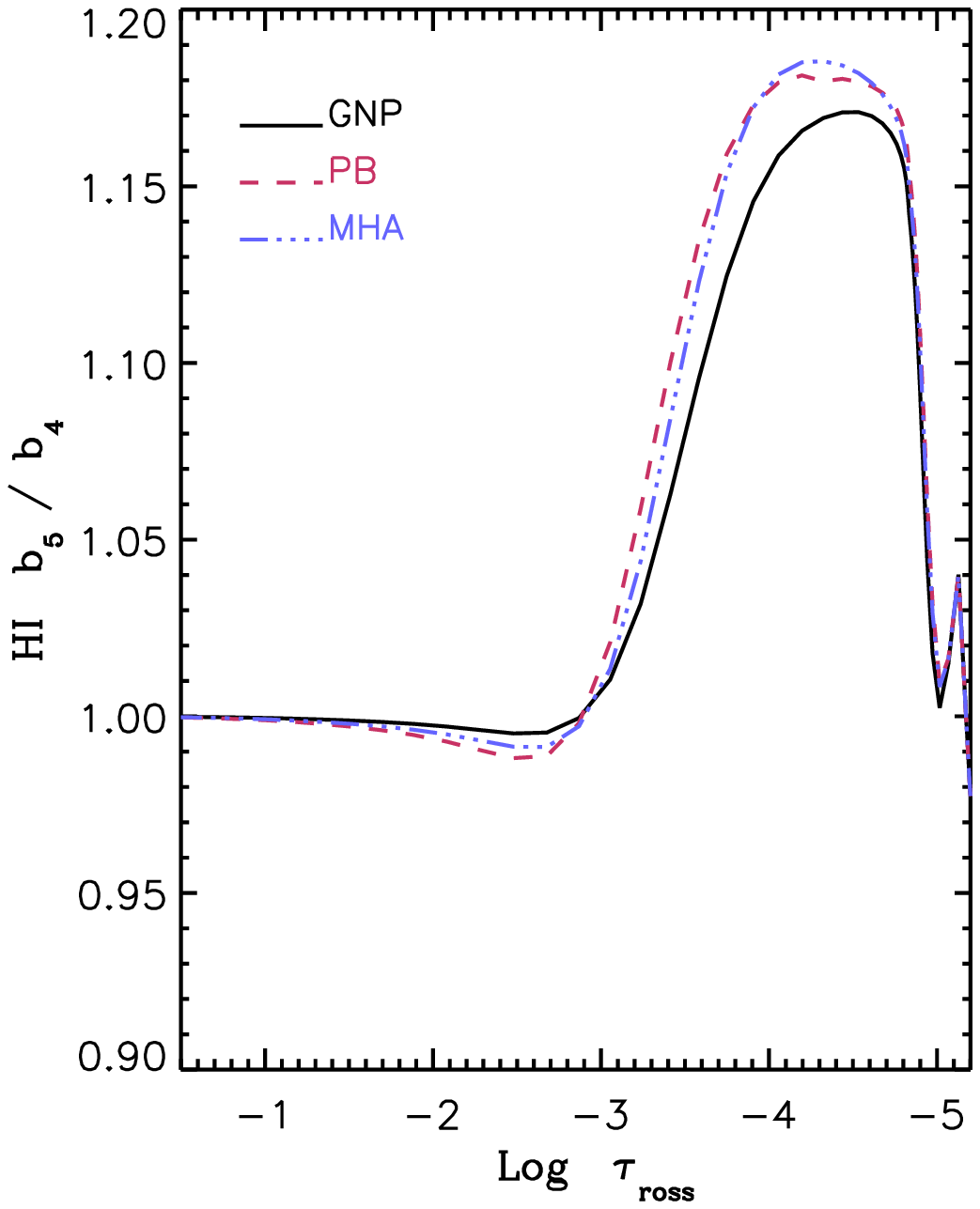}}
\end{minipage}
\hfill
\begin{minipage}{8.8cm}
\resizebox{\hsize}{!}
      {\includegraphics[angle=0]{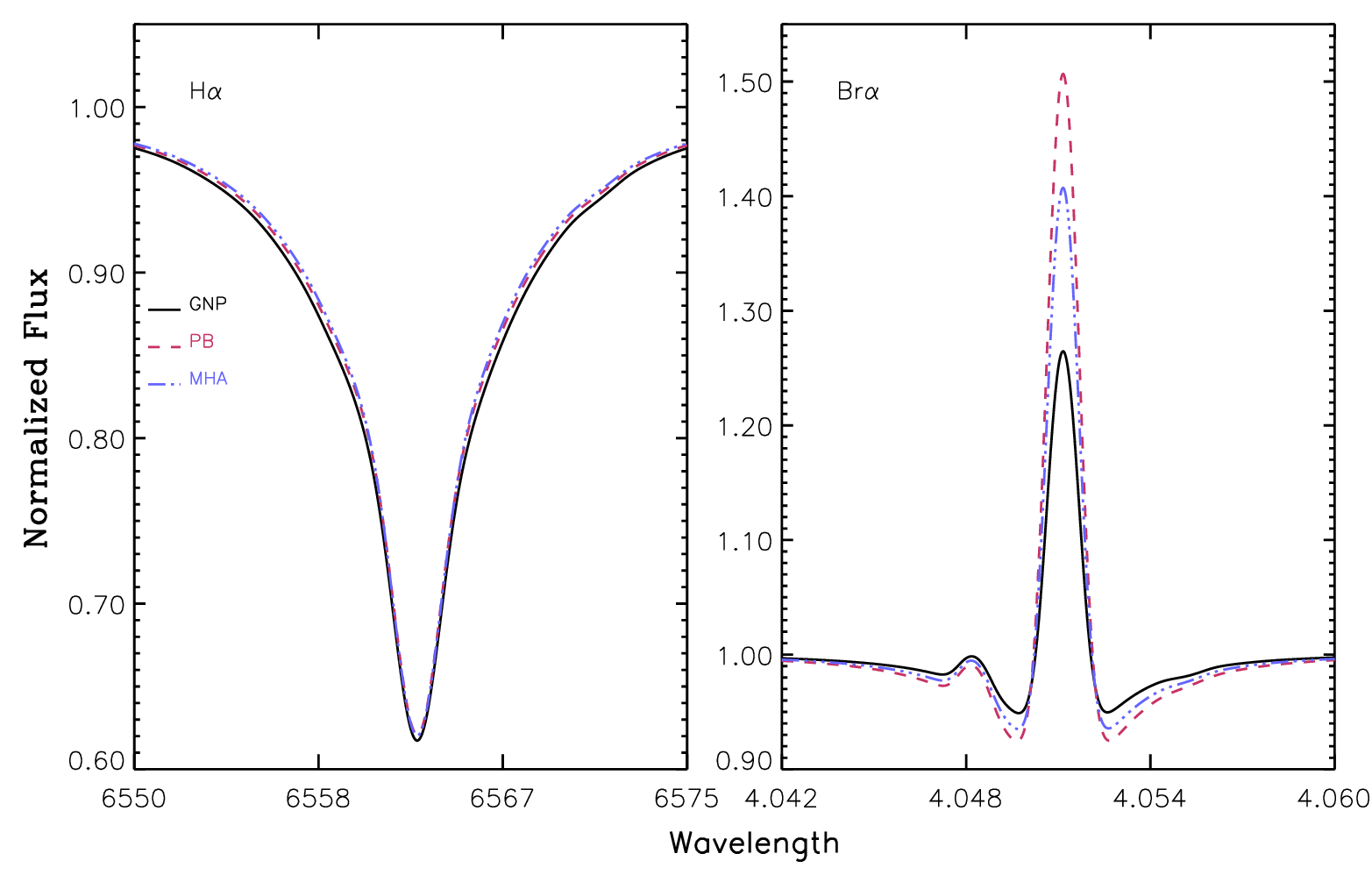}}
\end{minipage}
\caption{Influence of different collision strengths on the population of
$n=4/5$ (left) and the corresponding \Bra\ profile (right). For comparison,
we also display the changes in \Ha. See text.}
\label{colldata}
\end{figure*}

In summary, even in the outermost photosphere the actual value of $b_5/b_4$
is controlled by (almost) all radiative processes (with additional
collisional contributions in the lower photosphere), and thus depends on a
precise description of the continuum and line radiation field and to a lesser 
extent on the correct run of the temperature stratification (entering the
recombination coefficients). Interestingly,
however, we have also seen that in almost all of our simulations level 4 is
more underpopulated than level 5, independent of the various processes
considered. In Appendix~\ref{outerconditions} we will show that at least
this principal behavior can be regarded as the consequence of a typical
nebula-like situation, namely as due to the competition between
recombination and downwards transitions.

In the wind, on the other hand, a Case B like nebula approximation is able 
to explain the run of all hydrogen occupation numbers alone.  As it will be
shown in the appendix, the actual conditions in the {\it outer} wind depend
strongly on whether line-blocking/-blanketing is considered or not. In all
cases, however, the abrupt decrease of the \Bra\ source function in the
transition region between photosphere and wind is triggered by the onset of
dilution and the Doppler-effect shifting the (resonance-)lines into the
neighboring continuum, thus effectively pumping the excited levels.

\subsection{Influence of various parameters}
\label{influence}

\paragraph{Collision strengths.} As already outlined, collisions {\it do}
play a role in the formation of the emission peak of \Bra\ and, even more,
in the line wings. In particular, these are the collisional
ionization/recombination processes for $n \ge 3$ and the collisional
excitation/deexcitation processes within transitions $i \leftrightarrow j, i
\ge 3,j \ge i+1$, (strongest for $j=i+1$) which keep the occupation numbers
for levels $\ge 3$ and higher in of close to LTE than expected from considering
the radiative processes (for given radiation field) alone. From comparing
the departure coefficients calculated with and without collisional rates,
one might predict that a decrease of the collisional strengths in complete
models will increase the strengths of the absorptions wings (since, around
$\taur = 10^{-2}$, $b_4 \approx b_5 \approx 1$ compared to $b_4 > b_5 > 1$
with and without collisions, respectively), whereas the influence on the
emission is difficult to estimate. In Fig.~\ref{colldata}, we display the
result of three calculations using different sets of collision strengths
and its impact on the \Ha\ and \Bra\ profiles.

Our updated {\sc cmfgen} ``standard'' model utilizes the hydrogen collision
strengths from \citet{mih75} (MHA), which are compared to our previous data
set from \cite{giovanardi87} (GNP) and recent collision strengths
from \citet{PB04}\footnote{based on ab-initio calculations by Keith Butler}
(PB). The difference between these data sets is significant, particularly
for transitions with intermediate $i,j$ such as \Bra. Our ``standard'' MHA
collision strengths lie in between the GNP and PB data sets, the latter
being typically a factor of five smaller than the GNP implementation.
However, the reaction of the departure coefficients seems to be small. The
``only'' difference is a weak increase of $b_4$ in the lower part of the
line-forming region (as predicted above) and a weak decrease of $b_4$ in the
outer one, i.e., the NLTE effects become increased everywhere. Consequently,
the absorption wings of \Bra\ become deeper and the emission-peak higher,
when using the data set with reduced collision strengths (PB) by
\citet{PB04}\footnote{Note that a similar investigation performed by these
authors gave different results, due to numerical problems (N. Przybilla,
priv. comm.).} (Fig.~\ref{colldata}, right panel, dotted profile), where the
small differences in departure coefficients are (non-linearly) amplified
according to Eq.~\ref{sline}.
Since the changes in collision strengths affect both the peak of the line core
and the absorption wings of \Bra, they cannot be mapped directly onto \Mdot\
variations, as the latter only modifies the line core in the thin wind case.
Such changes may rather be accomplished by slightly modifying the 
gravity of the star.

\begin{figure}
\resizebox{\hsize}{!} 
   {\includegraphics{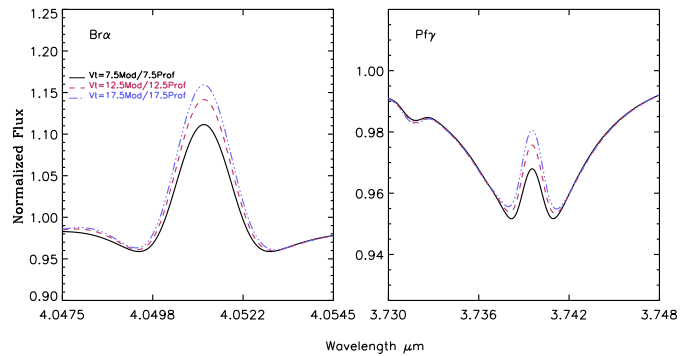}}
\caption{Influence of different microturbulent velocities on
\Bra\ (left) and \Pfg\ (right). See text.}
\label{compvturb}
\end{figure}

\paragraph{Microturbulence.} One of the basic ``unknowns'' in the
calculation of synthetic profiles based on model atmospheres is the
microturbulent velocity, \Vt\ (e.g., \citealt{sh98, villa2000, repo04,
hunter07}). Though it is possible to obtain a ``compromise'' estimate for
this quantity from a simultaneous fit to a multitude of different lines, it
is well known that different lines indicate different values, pointing to a
dependence on atmospheric height. As a rule of thumb, in the parameter range
considered here a value of \Vt\ on the order of 10~\kms\ seems to be
consistent with a variety of investigations. Fig.~\ref{compvturb} compares
the influence of this quantity on the synthetic \Bra\ profile, again by
means of our model of HD\,37468, and a typical mass-loss rate. Though a
small impact is visible in the blue wing (due to the reaction of the \HeI\
component), the major effect concerns the height of the emission peak, which
increases for increasing \Vt\ (varied between 7.5 and 17.5 \kms). By
comparing with Fig.~\ref{comphabra}, we see that for (very) thin winds (with
the line wings well in absorption), an uncertainty in \Vt\ of $\pm$5 \kms\
(which is a typical value) can easily induce uncertainties of a factor of
two in the deduced \Mdot. We note, however, that unlike \Mdot\ which changes
``only'' the height of the emission peak (see Fig.~\ref{comphabra}),
microturbulence modifies both its height and width
(see Fig.~\ref{compvturb}-left).
A similar effect as for \Bra\ but with lower amplitude is found
  for \Pfg\ (see Fig.~\ref{compvturb}-right).
Therefore, provided the spectral resolution is high
enough, in principle one could separate the effects of \Mdot\ and 
microturbulence and reduce the uncertainty in the final \Mdot\ estimate.

Most importantly, micro-turbulence affects level n= 4 (similar to the
influence of the collisional strengths): The higher the turbulence, the
sooner and the more effective this level becomes depopulated, increasing the
line source-function. Thus, it is important to adapt \Vt\ already in the
atmospheric model (\rarrow\ changes in the occupation numbers), and not only
in the formal integral, as it is often done with respect to metallic lines
and \HeI. 

\begin{figure}
\resizebox{\hsize}{!}
{\includegraphics
{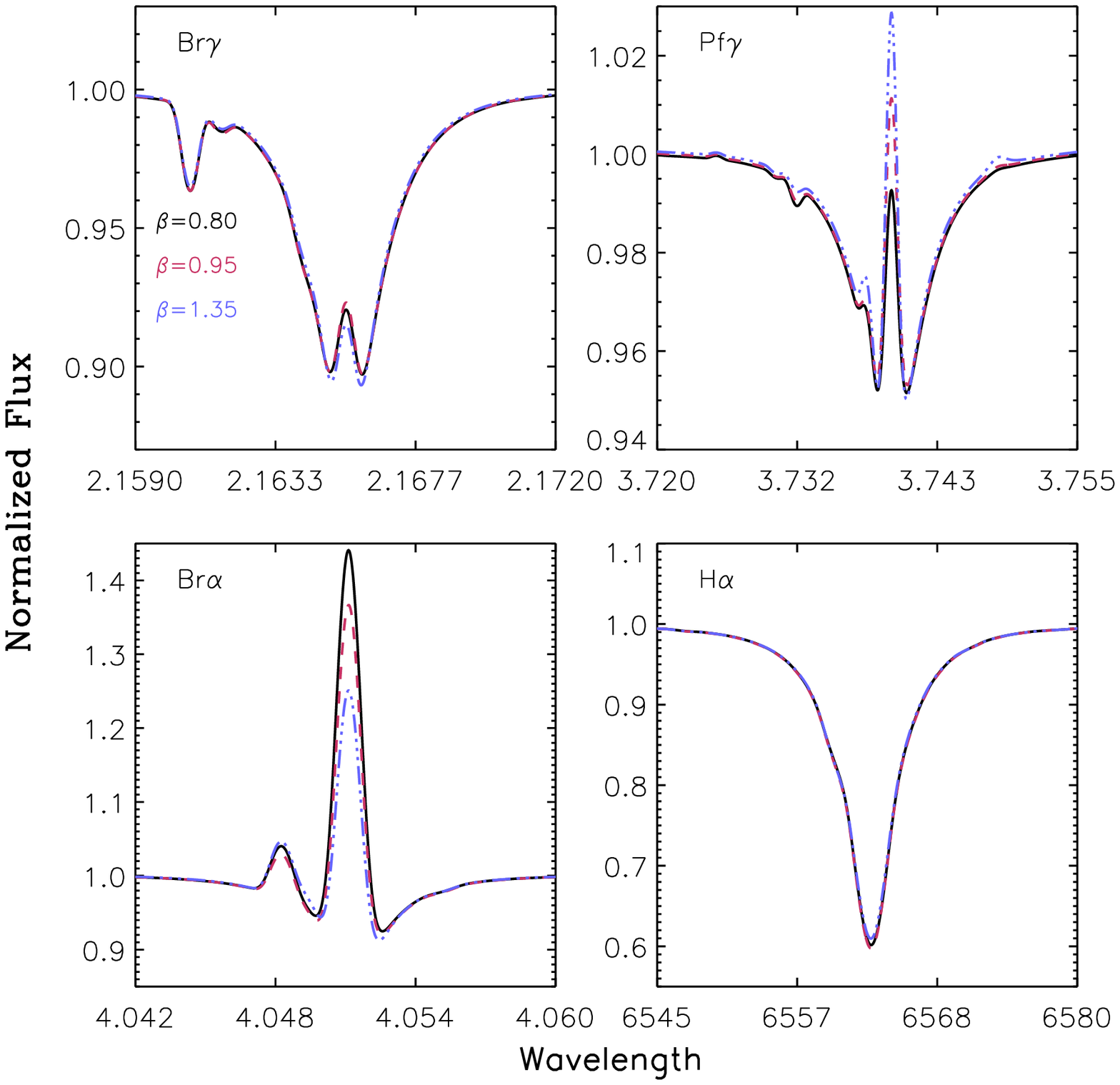}}

\caption{Influence of $\beta$ in thin wind objects (here: our model of
HD\,76341). Three combinations of $\beta$ and \Mdot\ which keep the \Ha\
core at the same depth. The different sensitivity of the cores of \Bra\ and
\Pfg\ to $\beta$ breaks the well-known \Mdot-$\beta$ degeneracy.
See text.}
\label{fibeta}
\end{figure}

\paragraph{$\beta$-law.} Here we concentrate on the effects of the steepness
of the wind velocity law (expressed in terms of $\beta$) on the line
profiles from thin wind objects. At first, for very thin (= weak) winds such
as the one from HD\,37468, there is almost no reaction at all, since the
profile, particularly the line core, is {\it not} formed in the wind. For
winds with a somewhat higher density (as HD\,76341) where the line core of
\Ha\ already reacts, the situation is somewhat different.
Figure~\ref{fibeta} shows corresponding $L$-Band hydrogen lines together
with \Ha, where both $\beta$ and \Mdot\ have been modified in opposite
directions to keep the \Ha\ core at the same depth. (For this model, such a
combination preserves the \Brg\ core as well). Interestingly, however, the
cores of \Bra\ and \Pfg\ still react strongly, showing that these lines,
together with \Ha\ and/or \Brg, can be confidently used to constrain both
$\beta$ and \Mdot\ in objects with (not too) thin winds and to break the
classical \Mdot-$\beta$ degeneracy. 

\paragraph{Transition Velocity.} A variation of the transition velocity
(which defines the transition from photosphere to wind) in the thin/weak wind case
shows only minor effects (not displayed here) in the emission core of \Bra,
provided such transition takes places at reasonable velocities (roughly between
0.05 {\ldots} 0.5 V$_{\rm sound}$). For the rest of the diagnostic lines
considered here, we find no effect at all. If the transition is moved deeper
into the photosphere, i.e., below 0.05 V$_{\rm sound}$, the
\Bra\ core starts to display a moderate sensitivity. 

Finally, we investigated how \Bra\ and \Pfg\ respond to clumping
in the thin wind regime. Similarly to the dependence on the transition velocity,
we found no effects, unless severe clumping is already present at the base of the
photosphere, in which case minor effects start to appear in the core of \Bra.

\section{Discussion}
\label{discussion}

\subsection{On the reliability of the derived mass-loss rates.}
\label{mdot_reliability} 
Since the number of \Mdot\ determinations for OB-stars has significantly
increased during the last decade, and largely differing values even for the
same objects can be found in the literature, it might be necessary to
comment on the reliability of the data provided here. Actually, the major
origin of differences in the mass-loss rates bases on different assumptions
regarding the clumping properties of the wind. These range from unclumped
media over microclumping (constant or stratified, with different
prescriptions on the clumping-law) to the inclusion of the effects from
macroclumping and vorosity. 

From the comparison provided in Appendix~\ref{detcomp}, it becomes obvious
that in most cases the {\it basic} quantity which can be deduced from the
observations, namely the optical depth invariant $Q$, is rather consistent
within a variety of studies, at least when concentrating on
$\rho^2$-dependent diagnostics. Diagnostics relying mostly on UV resonance
lines are more strongly affected, as can be seen, e.g., from the differences
in the corresponding $\Qr$ value derived here and by \citet{fulli06}, who
studied the \PV\ resonance line alone. Reasons for such discrepancy are (i)
the influence of X-ray/EUV emission on the ionization equilibrium,
particularly in the mid and outer wind (see below), and (ii) the strong
impact of macroclumping/vorosity on the formation of resonance lines \citep[
and references therein]{Sundqvist10a, Sundqvist10b}.

In this work, we have derived the stellar and wind parameters from a
consistent, multi-wavelength analysis, allowing for a rather general
clumping law. Since we obtained almost perfect simulations of the observed
energy distributions, from the UV to the IR (and sometimes even the radio
regime), we are quite confident on the quality of the provided values.
Corresponding error estimates have been quoted in Sect.~\ref{codes}. Even
admitting that the neglect of macroclumping/vorosity might influence the UV
resonance lines (particularly those of intermediate strength) to a certain
degree, and that our treatment of the X-ray emission is only a first
approximation, the ubiquity of excellent fits to features from a variety of
elements forming in different layers cannot be considered as pure
coincidence. We are optimistic that problems within individual features
(which can be disastrous in analyses concentrating on such features alone)
have, if at all, only a mild impact in a multi-wavelength study as performed
here.

One specific problem within our analysis which cannot be neglected is the
problem with the cores of the optical hydrogen and HeI lines, encountered for 
Cyg\,OB2~\#7 and \#8A as well as for $\alpha$ Cam (but see our corresponding
comment in Sect.~\ref{dense}). Briefly repeated, the problem reflects
the fact that within our analysis we were not able to obtain a simultaneous
fit for both the cores of these photospheric features and the IR-lines
formed in the lower/mid wind. For a perfect fit of the IR features, a low
\Mdot\ in parallel with a quite large clumping factor (CL$_1$ $\approx$ 0.01)
is needed, whereas a fit of the photospheric line cores requires a larger
\Mdot\ accompanied with moderate clumping (CL$_1$ $\approx$ 0.1). Further tests
are certainly necessary to clarify this problem (i.e., the shape of the
assumed clumping law might still not be optimum). In the meanwhile we
suggest that the mass-loss rates of the problematic objects should be
considered as a lower limit, and might need to be increased in future work.

\paragraph{Thin and weak winds.} 
As has been outlined in Sect.~\ref{intro}, the vast majority of mass-loss
determinations for weak winds and weak-wind candidates has been performed by
analyzing UV resonance lines, since \Ha\ is no longer usable at low \Mdot,
and the IR has not been invoked until now. 

A crucial point regarding the reliability of UV \Mdot-determinations for
weak-winded stars has been already provided by \citet[ see their
Fig. 20]{Puls08}. They calculated the diagnostic UV and \Ha\ profiles for a
set of thin and weak wind models where the mass-loss rate was varied by
almost two orders of magnitude whilst the X-ray luminosity was increased in
parallel to keep the ionization structure and thus the UV lines at the
observed level. The changes imposed on the wind did {\it
not} reach the photospheric levels, and thus the UV iron forest did not
change.  Most alarming, however, is their finding that essentially all
profiles could be equally well reproduced with {\it any} \Mdot\ combined
with an appropriate X-ray luminosity, $L_x$. Thus, no independent UV
mass-loss determination is feasible unless the X-ray properties of the star
are accurately known.

This problem, of course, is also present in our analysis. But here, we have
utilized the \Bra\ line as the primary \Mdot\ indicator, due to its high
sensitivity on \Mdot, being formed in the upper photosphere for weak winded
stars. A similar investigation of the same \Mdot-$L_x$ combinations as for
the UV, but using \Bra, lead to very promising result (see
\citealt{Puls08}, their Fig. 22): In the case of very low \Mdot, i.e., a
deep-seated line formation region, \Bra\ turned out to be basically 
unaffected by X-rays. Even with increasing mass-loss rate, the hydrogen core
of \Bra\ remains unaffected, at least for canonical X-ray luminosities ($L_{\rm
x}/L_{\rm bol} < 10^{-4})$. Changes, however, arise for the \HeII\
emission component of \Bra, caused by the large sensitivity of the
\HeII/\HeIII\ ionization equilibrium to X-rays. This problem needs to be
kept in mind for future analyses.

Finally, to assess the precision of the derived mass-loss rate for our
weak-wind object HD\,37468, we have to consider the major sources of error,
as discussed in Sect.~\ref{influence}, particularly the impact of the
hydrogen bound-bound collision strengths. Here we estimate a total error of
plus/minus 0.5 dex, which is irrelevant at such low mass-loss rates.

\subsection{Stratification of clumping factors}

Since our analysis comprises both $\rho$ and $\rho^2$ diagnostics (with all
the caveats regarding the potential impact of macroclumping/vorosity), we
are able to provide absolute values for $\fcl(r)$ as well as for \Mdot. This
is quite different from (and superior to) the investigation by
\citetalias{puls06a}, who could derive ``only'' {\it relative} values
(normalized to an outer wind assumed to be unclumped), because of relying on
$\rho^2$ diagnostics alone. 

\begin{figure}
\resizebox{\hsize}{!} {\includegraphics[angle=90]
{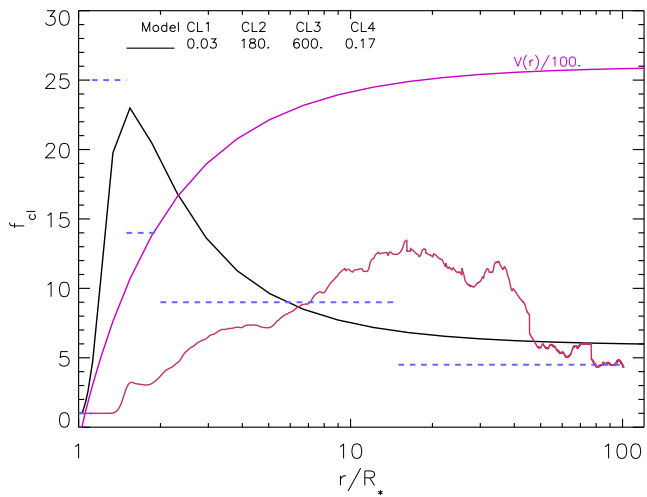}} \caption{Radial
stratification of the clumping factor, $f_{cl}$, for $\zeta$ Pup. Black
solid: clumping law derived from our model fits.  Red solid: Theoretical
predictions by \cite{run02} from hydrodynamical models, with self-excited
line driven instability. Dashed: Average clumping factors derived by
\cite{puls06a} assuming an outer wind matching the theoretical predictions.
Magenta solid: run of the velocity field in units of 100\,\kms. See also
Sect.~\ref{dense}.} 
\label{firuncl}
\end{figure}

\paragraph{Dense winds.} For most of our dense wind objects, we found rather
strong clumping close to the wind-base (which has a substantial impact on
the actual mass-loss rate, see next subsection), whereafter the clumping
factor decreases towards the mid/outer wind. This finding is similar to the
results from \citetalias{puls06a}, at least qualitatively. An even
quantitative comparison is possible for the case of $\zeta$~Pup, the only
star among our sample for which we could determine the run of the clumping
factor throughout the {\it entire} wind (Fig.~\ref{firuncl}). When
normalizing the results by \citetalias{puls06a} to a similar value in the
outer wind ($\fcl \approx 5$), the agreement with our results is excellent
and re-assuring, since both investigations are completely independent from
each other and rely on considerably different methods. Likewise re-assuring
is the fact that the derived stratification of $\fcl(r)$ is very similar to
recent results by \citet{Sundqvist10b}, who performed a consistent analysis
of \PV\ and hydrogen/helium recombination lines in the O6I(n)f supergiant
$\lambda$~Cep (a cooler counterpart of $\zeta$ Pup), {\it including the
consideration of macroclumping and vorosity effects}. Also for this object,
it turned out that clumping peaks close to the wind base, with a maximum
value of $\fcl \approx 28$, which is rather close to the value derived here
for $\zeta$ Pup (see Fig~\ref{firuncl}).

Accounting as well for the previous work by \citet{Crowther02, hil03} and
\citet{bouret03, bouret05} (see Sect.~\ref{intro}), there is now 
overwhelming evidence for the presence of highly clumped material close to
the wind base. These findings are in stark contrast to theoretical
expectations resulting from radiation hydrodynamic simulations
(Sect.~\ref{intro}), which predict a rather shallow increase of the clumping
factor, due to the strong damping of the line instability in the lower wind
caused by the so-called line-drag effect \citep{Lucy84, OR85}. In
Fig~\ref{firuncl} we compare our present results and those from
\citetalias{puls06a} with prototypical predictions from radiation
hydrodynamic simulations, here from the work by \citet{run02}. Though there
is a fair agreement for the intermediate and outer wind, the disagreement in
the lower wind is striking. A similar conclusion has been reached by
\citet{Sundqvist10b}, and further progress on the hydrodynamic modeling
seems to be necessary to understand this problem.

\paragraph{Thin winds.} Except for HD\,217086, all our sample stars with
thin winds (i.e., \Ha\ in absorption) require no clumping to yield excellent
fits, and for the former object the derived clumping is less ($\fv = 0.1$)
than for the typical dense wind case. Again, this finding is in agreement
with the results from \citetalias{puls06a}, who constrained the clumping
factors for thin winded objects to be similar in the inner and outer wind.
Thus and in connection with our present results, one might tentatively
conclude that most thin winds are rather unclumped, which immediately raises
the question about the difference in the underlying physics. Might it be
that the wind-instability is much stronger in objects with dense
winds, e.g., due to (stronger) non-radial pulsations, or is there a
connection with sub-surface convection, as suggested by
\citet{Cantiello09}?

\subsection{Wind-momentum rates}

\begin{figure}
\resizebox{\hsize}{!}
{\includegraphics{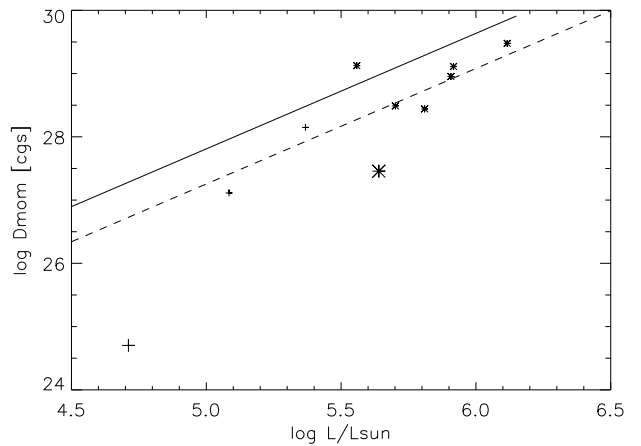}}
\caption{Wind-momentum rates for our sample stars, as a function of \LL.
Asterisks correspond to supergiants, crosses to other luminosity classes. The
two weak wind candidates HD\,76341 and HD\,37468 are denoted by large symbols.
The solid line shows the theoretical predictions for the WLR by
\citet{vink00}, whilst the dashed one has been shifted to lower values 
by 0.55 dex (see text).}
\label{wlr}
\end{figure}

In Fig.~\ref{wlr}, we plot the wind-momentum luminosity relation (WLR) for
our sample stars, i.e., the wind-momentum rates, modified by the factor
(\Rstar/\Rsun)$^{0.5}$, as a function of luminosity. The results of our
analysis 
 suggest
a well defined relation, if we exclude three outliers, the
binary Cyg\,OB2~\#8A well above and HD\,76341/HD\,37468 well below the
average relation. 
 Of course, a much larger sample needs to be investigated before a
final conclusion can be drawn.

From a linear regression to our results, we obtain an `observed'
relation in parallel to the theoretical predictions by \citet[
solid]{vink00}, but 0.55 dex (factor of 3.5) lower. Due to its large
deviation (2 dex!), HD\,37468 is certainly a weak winded star, whereas
HD\,76341 (with a deviation of 0.9 dex) might be considered as a
weak-wind candidate, interestingly a Ib supergiant.\footnote{So far,
only dwarfs and few giants have been suggested as weak-winded stars.}.

{\it On the assumption of an unclumped outer wind}, \citetalias{puls06a} 
found a good agreement between their ``observed'' and the theoretical WLR.
These results could be unified with ours on the hypothesis that the outer
regions of most winds are actually clumped, with a typical clumping factor
on the order of 10. With regard to our analysis of $\zeta$ Pup (and also the
hydro simulations), this seems to be a reasonable, though somewhat too
large value. 

A down-scaling of theoretical wind-momenta as obtained here is also
consistent with results from the analysis of $\lambda$ Cep by
\citet{Sundqvist10b}, who derived a mass-loss rate being a factor of two
lower than predicted by \citet{vink00}. Let us finally note that an
independent ``measurement'' of the mass-loss rate of $\zeta$ Pup based on
X-ray line emission by \citet{Cohen10} resulted in a value of 3.5 \mdu, with
a lower limit of 2.0 \mdu, if the abundance pattern would be solar (which is
rather unlikely). 

\section{Summary, conclusions and future perspectives}
\label{summary}

In this pilot study, we have investigated the diagnostic potential of $L$-band
spectroscopy to provide strong constraints on hot star winds, with
particular emphasis on the determination of their clumping properties and 
(actual) mass-loss rates, even for objects with very thin (=weak) winds. To
this end, we have secured $L^{(')}$ 
band spectra (featuring \Bra, \Pfg\ and
\HeI\,3.703) for a sample of ten O-/early B-type stars, by means of
ISAAC@VLT and SpeX@IRTF. 
The sample has been designed in such a way as to
cover objects with both dense and thin winds, with the additional
requirement that spectroscopic data in the UV, optical and $H/K$ band as well as
radio observations are present and that most of the objects have been
previously analyzed by means of quantitative spectroscopy.

For all stars, we performed a consistent multi-wavelength NLTE analysis by
means of {\sc cmfgen}, using the complete spectral information including our
new $L$-band data. We assumed a microclumped wind, with a rather universal
clumping law based on four parameters to be fitted simultaneously with the
other stellar and wind parameters. Moreover, we accounted for rotational and
macroturbulent broadening in parallel. 

For the objects with dense winds, we were able to derive absolute values for
mass-loss rates and clumping factors (and not only relative ones as in
\citetalias{puls06a}), where \Pfg\ and \Bra\ proved to be invaluable tools to
derive the clumping properties in the inner and mid wind, respectively. 

For our objects with thin winds, on the other hand, the narrow emission core
of \Bra (in combination with its line wings) proved as a powerful mass-loss
indicator, due to its strong reaction even at lowest \Mdot, its independence
of X-ray emission, 
and its only moderate contamination by additional effects
such as atomic data, microturbulence and velocity law.

Contrasted to what might be expected, the height of the \Bra\ line core
increases with decreasing mass-loss. This is a consequence of the transition
region between photosphere and wind ``moving'' towards lower $\taur$ for
decreasing \Mdot, so that more and more of the strong line source function
becomes ``visible'' when the wind becomes thinner. The origin of such a
strong source function is a combination of various effects in the upper
photosphere and the transonic region that depopulates the lower level of
\Bra, $n=4$, stronger than the upper one, $n=5$. By detailed simulations, we
explained this in terms of a nebula-like situation, due to the competition
between recombinations and downwards transitions, where the stronger decay 
from $n=4$ (compared to the decay from $n=5$) is decisive.
\footnote{though most other processes play an important role as
well, by establishing the degree of depopulation, which controls the peak
height.}

As an interesting by-product, the specific sensitivity of \Bra\ and \Pfg\ on
the velocity field exponent $\beta$ in (not too) thin winds allows a break in
the well-known \Mdot-$\beta$ degeneracy when using \Ha\ alone. On the other
hand, the dependence of \Bra\ on $\beta$ for weak winds is negligible, which
decreases the error bars of the derived \Mdot. A major point controlling the
depopulation of the lower level of \Bra\ and thus the height of the emission
peak are the hydrogen bound-bound collisional data which are used. Our
models utilize data by \citet{mih75}, which provide something of a compromise
regarding collisional strengths when comparing with
other data sets. Overall, we estimate the error in our \Bra-\Mdot\
determination of weak-winded stars by $\pm$ 0.5 dex.  This is certainly
better than UV diagnostics that strongly depend on the assumed description
of X-ray properties and are further hampered by the impact of macroclumping
and vorosity.

We compared the results of our analysis with those from previous work, in
particular the derived \Teff, \logg\ and $Q$-values (the optical depth
invariant(s)). With respect to \Teff, the (average) agreement is significantly
hampered due to sizable differences for Cyg\,OB2~\#8C ($\Delta \Teffe
\approx$ 4000~K), whereas the rest agrees within the conventional errors of
$\pm$ 1000~K. The large disagreement for the former object has been
attributed to differences in the derived rotational/macroturbulent
velocities. On the other hand, the agreement with respect to $Q$ is
satisfactory. 

Almost all of our dense wind objects require large clumping close to the
wind base, whereas for the thin winded stars we did not need to invoke
clumping at all. Our clumping factors for the best studied object,
$\zeta$~Pup, agree very well with the work by \citetalias{puls06a}, if the
clumping in the outer wind (which could not derived by the latter authors)
is scaled to similar values. Moreover, our results on strong clumping in the
lower wind are also consistent with other findings, particularly those by
\citet{Sundqvist10b}, and challenge present radiation hydrodynamic
simulations which predict a much shallower increase of the clumping factors.

Because of using \Bra\ as a mass-loss indicator, we were able to fully
characterize one weak-winded star, the O9.5 dwarf HD\,37468, and one weak
wind candidate, the O9Ib supergiant(!) HD\,76341. 

Finally, our results  suggest a well defined WLR (discarding the two
weak-winded objects and the binary Cyg\,OB2~\#8A) that is located 0.55
dex below the predictions by \citet{vink00}. From a comparison with
\citetalias{puls06a}, it seems likely that at least dense winds are
considerably clumped in their outer regions. We suggest that the mass-loss
rates from \citetalias{puls06a} are upper limits indeed, and that a
downscaling of their values by factors on the order of 2 to 3 seems likely. 

From all these results, we conclude that the diagnostic potential of IR
$L$-band spectroscopy for deriving clumping properties and mass-loss rates
of hot star winds is really promising. We suggest to extend this rather
small sample with further $L$-band observations for a carefully selected
sample of OB stars (with large and low \Mdot) {\it in order to derive
statistically conclusive results} on the ``true'' mass-loss rates from these
stars. Any result drawn from only a few objects suffers from a variety of
problems (e.g., the objects might be peculiar, or problems related to the
diagnostic tools might remain hidden), and only a careful analysis of a
large number of objects (performed with the same diagnostic tool) allows 
the identification of trends and outliers.

Moreover, there is the additional problem of variability.  Due to
its higher sensitivity to mass-loss, it might be expected that \Bra\
is an even better candidate than \Ha\ (e.g., \citealt{markova05} and
references therein) to study and to analyze wind variability in dense
winds. The impact of \Bra\ line profile variability on
mass-loss/clumping diagnostics
needs to be investigated as well. To our knowledge, corresponding
observations have not been carried out so far, and deserve future
interest. In thin winds, on the other hand, a {\it strong} variability
is not to be expected, because of the near-photospheric origin of \Bra\ and
assuming a stationary photosphere. This expectation needs to be
confirmed as well, and a contradicting outcome might point to
variability in the location of and the conditions in the transition
zone, e.g., related to pulsations.

Finally, and from our experience accumulated so far, we are confident that \Bra will
become {\it the} primary diagnostic tool to measure very low mass-loss
rates at unprecedented accuracy, thus clearly identifying weak-winded 
stars and quantifying the degree with which they lie below theoretical predictions.

\acknowledgements{ 
We would like to thank our anonymous referee for useful comments and
suggestions. We thank John Hillier for providing the {\sc cmfgen} code.
Likewise, we thank Artemio Herrero, Sergio Sim{\'o}n-D{\'{\i}}az and Norbert
Przybilla for providing the \Ha\ profiles for various objects. Financial
support from the Spanish Ministerio de Ciencia e Innovaci\'on under 
projects AYA2008-06166-C03-02 and AYA2010-21697-C05-01 is acknowledged. This material is based upon
work supported by the National Science Foundation under Grant No. 0607497 
and 1009550 to the University of Cincinnati.}

\Online
\appendix
\section{Detailed comparison of present results with other investigations} 
\label{detcomp}

The present work comprises a detailed analysis of a small sample of hot
stars, based on the combination of optical, NIR and UV spectra with one of
the most sophisticated NLTE atmosphere codes presently available, {\sc
cmfgen}. Thus, a comparison of the derived results with those from more
restricted investigations (with respect to wavelength range) based on alternative 
atmosphere codes provides an opportunity to address typical uncertainties inherent
to the spectroscopic analysis of such objects, caused by different data-sets
and tools. As outlined in Sect.~\ref{obs} (cf. Table~\ref{runs}), most
previous investigations of our targets have been analyzed by means of {\sc
fastwind} \footnote{which relies on certain approximations mostly related to
the treatment of (EUV-)line-blocking}, or by (quasi-) analytic methods designed
for specific diagnostics such as \Ha\ or the IR-/mm-/radio continuum (for an
overview of these methods, see \citealt{kp00, Puls08} and references
therein). 

Brief comments on important differences between our and those results have
been already given in Sects.~\ref{dense} and \ref{thin}, and the complete
set of the various stellar and wind-parameters is presented in
Table~\ref{table_comp}. In the following, the various investigations are
referred to following the enumeration provided at the end of this table
(ref\#). Note that a direct comparison of mass-loss rates is
still not possible, due to the uncertainties in distances (and thus radii)
for Galactic objects. Moreover, previous analyses were 
based on either unclumped models (ref\# 2-7) or that the derived
clumping factors had to be normalized to the clumping conditions in the
outermost wind, which are still unclear (ref\# 8). Thus, a {\it
meaningful} comparison is possible only for the optical depth invariants
\footnote{for a derivation, see \citealt{kp00} and \citealt{Puls08}},
\beq Q=\frac{\Mdote}{(\Rstare \vinfe)^{1.5} \sqrt{\fv}} \qquad \mbox{and}
\qquad \Qr=\frac{\Mdote}{\Rstare \vinfe^2}, 
\label{qdef} 
\eeq 
which describe the actual measurement quantities related to (i)
$\rho^2$-dependent processes ($Q$), if $\fv$ is the average clumping factor
in the corresponding formation region and the clumps are optically thin, and
(ii) $\rho$-dependent processes ($\Qr$), under the assumption that clumping
plays only a minor role (but see Sect.~\ref{intro}).

\begin{figure*}
\begin{center}
\begin{minipage}{14.5cm}
\resizebox{\hsize}{!} {\includegraphics[angle=90]{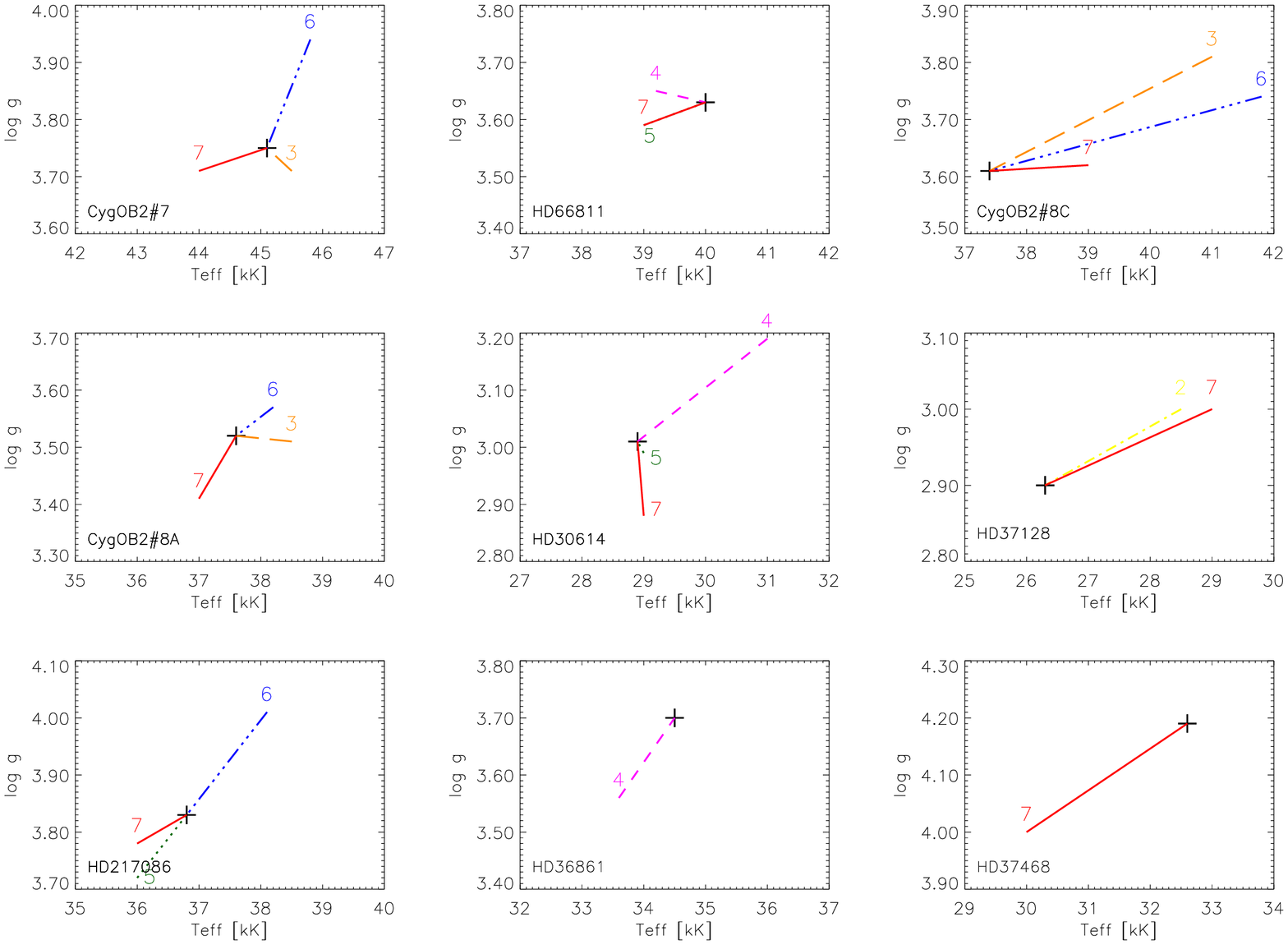}} 
\end{minipage}
\begin{minipage}{14.5cm}
\resizebox{\hsize}{!} {\includegraphics[angle=90]{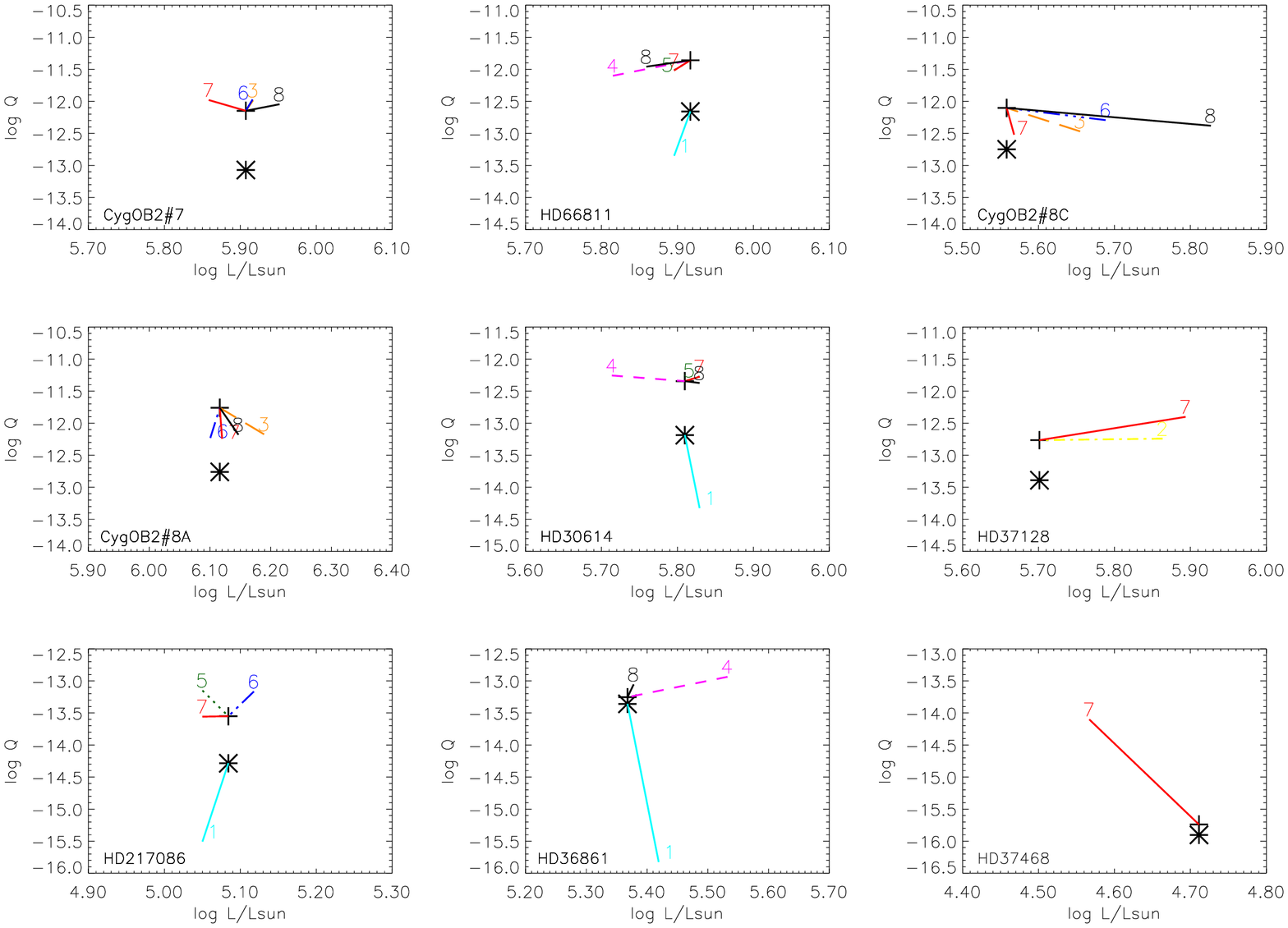}} 
\end{minipage}
\end{center}
\caption{Comparison of the results obtained in this work (crosses) with
results from other investigations (for data and reference identifiers, see 
Table~\ref{table_comp}). {\bf Upper} 3$\times$3 panels: \logg\ vs. \Teff;
all panels have the same scale, corresponding to an extent of 5000~K in
\Teff\ and 0.4~dex in \logg. {\bf Lower} 3$\times$3 panels: $\log Q$ vs.
\LL, with axes extending over 0.4 dex in \LL\ and 3.5 dex in $\log Q$. In
order to facilitate the comparison with $\rho^2$-diagnostics, all $Q$ values
have been normalized to $\fv=1$ (see Eq.~\ref{qdef}).  The asterisks provide
the $\Qr$-values which have to be compared with the corresponding values
from ref\#~1 (\citealt{fulli06}, $\rho$-diagnostics, but including the
product with the ionization fraction of \PV). Note that all $\Qr$ values have
been scaled by a factor of 10 to fit into the individual figures.} 
\label{fig_comp}
\end{figure*}

Fig.~\ref{fig_comp} provides an impression of the differences in the most
important parameters for the individual stars, by comparing the effective
temperatures and gravities (upper 3$\times$3 panels), and the optical depth
invariants and luminosities (lower 3$\times$3 panels). Note that all panels
provide identical scales, to enable an easy visualization. From the figure,
it is quite clear that {\it typical} differences in \Teff\ are of the order
of 1,000 to 2,000~K, with corresponding differences of 0.1 to 0.2 dex in
\logg. The largest differences are found for the effective temperature of
Cyg\,OB2~\#8C (roughly 4,000~K when comparing with ref\#~3 and 6/8), which has
been already discussed in Sect.~\ref{dense}, and most probably relates to
an underestimation of \vsini\ in these investigations. It is reassuring that
in most cases the connecting lines between our (crosses) and the other
results in the \Teff-\logg\ plane have a positive slope, indicating that
higher temperatures go in line with higher gravities and vice versa, which
is consistent with the behavior of the gravity indicators (usually, the
wings of the Balmer lines). 

\begin{table}
\caption{Mean difference of derived effective temperature, $\Delta \Teffe=
\Teffe({\rm ref\,\,i})-\Teffe({\rm this\,\, work})$ and derived optical depth
invariant, $\Delta \log Q$, for $N$ objects from reference $i,i=1,8$ in
common with our sample (cf.
Table~\ref{table_comp}). $\Delta \Teffe$ in kK, $\Delta \log Q$ in dex.
For ref\#~1, we compare the $\Qr$-values, whereas for the
rest we compare the $Q$-values normalized by $\fv$ (see Eq.~\ref{qdef}.)
Positive values indicate that the results from the specific work are on
average larger than the results derived here. $\sigma$ is the dispersion 
of these differences. No values are given for ref\#~2 \citep{kudetal99},
since there is only one object in common.}
\begin{center}
\begin{tabular}{l c r c c l}
\hline
ref\# & $N$  & $ \langle \Delta \Teffe \rangle$ & $\sigma(\Delta
\Teffe)$  & $ \langle \Delta \log Q \rangle$ & $\sigma(\Delta
\log Q)$ \\  
\hline
1  & 4 &      &      & -1.38 & 0.76 \\ 
3  & 3 & 1.63 & 1.72 & -0.21 & 0.31 \\
4  & 3 & 0.13 & 1.70 &  0.05 & 0.28 \\
5  & 3 & -0.57& 0.59 &  0.11 & 0.28 \\
6  & 4 & 1.75 & 1.79 & -0.02 & 0.38 \\
7  & 8 & -0.21 & 1.67& -0.06$^\ast$ & 0.31$^\ast$ \\
8  & 6 &       &     & -0.08 & 0.23 \\
\hline
\end{tabular}
\end{center}
\footnotesize{$^\ast$ mean difference and dispersion only for 7 objects,
excluding the weak-winded object HD\,37468.}
\label{deltatq}
\end{table}

\begin{table*}
\caption{Comparison of stellar and wind parameters as derived in the present
analysis with results from previous investigations. Units as in Table~\ref{para};
optical depth invariants, $Q$ and $\Qr$ (see Eq.~\ref{qdef}), 
calculated with \Mdot\ in \Msunyr, \Rstar\ in \Rsun\ and
\vinf\ in \kms. $\fv$ values from this work as in Table~\ref{para}, i.e.,
equal to CL1 (Eq.~\ref{eq:clump}). $\fv$ values from reference (8)
\citep{puls06a}) corresponding to the clumping factors within the innermost
clumped region (``region 2'') extending between $1.1 \le r/\Rstare \le 2$ and
assuming an unclumped outer wind.
For reference (1) \citep{fulli06}), the quoted results for \Mdot\ and $\log \Qr$ 
include the product with the ionization fraction of \PV.}
\tabcolsep1.8mm
\begin{center}
\begin{tabular}{l l c c c c r c c r r r c c c c}
\hline
star           & ref.   & \Teff & \logg & \Rstar & \YHe  &\vsini & \vmac &
\vinf & \Mdot & $\beta$ & $\fv$&     \LL   & $\log \Qr$ & $\log Q$\\
\hline
Cyg\,OB2 $\#$7 & t.w. & 45.1  & 3.75  & 14.7   & 0.13  &95     & 65    & 3100  & 1.2   & 1.05    &0.03 &       5.91 &   -14.07& -12.15  \\  
               & 3      & 45.5  & 3.71  & 14.6   & 0.30  &105    &       & 3080  & 9.86  & 0.9     &     &       5.92 &       &-11.99 \\  
               & 6      & 45.8  & 3.94  & 14.4   & 0.21  &105    &       & 3080  & 9.98  & 0.77    &     &       5.92 &       &-11.97 \\  
               & 7      & 44.0  & 3.71  & 14.6   & 0.10  &145    &       & 3080  & 10.0  &         &     &       5.86 &       &-11.98 \\  
               & 8      & 45.8  & 3.94  & 15.0   & 0.21  &105    &       & 3080  & $\le$4.0  & 0.9     &0.2  &       5.95 &   &-12.05 \\  
HD\,66811      & t.w.   & 40.0  & 3.63  & 18.9   & 0.14  &215    & 95    & 2250. & 2.1   & 0.90    &0.03 &       5.92 & -13.66&-11.86 \\  
               & 1      & 39.0  &       & 19.4   &       &219    &       & 2250  & 0.44  & 0.5     &     &       5.90 & -14.35 & \\  
               & 4      & 39.2  & 3.65  & 17.5   & 0.20  &203    &       & 2300  & 6.4   & 0.92    &     &       5.82 &        &-12.10 \\  
               & 5      & 39.0  & 3.59  & 19.4   & 0.20  &220    &       & 2250  & 8.8   & 0.9     &     &       5.90 &        &-12.02 \\  
               & 7      & 39.0  & 3.59  & 19.4   & 0.17  &220    &       & 2250  & 8.77  &         &     &       5.90 &        &-12.02  \\  
               & 8      & 39.0  & 3.6   & 18.6   & 0.20  &220    &       & 2250  & 4.2   & 0.70    &0.2  &       5.86 &        &-11.96  \\  
Cyg\,OB2 $\#$8C& t.w.   & 37.4  & 3.61  & 14.3   & 0.10  &175    & 90    & 2800  & 2.0   & 1.30    &0.10 &       5.56 & -13.75 &-12.10  \\  
               & 3      & 41.0  & 3.81  & 13.3   & 0.09  &145    &       & 2650  & 2.25  & 0.9     &     &       5.65 &        &-12.47  \\  
               & 6      & 41.8  & 3.74  & 13.3   & 0.13  &145    &       & 2650  & 3.37  & 0.85    &     &       5.69 &        &-12.29  \\  
               & 7      & 39.0  & 3.62  & 13.3   & 0.10  &145    &       & 2650  & 2.0   &         &     &       5.57 &        &-12.52  \\  
               & 8      & 41.8  & 3.81  & 15.6   & 0.13  &145    &       & 2650  & $\le$3.5  & 1.0     &1.0  &       5.83 &    &-12.38 \\  
Cyg\,OB2 $\#$8A& t.w.   & 37.6  & 3.52  & 26.9   & 0.10  &110    & 80    & 2700  & 3.4   & 1.10    &0.01 &       6.12 & -13.76 &-11.76 \\  
               & 3      & 38.5  & 3.51  & 27.9   & 0.10  &95     &       & 2650  & 13.5  & 0.7     &     &       6.19 &        &-12.17  \\  
               & 6      & 38.2  & 3.57  & 25.6   & 0.14  &130    &       & 2650  & 10.4  & 0.74    &     &       6.10 &        &-12.23  \\  
               & 7      & 37.0  & 3.41  & 27.9   & 0.10  &95     &       & 2650  & 11.5  &         &     &       6.12 &        &-12.24  \\  
               & 8      & 38.2  & 3.57  & 27.0   & 0.14  &130    &       & 2650  & $\le$8.0  & 0.74    &0.40 &       6.15 &    &-12.18  \\  
HD\,30614      & t.w.   & 28.9  & 3.01  & 32.0   & 0.13  &100    & 75    & 1550  & 0.50  & 1.60    &0.01 &       5.81 & -14.19 &-12.34  \\  
               & 1      & 29.0  &       & 32.5   &       &129    &       & 1550  & 0.037 & 1.0     &     &       5.83 & -15.32 & \\  
               & 4$^a$ & 31.0  & 3.19  & 24.9   & 0.10  &100    &       & 1550  & 4.2   & 1.05    &     &       5.71 &        &-12.26  \\  
               & 5      & 29.0  & 2.99  & 32.5   & 0.10  &100    &       & 1550  & 6.04  & 1.15    &     &       5.83 &        &-12.27  \\  
               & 7      & 29.0  & 2.88  & 32.5   & 0.20  &100    &       & 1550  & 6.0   &         &     &       5.83 &        &-12.28  \\  
               & 8$^b$ & 29.0  & 3.0   & 32.5   & 0.10  &100    &       & 1550  & 2.95  & 1.15    &0.38 &       5.83 &        &-12.37  \\ 
HD\,37128      & t.w.   & 26.3  & 2.90  & 34.1   & 0.13  &55     & 60    & 1820  & 0.46  & 1.60    &0.03 &       5.70 & -14.39 &-12.77  \\  
               & 2      & 28.5  & 3.00  & 35.0   & 0.1   &80     &       & 1600  & 2.40  & 1.25    &     &       5.86 &        &-12.74  \\  
               & 7  & $\le$29.0 & 3.0   & 35.0   & 0.1   &80     &       & 1600  & 5.25  &         &     &       5.89 &        &-12.40  \\  
\hline
HD\,217086     & t.w.   & 36.8  & 3.83  & 8.56   & 0.1   &350    & 80    & 2510  & 0.028 & 1.2     &0.10 &       5.08 & -15.28 &-13.55  \\   
               & 1      & 36.0  &       & 8.6    &       &332    &       & 2550  &$\le$0.00174&1.0     &     &       5.05 & -16.51&  \\  
               & 5      & 36.0  & 3.72  & 8.6    & 0.15  &350    &       & 2550  &$\le$0.23  & 0.8     &     &       5.05 &    &-13.15  \\  
               & 6      & 38.1  & 4.01  & 8.3    & 0.09  &350    &       & 2550  & 0.21  & 1.27    &     &       5.12 &        &-13.17  \\  
               & 7      & 36.0  & 3.78  & 8.6    & 0.15  &350    &       & 2550  & $\le$0.09 &         &     &       5.05 &    &-13.56  \\  
HD\,36861      & t.w.   & 34.5  & 3.70  & 13.5   & 0.11  &45     & 80    & 2175  & 0.28  &  1.3    &1.0  &       5.37 & -14.36 &-13.25  \\ 
               & 1      & 33.6  &       & 15.1   &       &74     &       & 2400  & 0.0013&  0.7    &     &       5.42 & -16.83 &  \\  
               & 4      & 33.6  & 3.56  & 17.2   & 0.1   &66     &       & 2400  & 0.97  &  0.8    &     &       5.53 &        &-12.94  \\  
               & 8      & 33.6  & 3.56  & 14.4   & 0.10  &66     &       & 2400  & $\le$0.4  &  0.9    &0.5  &       5.38 &    &-13.06  \\  
HD\,76341      & t.w.   & 32.2  & 3.66  & 21.2   & 0.1   &63     & 80    & 1520  & 0.065 &  1.2    &1.0  &       5.64 & -14.88 &-13.95  \\ 
HD\,37468      & t.w.   & 32.6  & 4.19  &  7.1   & 0.1   &35     & 100   & 1500  & 0.0002&  0.8    &1.0  &       4.71 & -16.90 &-15.74  \\ 
               & 7$^c$ & 30.0  & 4.0   &  7.1   & 0.1   &80     &       & 2300  & 0.0165&  1.0    &     &       4.57 &        &-14.10  \\ 
\hline
\end{tabular}
\end{center}
\footnotesize{
References: (1) \citet{fulli06}; (2) \citet[ unblanketed analysis]{kudetal99}; 
(3) \citet{herrero02}; 
(4) \citet[ based on stellar parameters calibrated to the results from
optical NLTE analyses by \citealt{repo04}]{markova04} 
(5) \citet{repo04}; (6) \citet{mokiem05}; 
(7) \citet{repo05}; (8) \citet{puls06a}.\newline
$^{a)}$ using the high luminosity solution; $^{b)}$ stellar radius and
corresponding quantities scaled to the solution by (5,7) to facilitate the
comparison; $^{c)}$ stellar radius from this work.}  
\label{table_comp}
\end{table*}

The average differences with respect to effective temperature, $ \langle
\Delta \Teffe \rangle$, are presented in Table~\ref{deltatq}, discarding
ref\#~1 and 8 who {\it adopted} the stellar parameters, mostly from
ref\#~2-7.  Major discrepancies seem to be present when comparing with
ref\#~3 and 6, who derived temperatures being on average 1,700 K higher than
our results. Note, however, that a large part of this discrepancy is caused
by the results obtained for Cyg\,OB2~\#8C (see above). The other three
investigations (ref\# 4, 5 and 7) deviate, at least on average, much less
from the present one, by a few hundreds of Kelvin. The dispersion of the
differences, however, is very similar in all cases, about 1,700~K, except
for ref\#~5, with a dispersion of 600~K. Thus, overall, the dispersion of 
$ \langle \Delta \Teffe \rangle$ is somewhat larger than to be expected from
the typically quoted individual uncertainties of 1,000~K, which should give
rise to a dispersion of 1,400~K.

Regarding the optical depth invariants, the situation is satisfactory. 
Except for ref\#~3, the mean differences are at or below 0.1 dex (25\%),
with a dispersion of typically 0.3 dex (factor of 2), which is consistent
with the typical individual errors (see \citealt{markova04} for a detailed 
analysis). This result is particularly obvious from Fig.~\ref{fig_comp} (lower
panels), where in most cases all investigations show rather similar $Q$
values. 

Regarding the $\Qr$ values which are relevant when comparing with ref\#~1
(the \PV\ investigation by \citealt{fulli06}), the discrepancy is still
large, particularly when accounting for the fact that our results indicate
considerable clumping, thus reducing the absolute mass-loss rate
significantly with respect to previous investigations. On average, we find
$\langle \Delta \log \Qr \rangle \approx$-1.4~dex, with a dispersion of 0.8~dex,
Thus, either (i) the actual mass-loss rates are even smaller than derived
here, or (ii) the ionization fraction of \PV\ (remember that the results
from Fullerton et al. include the product with this quantity) is very low,
of the order of 4\% (which would require extreme conditions in the wind,
e.g., a very strong X-ray/EUV radiation field), or (iii) the line formation
calculations of UV-resonance lines (both in the investigation by Fullerton
et al. and in our analysis) require some additional
considerations, such as the presence of optically thick clumps and/or the
inclusion of a porosity in velocity space, see
Sect.~\ref{intro}.\footnote{Another possibility, though less likely, is a
strong underabundance of phosphorus, as claimed by \citet{paul94, paul01}.}

In summary, we conclude that at least the analysis of $Q$ seems to be
well-constrained, and that different investigations give rather similar
results. The remaining problem is the determination of actual mass-loss
rates, which involves the ``measurement'' of (absolute) values of clumping
factors. As we have shown in this investigation, $L$-band spectroscopy turns
out to be a promising tool for this objective. Let us note that only a
measurement of actual mass-loss rates will enable a strict comparison with
theoretical predictions (as performed in Sect.~\ref{discussion}), to identify
present shortcomings and to provide ``hard numbers'' for evolutionary
calculations.

The precision of effective temperatures, on the other hand, is less
satisfactory. Irrespective of the fact that we did not find a real {\it
trend} in the average differences with respect to three from five
investigations, the dispersion is quite large, and individual discrepancies
amount to intolerable values. Because of our detailed analysis covering a
large range of wavelength domains and using a state-of-the-art model atmosphere
code based on an ``exact'' treatment of all processes, we are quite
confident that the \Teff-errors in our work are of the order of 1,000~K or
less, which means that the corresponding errors in the previous
investigations must be of the order of 1,400~K or more. Additionally, two
from five investigations gave a rather large average difference with respect
to our results, which is alarming since all five investigations have been
performed with the same NLTE atmosphere code. Insofar, recent attempts to
provide reliable spectral-type-\Teff-calibrations have to be augmented by
results from large samples to decrease the individual scatter in a 
statistical way.

\section{Occupation numbers of the hydrogen $n=4$ and $n=5$ level in the
outer atmospheres of late O-type stars with thin winds} 
\label{outerconditions}

\paragraph{Conditions in the outer photosphere.} As we have seen from
Fig.~\ref{nebula}, almost all of our simulations (and many more which have
not been displayed) resulted in a stronger depopulation of level 4 compared
to level 5 in the outer atmosphere, {\it independent of the various
processes considered}.  One might question how far this result can be
explained (coincidence or not?). To obtain an impression on the relevant
physics, we write the rate equations for level $i>1$ in the following,
condensed form\footnote{ The case $i=1$ will be considered separately
below.}, again neglecting collisions, and assuming that ionization is only
possible to the ground-state of the next higher ion (as it is the case for
hydrogen): 
\beq 
n_i \sum_{i > j} A_{ij} Z_{ij} - \sum_{i < j} n_j A_{ji}
Z_{ji} + n_i R_{ik} = n_i^* R_{ki}, 
\label{rate1} 
\eeq 
where $A_{ij}$ are the Einstein-coefficients for spontaneous decay, $R_{ik}$
and $R_{ki}$ the rate-coefficients for ionization/recombination, and
$Z_{ij}$ the net radiative brackets for the considered line transition,
$Z_{ij} = (1- \frac{\bar J_{ij}}{S_{ij}})$ (the fraction denotes the ratio
of mean line intensity and line source function).  $n_i^*$ = $n_k n_e
\Phi_{ik} (T_e$) denotes the LTE population of level $i$, accounting for the
actual electron and ion density, $n_e$ and $n_k$ (for further details see,
e.g., \citealt{mih78}), such that the departure coefficients are given by
$b_i = n_i/n_i^*$. Note that for purely spontaneous decays $Z_{ij} = 1$, for
lines which are in detailed balance, $Z_{ij} = 0$, and for levels which are
strongly pumped (e.g., by resonance lines with a significantly overpopulated
lower level), $Z_{ij} < 0$. Solving for the departure coefficients,
Eq.~\ref{rate1} results in 
\beq 
b_i= \frac{\sum_{i < j} n_j/n_i^* A_{ji} Z_{ji} +
R_{ki}}{\sum_{i > j} A_{ij} Z_{ij} + R_{ik}}. 
\label{depart_excited} 
\eeq 
The sum in the nominator corresponds to the net-contribution of lines from
``above'' (i.e., with upper levels $j > i$), normalized to the LTE
population of the considered level, whereas the sum in the denominator is
the net-contribution of lines to lower levels ($j < i$). The complete
fraction can be interpreted as the ratio of populating and depopulating
rates, which can be split into the contributions from bound-bound and
bound-free processes, 
\beq 
b_i= \frac{\sum_{i < j}({\ldots})}{\sum_{i > j}({\ldots})+ R_{ik}} \,+\, 
\frac{R_{ki}}{\sum_{i > j}({\ldots})+ R_{ik}}.
\label{2comp} 
\eeq 
For all our simulations {\it 1-6} we have now calculated those two terms 
which determine $b_4$ and $b_5$. At first, let us concentrate on the outer
photosphere, as on the right of Fig.~\ref{nebula}. In almost all cases
(except for simulation {\it 5}), the 2nd term dominates the departure
coefficient, and, moreover, the first term (the ratio!, not the individual
components) remains rather similar, of order 0.2. Consequently, the stronger
depopulation of level 4 compared to level 5 is due to the fact that the
quantity 
\beq 
\frac{R_{ki}}{\sum_{i > j} A_{ij} Z_{ij} + R_{ik}} 
\eeq 
is usually larger for level 5 than for level 4, even though $R_{k5} <
R_{k4}$: the accumulated transition probability from level 4 to lower levels
($A_{41}Z_{41} + A_{42}Z_{42} + A_{43}Z_{43}$) is {\it much} larger than the
corresponding quantity from level 5 to lower levels ($A_{51}Z_{51} +
A_{52}Z_{52} + {\ldots}$). This behavior, finally, can be traced down to
the run of the oscillator-strengths in hydrogen: On the one side, e.g.,
$A_{41}$ is larger than $A_{51} $, etc., whereas, on the other,
the corresponding net radiative brackets ($Z_{4j}$ vs. $Z_{5j}$ etc.) do not
differ too much.

Two examples shall illustrate our findings. For the {\it complete}
solution, the first term in Eq.~\ref{2comp} is roughly 0.22, whereas the 2nd
term amounts to 0.62 for level 5 and to 0.53 for level 4. Thus, $b_4 \approx
0.75$ and $b_5 \approx 0.84$. For simulation {\it 2}, with $Z_{ij}=Z_{ji} =
1$ and $R_{ik} = 0$ for $i>1$, the first term $\approx 0.18$, and the 2nd
one is 0.41 and 0.3, respectively, such that $b_4 \approx 0.48$ and $b_5
\approx 0.6$ (cf. Fig.~\ref{nebula}).

In conclusion, the stronger depopulation of level 4 compared to level 5 in
the outer photospheres of hot stars can indeed be regarded as the
consequence of a typical nebula-like situation, namely as due to the
competition between recombination and downwards transitions. Different
approximations regarding the contributing lines do control the absolute size
of the departures, but not the general trend.

\paragraph{Conditions in the wind.} Though the formation of the emission
peak of \Bra\ for objects with thin winds is controlled by the processes in
the upper part of the photosphere, it is also important to understand the
conditions in the wind, since, as we have seen in Fig.~\ref{compsltaur}, the
onset of the wind prohibits a further growth of the corresponding source
function: Immediately after the transition point between photosphere and
wind, the source function drops to values corresponding to the local
Planck-function (i.e., the departure coefficients of $n_4$ and $n_5$ become
similar). Only in the outer wind the source function increases again (in
contrast to the predictions of the pure H/He model, see below), which
remains invisible in the profile, due to very low line optical depths. {\it
If} this abrupt decrease would not happen, the monotonic behavior of the
strength of the emission peak (Fig.~\ref{comphabra}) would no longer be
warranted for mass-loss rates at the upper end of the scale considered here,
and an important aspect of its diagnostic potential would be lost.

Let us first concentrate on the conditions in the {\it outer} wind, where
the ground-state has a major impact. We stress again that we
are dealing here with (very) weak winds, i.e., the continuum-edges are formed
deep in the photosphere, whereas the wind and the transition region are
already optically thin. Otherwise, we could no longer assume a ``given''
radiation temperature in the continuum, but would have to account for a
simultaneous solution of radiation field and occupation numbers, as it was
done, e.g., to explain the ground-state depopulation of \HeII\ in {\it
dense} hot star winds by \citet{gabler89}.

Within our assumptions of ionizations to the ground-state of the next higher
ion only and neglecting collisional processes, we obtain an alternative
formulation of the rate equation for the ground-state, by summing up the rate
equations for {\it all} levels $i$ (Eq.~\ref{rate1}),
\beq 
n_1 R_{1k} + \sum_{i>1}  n_i R_{ik} = n_1^* R_{k1} + \sum_{i>1} n_i^* R_{ki}, 
\label{gs}
\eeq 
since the line contributions cancel out. Solving for the ground-state
departure coefficient, we find
\beq
b_1= \frac{\sum_{i > 1} n_i^* (R_{ki} -b_i R_{ik})/n_1^*\, +\, R_{k1}}
{R_{1k}},
\eeq 
which can be approximated by the well known expression
\beq
b_1 \approx  \frac{1}{W} \frac{T_{\rm e}}{T_{\rm rad}} 
\exp \bigl[-\frac{h\nu_0}{k}\bigl(\frac{1}{T_{\rm e}}-\frac{1}{T_{\rm rad}}\bigr)\bigr]
\frac{1}{\rm corr.fac.},
\label{bgs}
\eeq
where $T_{\rm rad}$ is the radiation temperature in the ground-state (=
Ly\-man) continuum, $\nu \ge \nu_0$, and $W$ the dilution factor. A correction
factor of order unity accounts for the ionization/recombination from the
excited levels (for details, see, e.g., \citealt{puls05}).

From here on, we have to divide between line-blocked and unblocked (e.g.,
pure H/He-) models, as they behave different in the wind (though similar in
the outer photosphere), due to a considerably different run of electron
temperature and radiation-field on both sides of the Lyman edge.

For non-blocked models (see Fig.~\ref{nebula}), the wind-temperature is not
too different from $T_{\rm rad}$, and $b_1$ becomes strongly overpopulated
$\propto 1/W = r^2$. Moreover, all net radiative brackets coupled to the
ground-state,
\beqa
Z_{j1} & \approx& 1-W \frac{b_1}{b_j}
\exp\bigl[\frac{h\nu_{j1}}{k}\bigl(\frac{1}{T_{\rm e}}-\frac{1}{T_{\rm
rad,j1}}\bigr)\bigr] \nonumber \\
& \approx& 1- \frac{T_{\rm e}}{T_{\rm rad}}
\exp\bigl[\frac{h\nu_{0}}{k}\bigl(\frac{1}{T_{\rm rad}}-\frac{1}{T_{\rm
rad,j1}}\bigr)\bigr]\frac{1}{b_j \, \rm{corr.fac.}}
\eeqa   
become strongly negative, since (i) $b_1$ is severely overpopulated and (ii) the
Doppler effect in the wind allows for an illumination by the continuum
bluewards from the resonance-line rest-frame frequencies $\nu_{j1}$, i.e.,
$\bar J \approx W B_{\nu} (T_{\rm rad,j1})$ (optically thin
(Sobolev-)approximation), with $T_{\rm rad,j1} \gg T_{\rm rad}$ due to
missing line-blocking.

The consequence for the population of the excited levels
(Eq.~\ref{depart_excited}) is twofold. Because of the strong pumping by the
resonance lines, the (normalized) population of the higher levels
($n_j/n_i^*, j>i$) is much larger than in the photosphere, and the line term
becomes larger than the recombination term. Second, the denominator
decreases significantly, due to the direct effect of $Z_{j1}$ and since the
the ionization rates $\propto W$ become negligible.

In total, now the first term dominates in Eq.~\ref{2comp}, and the situation
is just opposite to the conditions in the outer photosphere: The lower the
considered level $i$, the larger is the nominator and the smaller the
denominator, such that we obtain the sequence $b_2 > b_3 > b_4{\ldots} > 1$
(cf. Fig.~\ref{nebula}.)

For line-blocked models, on the other hand, the cooling by the enormous number of
metallic lines leads to a strong decrease of the electron temperature in the
outer wind, and $T_{\rm e}$ becomes much smaller than the radiation
temperature in the Lyman continuum (for our late O-type model, 10,000~K vs.
25,000~K). In this case, ionization, though diluted, outweighs recombination
(the exponential term in Eq.~\ref{bgs}),
and the ground-state even becomes underpopulated ($b_1$ \rarrow 0.5).
Consequently, the resonance lines can no longer pump the excited levels
(even more, since for blocked models the radiation temperatures close to the
resonance lines, $T_{\rm rad,j1}$, are much smaller than in the unblocked
case). Thus, we find a situation similar to that in the outer photosphere,
namely that the 2nd term in Eq.~\ref{2comp} is the decisive one, and $n_5 >
n_4$, which is obvious also from the final increase of the line-source
function for \Bra\ in Fig.~\ref{comphabra} for all mass-loss rates
considered.

Finally, in the region between the outer photosphere and the outer wind, 
the dilution of the radiation field is faster or similar to the decrease of
$T_{\rm e}$, both for the blocked and the unblocked models. Thus,
the departure coefficients of level 4 and 5 increase in this region (due to an
overpopulated ground-state, and effective pumping due to the onset of the
Doppler-shift), though at a rather similar rate, with $b_4 \ga
b_5$. Consequently, the source-function approaches the LTE level, which
explains its abrupt decrease in the transition region.

\end{document}